\documentclass[ALICE,manyauthors]{cernphprep}
\usepackage[comma,square,numbers,sort&compress]{natbib}
\usepackage{hyperref}
\usepackage{lineno}
\usepackage{xspace}
\usepackage{nicefrac}
\usepackage{physics}
\usepackage{cleveref}

\usepackage[T1]{fontenc}
\usepackage{orcidlink} 

\crefname{figure}{Fig.}{Figs.}
\Crefname{figure}{Figure}{Figures}
\crefname{section}{Sec.}{Secs.}
\Crefname{section}{Section}{Sections}
\crefname{table}{Tab.}{Tabs.}
\Crefname{table}{Table}{Tables}
\crefname{equation}{Eq.}{Eqs.}
\Crefname{equation}{Equation}{Equations}

\usepackage{multirow}
\usepackage{graphicx}

\usepackage{siunitx}
\DeclareSIUnit\gevc{\GeV\per\clight}
\DeclareSIUnit\gevcc{\GeV\per\square\clight}
\DeclareSIUnit\mevc{\MeV\per\clight}
\DeclareSIUnit\clight{\text{\ensuremath{c}}}

\begin{document}
%

\newcommand{\pp}           {pp\xspace}
\newcommand{\ppbar}        {\mbox{$\mathrm {p\overline{p}}$}\xspace}
\newcommand{\XeXe}         {\mbox{Xe--Xe}\xspace}
\newcommand{\PbPb}         {\mbox{Pb--Pb}\xspace}
\renewcommand{\pA}           {\mbox{pA}\xspace}
\newcommand{\pPb}          {\mbox{p--Pb}\xspace}
\newcommand{\AuAu}         {\mbox{Au--Au}\xspace}
\newcommand{\dAu}          {\mbox{d--Au}\xspace}

\newcommand{\snn}          {\ensuremath{\sqrt{s_{\mathrm{NN}}}}\xspace}
\newcommand{\pt}           {\ensuremath{p_{\rm T}}\xspace}
\newcommand{\meanpt}       {$\langle p_{\mathrm{T}}\rangle$\xspace}
\newcommand{\ycms}         {\ensuremath{y_{\rm CMS}}\xspace}
\newcommand{\ylab}         {\ensuremath{y_{\rm lab}}\xspace}
\newcommand{\etarange}[1]  {\mbox{$\left | \eta \right |~<~#1$}}
\newcommand{\yrange}[1]    {\mbox{$\left | y \right |~<~#1$}}
\newcommand{\dndy}         {\ensuremath{\mathrm{d}N_\mathrm{ch}/\mathrm{d}y}\xspace}
\newcommand{\dndeta}       {\ensuremath{\mathrm{d}N_\mathrm{ch}/\mathrm{d}\eta}\xspace}
\newcommand{\avdndeta}     {\ensuremath{\langle\dndeta\rangle}\xspace}
\newcommand{\dNdy}         {\ensuremath{\mathrm{d}N_\mathrm{ch}/\mathrm{d}y}\xspace}
\newcommand{\Npart}        {\ensuremath{N_\mathrm{part}}\xspace}
\newcommand{\Ncoll}        {\ensuremath{N_\mathrm{coll}}\xspace}
\newcommand{\dEdx}         {\ensuremath{\textrm{d}E/\textrm{d}x}\xspace}
\newcommand{\RpPb}         {\ensuremath{R_{\rm pPb}}\xspace}

\newcommand{\nineH}        {$\sqrt{s}~=~0.9$~Te\kern-.1emV\xspace}
\newcommand{\seven}        {$\sqrt{s}~=~7$~Te\kern-.1emV\xspace}
\newcommand{\onethreesix}  {$\sqrt{s}~=~13.6$~Te\kern-.1emV\xspace}
\newcommand{\onethree}  {$\sqrt{s}~=~13$~Te\kern-.1emV\xspace}
\newcommand{\twoH}         {$\sqrt{s}~=~0.2$~Te\kern-.1emV\xspace}
\newcommand{\twosevensix}  {$\sqrt{s}~=~2.76$~Te\kern-.1emV\xspace}
\newcommand{\five}         {$\sqrt{s}~=~5.02$~Te\kern-.1emV\xspace}
\newcommand{\twosevensixnn}{$\sqrt{s_{\mathrm{NN}}}~=~2.76$~Te\kern-.1emV\xspace}
\newcommand{\fivenn}       {$\sqrt{s_{\mathrm{NN}}}~=~5.02$~Te\kern-.1emV\xspace}
\newcommand{\LT}           {L{\'e}vy-Tsallis\xspace}
\newcommand{\GeVc}         {\ensuremath{\mathrm{Ge\kern-.1emV}/c}\xspace}
\newcommand{\MeVc}         {\ensuremath{\mathrm{Me\kern-.1emV/}c}\xspace}
\renewcommand{\TeV}          {Te\kern-.1emV\xspace}
\renewcommand{\GeV}          {Ge\kern-.1emV\xspace}
\renewcommand{\MeV}          {Me\kern-.1emV\xspace}
\newcommand{\GeVmass}        {\ensuremath{\mathrm{Ge\kern-.1emV}/c^2}\xspace}
\newcommand{\MeVmass}        {\ensuremath{\mathrm{Me\kern-.1emV/}c^2}\xspace}
\newcommand{\lumi}         {\ensuremath{\mathcal{L}}\xspace}

\newcommand{\ITS}          {\rm{ITS}\xspace}
\newcommand{\FIT}          {\rm{FIT}\xspace}
\newcommand{\TOF}          {\rm{TOF}\xspace}
\newcommand{\ZDC}          {\rm{ZDC}\xspace}
\newcommand{\ZDCs}         {\rm{ZDCs}\xspace}
\newcommand{\ZNA}          {\rm{ZNA}\xspace}
\newcommand{\ZNC}          {\rm{ZNC}\xspace}
\newcommand{\SPD}          {\rm{SPD}\xspace}
\newcommand{\SDD}          {\rm{SDD}\xspace}
\newcommand{\SSD}          {\rm{SSD}\xspace}
\newcommand{\TPC}          {\rm{TPC}\xspace}
\newcommand{\TRD}          {\rm{TRD}\xspace}
\newcommand{\VZERO}        {\rm{V0}\xspace}
\newcommand{\VZEROA}       {\rm{V0A}\xspace}
\newcommand{\VZEROC}       {\rm{V0C}\xspace}
\newcommand{\Vdecay} 	   {\ensuremath{V^{0}}\xspace}

\newcommand{\ee}           {\ensuremath{e^{+}e^{-}}}
\newcommand{\pip}          {\ensuremath{\pi^{+}}\xspace}
\newcommand{\pim}          {\ensuremath{\pi^{-}}\xspace}
\newcommand{\kap}          {\ensuremath{\rm{K}^{+}}\xspace}
\newcommand{\kam}          {\ensuremath{\rm{K}^{-}}\xspace}
\newcommand{\pbar}         {\ensuremath{\rm\overline{p}}\xspace}
\newcommand{\kzero}        {\ensuremath{{\rm K}^{0}_{\rm{S}}}\xspace}
\newcommand{\lmb}          {\ensuremath{\Lambda}\xspace}
\newcommand{\almb}         {\ensuremath{\overline{\Lambda}}\xspace}
\newcommand{\Om}           {\ensuremath{\Omega^-}\xspace}
\newcommand{\Mo}           {\ensuremath{\overline{\Omega}^+}\xspace}
\newcommand{\X}            {\ensuremath{\Xi^-}\xspace}
\newcommand{\Ix}           {\ensuremath{\overline{\Xi}^+}\xspace}
\newcommand{\Xis}          {\ensuremath{\Xi^{\pm}}\xspace}
\newcommand{\Oms}          {\ensuremath{\Omega^{\pm}}\xspace}
\renewcommand{\degree}       {\ensuremath{^{\rm o}}\xspace}

\newcommand{\pP}{\ensuremath{\mbox{p--p}}\xspace}
\newcommand{\ApAp} {\ensuremath{\mbox{\ensuremath{\overline{\mbox{p}}}--\ensuremath{\overline{\mbox{p}}}}}\xspace}
\newcommand{\pL}{\ensuremath{\mbox{p--$\Lambda$}}\xspace}
\newcommand{\ApAL}{\ensuremath{\mbox{\overline{p}--\overline{\Lambda}}}\xspace}
\newcommand{\pAL}{\ensuremath{\mbox{p--$\overline{\Lambda} $}}\xspace}
\newcommand{\pSizero}{\ensuremath{\mbox{p--$\Sigma^0$}}\xspace}
\newcommand{\pSiplus}{\ensuremath{\mbox{p--$\Sigma^{+}$}}\xspace}
\newcommand{\pXim}{\ensuremath{\mbox{p--$\Xi^{-}$}}\xspace}
\newcommand{\pXip}{\ensuremath{\overline{\mathrm{p}}$\mbox{--}$\xip}\xspace}
\newcommand{\pXi}{\ensuremath{\mbox{p--$\Xi$}}\xspace}
\newcommand{\pipi}{\ensuremath{\mbox{$\pi$--$\pi$}}\xspace}
\newcommand{\ppi}{\ensuremath{\mbox{p--$\pi^\pm$}}\xspace}
\newcommand{\Kp}{\ensuremath{\mbox{K--p}}\xspace}

\renewcommand{\mT}{\ensuremath{m_\mathrm{T}}\xspace}
\newcommand{\pT}{\ensuremath{p_\mathrm{T}}\xspace}
\newcommand{\kstar}{\ensuremath{k^{*}}\xspace}
\newcommand{\source}{\ensuremath{S(r^*)}\xspace}
\newcommand{\wavefunction}{\ensuremath{\psi(k^*, \Vec{r}^*)}\xspace}
\newcommand{\reff}{\ensuremath{r_{\mathrm{0}}}\xspace}
\newcommand{\rcore}{\ensuremath{r_{\mathrm{core}}}\xspace}

\newcommand{\plmb}  {\ensuremath{\mathrm{p}_\lmb}\xspace}
\newcommand{\psig}  {\ensuremath{\mathrm{p}_{\Sigma^{+}}}\xspace}

\newcommand{\ck}    {\ensuremath{C(k^*)}\xspace}
\newcommand{\cbk}   {\ensuremath{C(\mathbf{k}^*)}\xspace}

\newcommand{\sr}    {\ensuremath{S(r^*)}\xspace}
\newcommand{\sbr}   {\ensuremath{S(\mathbf{r}^*)}\xspace}

\newcommand{\p}     {\ensuremath{\mathrm{p}}\xspace}
\newcommand{\bp}    {\ensuremath{\mathrm{\textbf{p}}}\xspace}
\newcommand{\bk}    {\ensuremath{\mathrm{\textbf{k}}}\xspace}
\newcommand{\br}    {\ensuremath{\mathrm{\textbf{r}}}\xspace}
\newcommand{\fp}    {\ensuremath{\Tilde{\mathrm{p}}}\xspace}

\newcommand{\re}    {\ensuremath{r_\mathrm{eff}}\xspace}
\newcommand{\rc}    {\ensuremath{r_\mathrm{core}}\xspace}

\newcommand{\avgm}  {\ensuremath{\langle m_\mathrm{res}^\mathrm{eff}\rangle} \xspace}
\newcommand{\avgt}  {\ensuremath{\langle c\tau_\mathrm{res}^\mathrm{eff}\rangle} \xspace}

\newcommand{\avgdndeta}{\ensuremath{\expval{\dd N_{\mathrm{ch}} / \dd\eta}_{\abs{\eta}<0.5}}\xspace}

\newcommand{\avgdndetaMBzeroTen}{\ensuremath{17.42\pm0.29}}
\newcommand{\avgdndetaMBtenFifty}{\ensuremath{10.15\pm0.17}}
\newcommand{\avgdndetaMBfiftyHundred}{\ensuremath{4.48\pm0.09}}
\newcommand{\avgdndetaMBinclusive}{\ensuremath{7.10\pm0.18}}

\newcommand{\avgdndetaPairzeroTen}{\ensuremath{23.95\pm0.34}}
\newcommand{\avgdndetaPairtenFifty}{\ensuremath{16.72\pm0.24}}
\newcommand{\avgdndetaPairfiftyHundred}{\ensuremath{9.73\pm0.14}}
\newcommand{\avgdndetaPairinclusive}{\ensuremath{17.37\pm0.25}}


\begin{titlepage}
	\PHyear{2026}       
	\PHnumber{180}      
	\PHdate{18 June}  

	\title{Multiplicity dependence of the size of the common hadron emission source in pp collisions at the LHC
	}
	\ShortTitle{Multiplicity Dependence of the Hadron Emission Source}   

	\Collaboration{ALICE Collaboration\thanks{See Appendix~\ref{app:collab} for the list of collaboration members}}
	\ShortAuthor{ALICE Collaboration} 

	\begin{abstract}
		Femtoscopic analysis can shed light on hadron production in \pp collisions. In this paper, proton-proton correlations measured in collisions at \onethreesix recorded with the ALICE detector at the LHC are presented.
		The analysis is based on the minimum bias dataset collected in 2022 following the upgrade of the ALICE detector and corresponds to an integrated luminosity of $\SI{19.3}{pb^{-1}}$. The
		increased integrated luminosity allows us, for the first time, to simultaneously measure the multiplicity and transverse-mass (\mT) dependence of the size of the hadron-emitting source. Precise knowledge of the femtoscopic source size in \pp collisions is a crucial ingredient for using femtoscopy to study the residual strong interaction among stable and unstable hadrons at the LHC.
		In this light, the source radius was determined from the measured correlation functions by assuming several state-of-the-art models of the nucleon--nucleon interactions.
		The consistency among the extracted radii demonstrates the robustness of the measurement with respect to interaction model assumptions. A comparison to femtoscopic radii measured in Pb--Pb collisions at \five reveals a markedly different multiplicity dependence in similar \mT intervals, providing new insight into the system-size dependence of particle emission dynamics.

	\end{abstract}
\end{titlepage}

\setcounter{page}{2} 


\section{Introduction}
In recent years, femtoscopy has been successfully used to perform precision studies of the strong interaction among hadrons \cite{FemtoReview}.
Ultra-relativistic \pp collisions, in particular, are a very suitable environment for such studies because of the small system size, and therefore the small relative distances at which the particle pairs are emitted ($\sim1.5-2$ fm). As a result, the measured correlations are sensitive to the short-range component of the interaction.
Especially in the strangeness sector, this method delivered high-precision results which allowed us to benchmark Lattice QCD calculations for the first time \cite{p-Omega_nature, Lambda-Xi_plb}, and access inelastic channels of the nucleon--hyperon interactions \cite{NLambda-NSigma_cc_plb}. Recently, interactions in the charm sector have been measured for the first time \cite{p-D_prd, pi_K_charm_prd}.

The success of studying hadronic interactions with femtoscopy relies on the precise control of the source function \source, which provides the probability density distribution of the distance $r^{*}$ in the pair rest frame at which hadrons are produced in collisions \cite{KOONIN197743}.
Several studies have been carried out in the past exploiting \pipi correlations measured in pp or p--Pb collisions at the LHC by the ALICE~\cite{ALICE_pipi_run1}, ATLAS~\cite{ATLAS:2017shk,ATLAS:2022wvk}, CMS~\cite{CMS:2019fur} and LHCb~\cite{LHCb:2017pnz} Collaborations to characterize the particle emission source in small colliding systems.
Similar measurements were also performed at RHIC by the STAR Collaboration, which investigated pion femtoscopy in pp and heavy-ion collisions over a broad range of collision energies \cite{STAR_pp_HBT_2011,STAR_Kaon_HBT_2013}.
Such studies employed analytic forms like the Cauchy and Lévy distributions \cite{LevySource}, whose long tails can effectively absorb contributions from strongly decaying resonances, but without precise control over them. As a result, while these functions reproduce the data, they do not allow one to isolate the resonance contribution or assess the existence of a common hadron emission source.

Studies carried out by the ALICE Collaboration in pp collisions first employing the femtoscopy analysis of baryon--baryon~\cite{CommonSource_pp,Erratum_CommonSourceRun2}, and then also extending to \pipi, \Kp~\cite{CommonSource_2023} and \ppi pairs~\cite{pi-p_analysis}, found evidence for a common emission source for all hadron pairs, if the contribution of strongly decaying resonances is correctly taken into account.
In these studies, the primordial source is described by a Gaussian profile, and it is found that the source size decreases with the increasing average transverse mass $m_\mathrm{T}$ of the pair~\cite{CommonSource_pp, CommonSource_2023, Erratum_CommonSourceRun2}.
The latter is defined as $\mT = \sqrt{\vec{k}_\mathrm{T}^2+\left<m\right>^2}$, where $\vec{k}_{\mathrm{T}}=\frac{1}{2}(\vec{p}_{\mathrm{T,1}}+\vec{p}_{\mathrm{T,2}})$ the average transverse momentum and $\left<m\right> = \frac{m_1+m_2}{2}$ the average mass of the pair.
Also in heavy-ion collisions at RHIC and LHC, a \mT dependence of the source size has been observed, albeit different for different particle species~\cite{ALICE:2011dyt, CMS:2023xyd}. In these systems, this scaling is generally interpreted as evidence of collective effects such as radial flow~\cite{Pratt:2008qv, Kisiel_mT_RHIC}, and it is well reproduced by event generators that include such phenomena. Conversely, in \pp collisions, the origin of the \mT scaling is still under debate; there is no consensus on whether collective effects are responsible for this effect or not.
On the one hand, event generators do not quantitatively reproduce the \mT scaling observed in \pp collisions \cite{WignerDeuteron, CommonSource_2023}.
On the other hand, first phenomenological models emerged, which parametrize the observed \mT scaling~\cite{CECApaper,Wang2025}.
Both highlight the importance of precise and multi-differential measurements of source properties in small systems to better understand whether collective behavior plays a role.

This paper presents a measurement of the femtoscopic source size in \pp collisions using \pP and \ApAp pairs, differential for the first time in both \mT{} and event multiplicity, extending previous studies \cite{CommonSource_pp, Erratum_CommonSourceRun2} that were differential only in \mT{}.
This is possible thanks to the upgrade of the ALICE detector \cite{ALICETPC:2020ann, Reidt:2021tvq} and the switch to a continuous readout \cite{ALICE_LS2:2023}, which enabled the collection of an extensive Minimum-Bias (MB) dataset in the year 2022 with an integrated luminosity of \SI{19.3}{pb^{-1}}.
For the first time, this measurement extends into the low-multiplicity regime, reaching events with an average of \avgdndetaPairfiftyHundred{} charged tracks in $|\eta|<0.5$, compared to roughly thirty charged particles in previous studies~\cite{CommonSource_pp}.
In addition, we investigate the robustness of the measured radii with respect to different state-of-the-art nucleon--nucleon interaction models.

These results provide necessary input for coalescence models \cite{WignerDeuteron,ToMCCA,ToMCCA3}, which estimate (anti)nuclei yields by evaluating intra-nucleon distances via wave function overlaps. Such models are crucial for predicting (anti)nuclei fluxes in cosmic-ray interactions with the interstellar medium, where most (anti)nuclei are produced at low collision energies $(\approx\SI{20}{\GeV})$~\cite{Serksnyte:2022onw}, and for refining background estimates in dark matter searches \cite{ALICE:2022zuz}. The access to low multiplicities is particularly relevant in this context, as it corresponds to the conditions under which these processes occur. More broadly, the multiplicity-differential nature of this measurement bridges the femtoscopic source size across \pp and \PbPb systems, providing crucial input for future femtoscopic studies.
\section{Data Analysis}
\label{sec:dataAnalysis}

This paper presents the measurement of the \pP and \ApAp correlation functions in MB \pp collisions at \onethreesix, performed with the upgraded ALICE detector~\cite{ALICE_LS2:2023}.
The correlation function $\ck$ is an observable that depends solely on the relative momentum \kstar of the two particles in their pair rest frame. Here, $k^*$ is defined as $k^* = \frac{1}{2} \left| \bp^*_1 - \bp^*_2 \right|$, where $\bp^*_1$ and $\bp^*_2$ are the particle momenta in the pair rest frame.
The correlation function provides information about both the interaction of the two particles and the space--time characteristics of the particle-emitting source.
The following ALICE subdetectors were used in this analysis: the FT0 component of the Fast Interaction Trigger (\FIT)~\cite{ALICE_FIT}, the Inner Tracking System (\ITS)~\cite{ALICE_ITS}, the Time Projection Chamber (\TPC)~\cite{ALICE_TPC}, and the Time-Of-Flight (\TOF) detector~\cite{ALICE_TOF1,ALICE_TOF2}.
Minimum-bias events were selected based on timing signals from the FT0 detector, which consists of two modules (FT0-A and FT0-C) located on opposite sides of the interaction point, covering the pseudorapidity ranges $3.5 < \eta < 4.9$ and $-3.3 < \eta < -2.1$, respectively.
The MB selection required coincident signals in FT0-A and FT0-C consistent with beam--beam interaction timing, and a reconstructed primary vertex in the central barrel with at least two associated tracks. The summed amplitudes from FT0-A and FT0-C were additionally used to classify events into forward multiplicity classes.

Using the total FT0 signal, three multiplicity classes were defined: 0--10\% (highest multiplicity), 10--50\%, and 50--100\% (lowest multiplicity). These intervals correspond to the respective percentiles of the FT0 amplitude distribution and thus reflect different levels of event activity. Charged-particle multiplicities at midrapidity were then estimated using tracking information from the ITS and TPC. The measured charged-track densities at $\eta=0$, within a unit pseudorapidity interval, were corrected for detector acceptance and efficiency using correction factors derived from Monte Carlo simulations by comparing reconstructed and generated track multiplicities. These corrected values provide the average charged-particle density, \avgdndeta, for each multiplicity class. The \avgdndeta\ values for the minimum bias event samples are \avgdndetaMBzeroTen, \avgdndetaMBtenFifty, and \avgdndetaMBfiftyHundred\ for the 0--10\%, 10--50\%, and 50--100\% classes, respectively. For the full minimum bias sample (0--100\%), the inclusive average charged-track multiplicity is \avgdndetaMBinclusive. Since the analysis focuses on two-particle correlations between (anti)protons, only events containing at least one reconstructed (anti)proton--(anti)proton pair can contribute, which introduces a bias towards higher charged-particle multiplicities. This is particularly relevant in \pp\ collisions with the MB selection, where the overall multiplicity is generally low and a significant fraction of events does not contain two (anti)protons. The resulting \avgdndeta\ values for events satisfying this requirement are \avgdndetaPairzeroTen, \avgdndetaPairtenFifty, and \avgdndetaPairfiftyHundred\ for the 0--10\%, 10--50\%, and 50--100\% classes, respectively, and \avgdndetaPairinclusive\ for the full minimum bias sample, reflecting the enhanced event activity imposed by the pair selection.

The tracking of charged particles and the reconstruction of the primary vertex (PV) is done by combining the track information from the \ITS and \TPC, both located inside a uniform magnetic field directed along the beam axis.
The quality of the tracks used for the construction of the \pP and \ApAp correlation functions is ensured by requiring that each track lies within the pseudorapidity range of $\abs{\eta}<0.8$. To ensure a uniform detector coverage, only events with a PV within $\pm\SI{10}{\cm}$ from the nominal interaction point are selected. A minimum of 80 clusters in the \TPC assigned to each track is required as well.
Additionally, only tracks that cross at least 70 out of the available 152 rows of the \TPC readout pads are selected.
Particle identification (PID) for the protons is performed employing the \TPC and \TOF detectors. The specific energy loss, $\dd E/\dd x$, of charged particles traversing the \TPC{} gas is described by the Bethe--Bloch parametrization. The deviation of the measured signal from the expected mean $\expval{\dd E / \dd x}$ is expressed as $\textrm{n}_{\sigma,\TPC}$, where n is the number of standard deviations based on the \TPC{} energy resolution. In this analysis, tracks with momentum below \SI{0.75}{\gevc} are accepted if they satisfy the condition $\abs{\textrm{n}_{\sigma,\TPC}} < 3$.
At higher momenta, the proton energy loss in the \TPC begins to overlap with that of other particle species; therefore, \TOF information is used in addition. The difference between the measured and expected time-of-flight (assuming the proton mass) is expressed in units of the time resolution as $\textrm{n}_{\sigma,\TOF}$. The \TPC and \TOF signals are combined into a single PID variable defined as $\textrm{n}_{\sigma,\text{combined}} = \sqrt{\textrm{n}_{\sigma,\TPC}^2 + \textrm{n}_{\sigma,\TOF}^2}$.
Tracks are accepted if they satisfy $\textrm{n}_{\sigma,\text{combined}} < 3$.
These selections are applied to enhance the purity, which shows a dependence on the transverse momentum \pT of the tracks. The protons are required to have a transverse momentum \pT between \SI{0.5}{\gevc} and \SI{2.2}{\gevc}, since it was observed that in this interval the purity is larger than 80\%. The \pT dependence of the purity leads to a \kstar dependence of the corrections, which will be discussed further in \cref{sec:CF_modeling}, the averaged purity is approximately 96\%.

For femtoscopic studies, only primary particles---those that participate in final-state interactions---are of interest. Therefore, contamination from secondary particles, originating from weak decays or interactions with detector material, must be suppressed. This is achieved by applying a strict selection on the track’s distance of closest approach (DCA) to the primary vertex.
For the DCA in the transverse plane ($\text{DCA}_{xy}$), a momentum-dependent selection is applied: $\abs{\text{DCA}_{xy}} < 0.0105 + 0.035 \cdot \left(p_{\mathrm{T}}/\SI{1}{\gevc}\right)^{-1.1} \, \si{\cm}$.
The absolute value of the DCA along the beam direction ($\text{DCA}_{z}$) must be less than \SI{0.1}{\cm}.
With these selections, the momentum-averaged fraction of primary protons reaches 88\%, while the contribution from secondaries from weak decays is approximately 4\%.
The remaining fraction consists of protons produced in interactions with detector material or incorrectly associated with the primary collision. Also the fractions show a \pT dependence which will be considered in the \kstar dependent corrections together with the purity in \cref{sec:CF_modeling}.

The systematic uncertainties of the correlation function are driven by the proton selection criteria and are estimated by tightening and relaxing the aforementioned selections and randomly combining them into 44 sets of variations. The varied selections are the minimum \pT, the number of TPC clusters and shared clusters, the $\eta$ range, and the $n_\sigma$ for the PID selections. To ensure that the observed deviations are not driven by statistical fluctuations, only those variations for which the pair yield in the signal region (i.e.  $\kstar < 200$ MeV/c) differs by less than 20\% from the default selection are retained. The magnitude of the systematic uncertainty is then evaluated as the root-mean-square deviation in each interval, assuming an underlying uniform distribution of the accepted variations.
\section{Data Modeling}
Experimentally, the correlation function is measured as the ratio
of correlated to uncorrelated pair distributions \cite{LisaReview}
\begin{equation}
	\ck = \mathcal{N} \frac{A(k^*)}{B(k^*)},
	\label{eq:ck_norm}
\end{equation}
where $\mathcal{N}$ is a normalization factor whose value is obtained in this analysis by forcing \ck to unity for $k^* \in [240, 340]$ \MeVc, where the femtoscopic signal is expected
to vanish~\cite{CommonSource_pp, FemtoReview}. The numerator $A(k^*)$ is the pair momentum distribution of particles from the same event (SE) and the denominator $B(k^*)$ the distribution of particles from different (mixed) events (ME).
The mixed-event distribution $B(k^*)$ is constructed by pairing particles from different events to approximate the underlying phase space.
To ensure that the mixed events reflect similar event characteristics and thus provide a meaningful baseline, events are grouped into intervals based on shared properties such as primary vertex position, multiplicity class, and number of charged tracks used for the vertex reconstruction.
Within each of these pairing classes, tracks from different events are combined to form mixed pairs, preserving the global event features while removing genuine correlations present in the same-event pairs.
This procedure effectively models the distribution expected in the absence of femtoscopic correlations and allows for an accurate extraction of the correlation function.

Theoretically, the correlation function is given by the Koonin--Pratt equation and is defined as the integral of the product of the source function \sbr and the wave function of a particle pair $\Psi(\br^*,\bk^*)$ \cite{LisaReview}
\begin{equation}
	\cbk = \int \mathrm{d}^3r^* \ S(\br^*) \ | \Psi(\br^*, \bk^*) |^2 ,
	\label{eq:ck_int}
\end{equation}
where \sbr may be understood as an effective parametrization of the properties of the particle source in terms of the relative distance $\br^*$ of the particle pair, and $\Psi(\br^*, \bk^*)$ includes the interaction term, therefore determining the shape of the correlation function.

\subsection{Modeling the Source Function}
In femtoscopic analyses, the spatial distribution of particle emission is often described using a source parametrization. A single Gaussian source provides an effective description of the emission region with one characteristic length scale, while more refined models account explicitly for contributions from short-lived resonances \cite{ALICE:2018ysd}.
Assuming an isotropic pair emission, the effective source function can be modeled with a spherical symmetric Gaussian profile
\begin{equation}
	S(r^*) = \frac{1}{(4\pi \reff^2)^{3/2}} \mathrm{exp} \left( - \frac{r^{*2}}{4\reff^2} \right),
	\label{eq:source_gauss}
\end{equation}
where the only parameter is the effective source size \reff.
However, this simplified description of the source does not fully capture the contribution of short-lived resonances to the particle yield. These resonances can decay into the particles of interest with decay length similar to the source size ($c\tau \sim$ 1--2~fm) and thus effectively enlarge the primordial source.
This is addressed by applying the Resonance Source Model (RSM) \cite{CommonSource_pp,Erratum_CommonSourceRun2}. It assumes an isotropic emission of primordial pairs, modeled by the Gaussian profile defined in \cref{eq:source_gauss} with the size parameter \rcore instead of \reff, and a halo produced by the decay of the resonances effectively leading to exponential tails in the source function.

The RSM requires the specification of two main components, the primordial yields and the decay kinematics of the resonances. The primordial yields are obtained from the statistical hadronization model (SHM), as implemented in the Thermal-FIST package \cite{thermalFist, shm_paper}, which returns a primordial fraction for protons of 35.8\% \cite{CommonSource_pp}. It is assumed that the primordial fraction does not depend on the multiplicity.
The resonance decay kinematics are derived from the EPOS model by analyzing the angular
distribution of daughter particles originating from strongly decaying baryonic resonances.
Rather than modeling the decays of all contributing resonances individually, resonances with similar masses and lifetimes (such as the $\Delta$ and $N^{*}$ states) are grouped into one effective resonance. The resulting effective resonance mass, $\langle m_{\mathrm{res}}^{\mathrm{eff}} \rangle = 1.36~\GeVmass$, and effective lifetime, $\langle c\tau_{\mathrm{res}}^{\mathrm{eff}} \rangle = 1.65~\mathrm{fm}$, characterize the average decay kinematics of this resonance ensemble.
This approximation is justified by the similarity of the resonance masses, which leads to only weak variations in the decay kinematics.
The impact of this simplification is accounted for by varying the effective resonance mass and lifetime by $\pm 10\%$, and the resulting variation is included in the
systematic uncertainties~\cite{CommonSource_pp, EPOS}.
Both approaches are employed in this work: the RSM serves as the baseline source measurement for subsequent femtoscopic analyses, while the effective Gaussian parametrization offers a practical modeling framework.

\subsection{Modeling the Correlation Function}
\label{sec:CF_modeling}
The theoretical correlation function requires careful consideration of the underlying forces and the different contributions to the desired particle system.
The (anti)protons, carrying the same charge, repel each other via the Coulomb force. Furthermore, since they are both spin-1/2 particles, they experience Pauli blocking at short distances. Additionally, the proton--proton strong interaction has a repulsive core, which, together with the other two repulsive contributions, leads to a depletion of the proton--proton correlation function at \kstar below \SI{10}{\MeVc}.
The attractive part of the strong interaction, on the other hand, leads to the enhancement of the correlation function above unity in the intermediate \kstar range ($\SI{10}{\MeVc} < \kstar < \SI{100}{\MeVc}$) \cite{CommonSource_pp}.

In this work, the strong interaction is modeled using the Norfolk potential~\cite{Norfolk_01,Norfolk_02}. Its systematically varied versions (see \cref{sec:norfolk_details}) enable a controlled assessment of theoretical uncertainties. A detailed comparison with other modern nucleon--nucleon potentials is presented in \cref{fitpotentials}, highlighting the robustness of the interpretation of our experimental data independently of the choice of the interaction model.

Apart from the genuine $\mathrm{p\text{--}p}$ interaction, the femtoscopic signal can include additional contributions, where the protons are either decay daughters of weakly decaying particles or misidentified particles of another species.
The former cases are referred to as feed-down contributions because the interaction of the mother particle is inherited by the daughter proton and the signal of that interaction contributes to the measured correlation function, albeit smeared due to the boost into the pair rest frame of the two protons. The relevant femtoscopic contributions for the studied system are calculated following the procedure of Ref. \cite{res_contribution},
\begin{equation}
	\begin{aligned}
		\{ \mathrm{p\text{--}p} \} = & ~ \p\text{--}\p + \p\text{--}\plmb + \plmb\text{--}\plmb + \p\text{--}\psig + \psig\text{--}\psig \\
		                             & + \plmb\text{--}\psig + \fp\text{--}\p + \fp\text{--}\plmb + \fp\text{--}\psig + \fp\text{--}\fp.
	\end{aligned}
	\label{eq:pp_contr}
\end{equation}
where $\mathrm{p}$ denotes genuine protons, $\Tilde{\mathrm{p}}$ denotes misidentified protons, $\mathrm{p}_\Lambda$ denotes protons that originate from $\Lambda$ feed-down, and $\mathrm{p}_{\Sigma^+}$ denotes protons that originate from $\Sigma^+$ feed-down, while each pair represents the corresponding correlation.
However, only the shape of the genuine correlation, \pP, and the correlation between a genuine proton and a proton from $\Lambda$ feed-down, \mbox{p--\plmb}, and $\Sigma^{+}$ feed-down, \mbox{p--\psig}, are considered explicitly for the fit. The other contributions are modeled as flat functions of \kstar because of their small relative weights.
These weights are given by the $\lambda$ parameter formalism. With them, all the contributions are added together with their respective $\lambda$ parameters and form the modeled correlation function $C_\mathrm{model}(k^*)$:

\begin{equation}
	\begin{aligned}
		C_\mathrm{model}(k^*) =
		 & ~ \lambda_\mathrm{p\text{--}p}(k^*) \ C_\mathrm{p\text{--}p}(k^*)
		+ \lambda_\mathrm{p\text{--}p_\Lambda}(k^*) \ C_\mathrm{p\text{--}p_\Lambda}(k^*)
		+ \lambda_\mathrm{p\text{--}p_{\Sigma^+}}(k^*) \ C_\mathrm{p\text{--}p_{\Sigma^+}}(k^*) \\
		 & + \lambda_\mathrm{feed}(k^*) + \lambda_\mathrm{misid}(k^*).
	\end{aligned}
	\label{eq:ck_model}
\end{equation}

Here, $\lambda_\mathrm{p\text{--}p}(k^*)$ refers to the $\lambda$ parameter of genuine contribution, $\lambda_\mathrm{feed}(k^*)$ to all other feed-down combinations except of $\p\text{--}\plmb$ and $\p\text{--}\psig$, and $\lambda_\mathrm{misid}(k^*)$ to the misidentified particles.
The $\lambda$ parameters for each contribution are obtained with a data-driven method by multiplying the purity and the feed-down fractions, which are single-particle properties.
They are determined separately for proton--proton and antiproton--antiproton pairs, since the detector response for positive and negative particles differs slightly, even though the underlying physical interaction is identical.
The purity is obtained by fitting the $\mathrm{n}_\sigma$ distributions of deviations of the detector responses from the theoretical parametrization.
Depending on the momentum, different $\mathrm{n}_\sigma$ distributions have to be used, due to the different detectors employed (i.e. only TPC PID for momenta below $\SI{0.75}{\gevc}$ and combined TPC and TOF PID for momenta above \SI{0.75}{\gevc}).
The signal is fitted with a Gaussian in both cases and the background is, respectively, described by a linear and an exponential function.
Subsequently, the purity is determined by the ratio of the integral of the signal function to the integral of the total function within the $\mathrm{n}_\sigma = \pm 3$ window as a function of \pT. The purity is around 99\% for protons with a momentum below 0.75~\GeVc and above 97\% for up to 1.7~\GeVc, after which it drops quickly to 80\% at 2.2~\GeVc, which motivates the aforementioned cut in the \pT at that value. The fraction of misidentified particles corresponds to the impurity of the sample, i.e. $1-\mathrm{purity}$.

The feed-down fraction is obtained by fitting template distributions of the DCA to the primary vertex of each contribution to the measured DCA distribution of the proton candidates. The templates are obtained from Monte Carlo simulations with the Pythia event generator (version 8.3, Monash tune) \cite{pythia8} and propagated through a simulation of the ALICE detector in GEANT 4 \cite{geant4} and the same
reconstruction algorithm used for data reconstruction.
More details on the procedure of the template fit can be found in~\cite{CommonSource_pp,ALICE:2018ysd}.
In contrast to the previously used implementation, this analysis uses two-dimensional DCA distributions comprising DCA\textsubscript{xy} and DCA\textsubscript{z}.
Both the purity and the feed-down fractions are obtained as a function of the transverse momentum \pT by splitting the sample into multiple \pT intervals. With projection maps, which correlate the \kstar with the \pT of both tracks and can be obtained in specific \mT and multiplicity intervals, it is possible to obtain the $\lambda$ parameters as a function of the pair \kstar.
The shapes of the contributions of $\p\text{--}\plmb$ and $\p\text{--}\psig$, as well as the evaluation of the $\lambda$-parameters, are discussed in \cref{sec:decompCor}.

Finally, the experimental correlation function can be fitted by multiplying \cref{eq:ck_model} by a background function to account for a non-femtoscopic background, which is attributed to energy- and momentum conservation and leads to an enhancement of the measured correlation function above unity for large \kstar.
The chosen function $B_\mathrm{non-femto}(k^*)$ is a cubic polynomial, for which the linear term is set to zero in order to ensure a flat behavior of the correlation function at $\kstar = 0$, as it is not expected that the non-femtoscopic correlations influence the shape of the correlation function in the signal region. This was demonstrated in previous studies for baryon pairs and triplets \cite{ALICE:2018ysd,ThreeBodyFemto_Run2}.
The final correlation function thus becomes
\begin{equation}
	C_\mathrm{fit}(k^*) = B_\mathrm{non-femto}(k^*) \times C_\mathrm{model}(k^*)\,.
	\label{eq:ck_total}
\end{equation}
It is used in a combined fit to the \pP and \ApAp correlation functions, in which a single source-size parameter \reff is determined simultaneously for both datasets, while the baseline parameters are fitted independently for each.
The strong-interaction contribution to the correlation function is explicitly modeled up to a relative momentum of $k^* = 280$~\MeVc. At larger values of $k^*$, where the effects of the strong interaction are negligible, the genuine correlation function is smoothly interpolated to unity in order to reduce computational cost.
A variation of the strong-interaction cutoff by $\pm 40$~\MeV is included in the evaluation of the systematic uncertainties.
Furthermore, all $\lambda$ parameters except for the genuine contribution are varied by $\pm 10\%$. Since the $\lambda$ parameters of all contributions have to add up to unity, any change of the non-genuine contributions is compensated by rescaling the genuine contribution accordingly.
The fit range of the total correlation function including the baseline (i.e. \cref{eq:ck_total}) is $k^* \in [0, 400]$~\MeVc with a variation of the upper limit by $\pm10\%$. The systematic uncertainty of the $C_\mathrm{model}(k^*)$ is discussed in \cref{fitpotentials}. The fitting is done using the Correlation Analysis Tool using the Schr\"odinger equation (CATS) framework \cite{Mihaylov_2018}, which can compute the correlation function from wave functions and potentials given as input as well as correctly accounting for the feed-down contributions.

\begin{figure}[h!]
	\centering
	\includegraphics[width=0.82\linewidth]{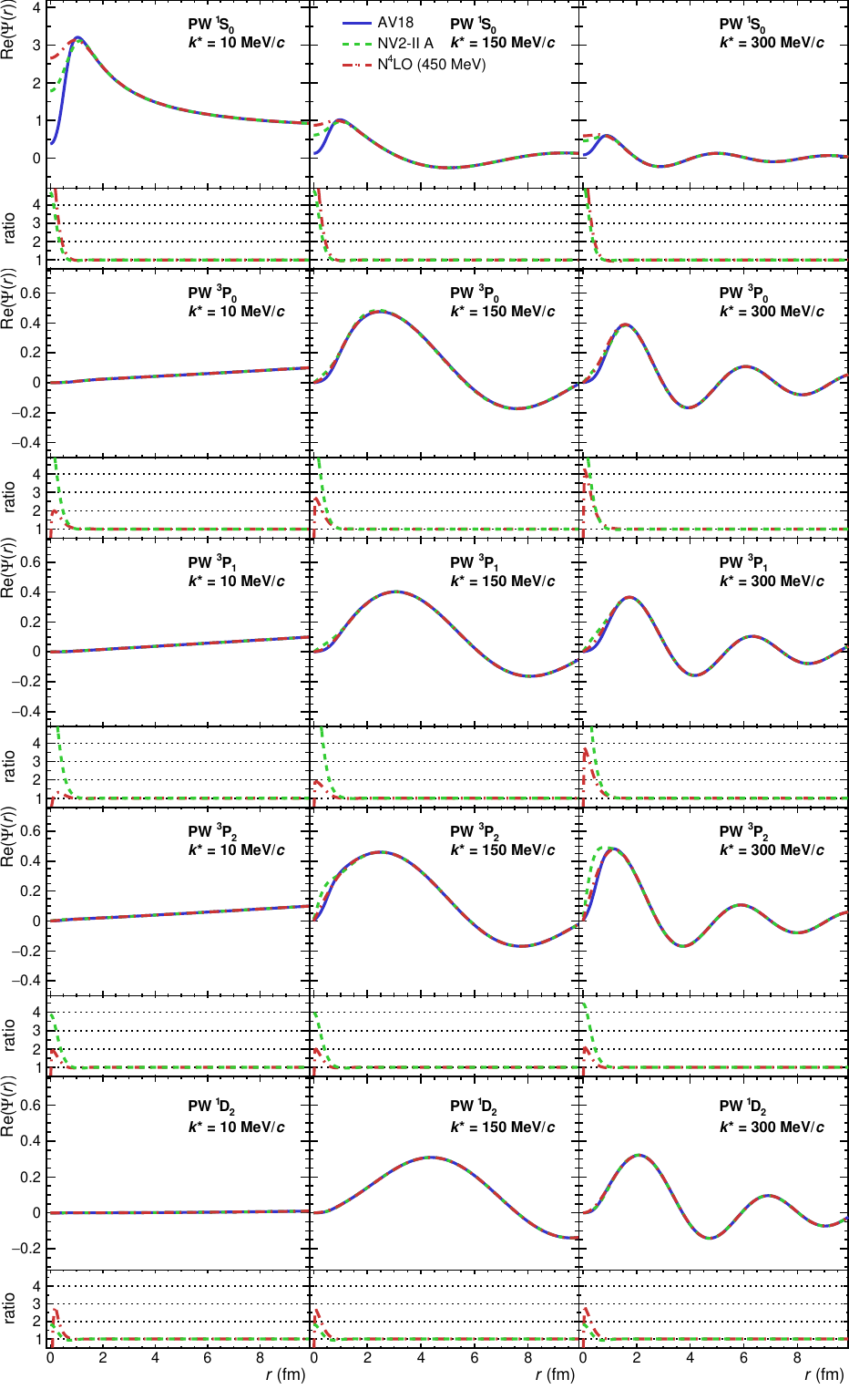}
	\caption{Real part of the radial \pP wave functions for Argonne \(v_{18}\) (blue), Norfolk NV2-II A (green), and ChEFT potentials (red) as a function of the relative nucleon-nucleon distance $r$. The ratio in the lower panels is obtained w.r.t. Argonne \(v_{18}\).}
	\label{fig:part_waves_comp}
\end{figure}

\begin{figure}[h!]
	\centering
	\includegraphics[width=0.98\linewidth]{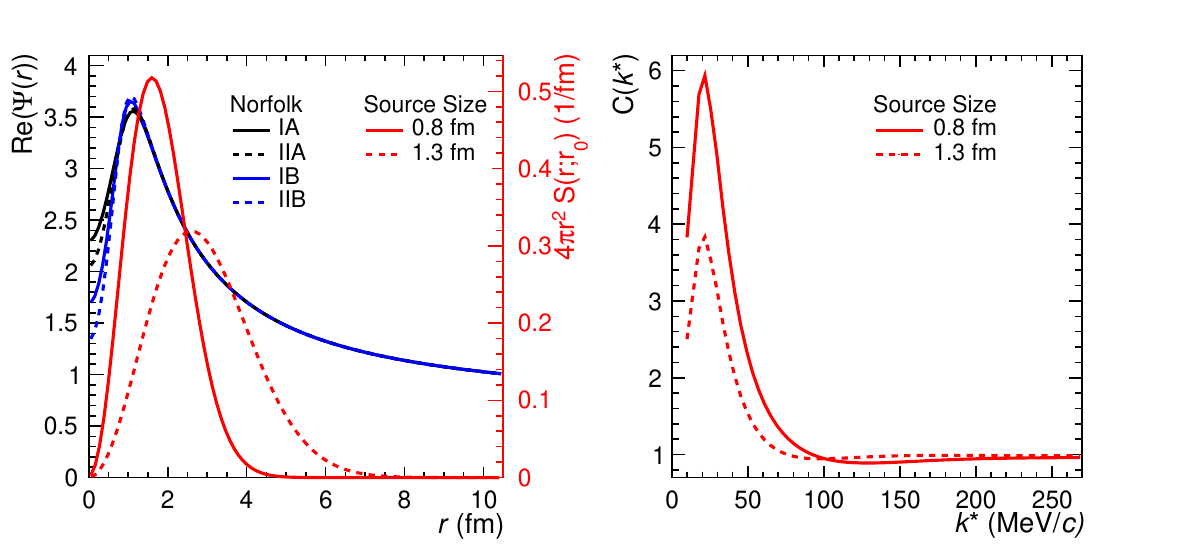}
	\caption{Left panel: the real part of the \pP wave functions obtained from the four variations of the Norfolk potential (black and blue lines together with black axis) and the two Gaussian source distributions (red lines with red axis) as a function of the relative nucleon--nucleon distance $r$. Right panel: the calculated \pP correlation functions for all the potential variations and the two source radii. The correlation functions for the four variations of the Norfolk potential are indistinguishable. See text for details.}
	\label{fig:norfolk_and_source_comparison}
\end{figure}

To simplify comparisons with theoretical calculations, the fully corrected correlation functions---tabulated as data points--- are provided via HEPData [TODO: include link], in which all non-genuine contributions are removed, leaving only the genuine femtoscopic correlation. These corrected correlation functions are tabulated as data points and cover the same $k^*$ range, as shown in the measured correlation functions in the figures. Since the corrected data do not provide additional information beyond what is already contained in the fits and plots presented in this paper, no additional figure is shown. The correction procedure is described in detail in \cref{sec:corrected_data}.

\subsection{Comparison of nuclear potentials}
\label{fitpotentials}
The modeling of the proton--proton correlation function is based on the Koonin--Pratt equation (\cref{eq:ck_int}), which requires as input the relative wave function of the two protons, obtained by solving the Schrödinger equation for a given nucleon--nucleon interaction. In previous femtoscopic analyses, the Argonne \(v_{18}\) potential~\cite{av18_paper} has been the standard choice due to its accuracy in describing nucleon--nucleon scattering data \cite{CommonSource_pp, Erratum_CommonSourceRun2, FemtoReview}. In this subsection, we compare this conventional approach to modern high-precision potentials derived from chiral effective field theory (ChEFT), which provide a more systematic and theoretically grounded framework. Specifically, we consider the Norfolk potential~\cite{Norfolk_01,Norfolk_02} and a semilocal momentum-space regularized chiral potential at N\(^4\)LO with a 450\,MeV cutoff~\cite{Reinert2018}.
To assess the impact of the interaction model on the correlation function, the two-proton wave functions from each potential are examined separately for each partial wave. \Cref{fig:part_waves_comp} shows the real part of each partial wave as a function of the distance $r$ between the two protons, for all three potentials considered. For the Norfolk potential, the NV2-II A variant has been used, which will be discussed in the next paragraph. Each row shows a different partial wave starting with the $^{1}S_{0}$ in the first row up to the $^{1}D_{2}$ wave in the last row. The three columns show each partial wave evaluated at a relative momentum $k^* = 10, 150$, and $300~\MeVc$, respectively. The ratio of all variations with respect to Argonne \(v_{18}\) is shown in the lower sub-panels of each panel.
All potentials are compatible with each other and deviations are seen only at distances below 1 fm for all partial waves.

To account for the systematic uncertainties present in the theoretical calculations, we have considered two different implementations of the Norfolk nucleon–nucleon potentials (NV2-I and NV2-II) and two variants of the interactions (A and B) \cite{Norfolk_01, Norfolk_02}. Details about the models are provided in the \cref{sec:norfolk_details}.
The real part of the four wave functions corresponding to the two versions NV2-I and NV2-II and the A and B variants are shown in the left panel of Fig.~\ref{fig:norfolk_and_source_comparison} (black and blue lines), together with the source distributions obtained for radii of 0.8 and 1.3~fm (full and dashed red lines, respectively). One observes that variations of the Norfolk interaction lead to small differences in the wave function only at short distances, for $r \lesssim 1.2$~fm, while the wave functions become nearly identical at larger separations.

The right panel of \cref{fig:norfolk_and_source_comparison} displays the resulting correlation functions for all Norfolk potential variations and for the two source radii considered.
For each source radius, the four evaluated wave functions yield nearly identical correlation functions, which collapse onto a single visible curve in the right panel. This behavior can be understood by noting that the dominant contribution to the measured correlation function arises from pair separations around and above $1.5$~fm, where all potential variants produce essentially the same wave function.
Even for the smallest source radius of 0.8~fm, the probability for particle pairs to be emitted at separations below 1.2~fm—where the wave functions differ—is strongly suppressed.

This is further illustrated by examining the overlap between the source distributions and the wave functions, shown in the left panel of Fig.~\ref{fig:norfolk_and_source_comparison}.
The differences observed in the short-range part of the wave function, arising from different interaction model choices, have a negligible impact on the resulting correlation function, as this distance region overlaps only a little with the source.
Since all four interaction variants fully overlap within the correlation function, the associated contribution to the systematic uncertainty of the fit component $C_\mathrm{model}(k^*)$ is negligible, at the permille level. The same conclusion is reached when considering systematic variations of the different interaction models.
\section{Results}
\begin{figure*}[ht]
	\centering
	\includegraphics[width=0.98\linewidth]{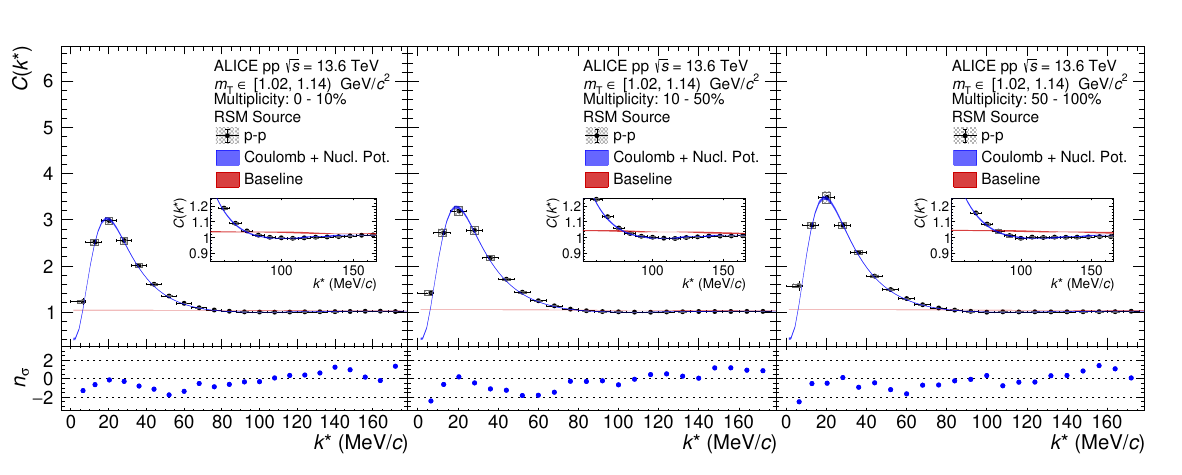}
	\includegraphics[width=0.98\linewidth]{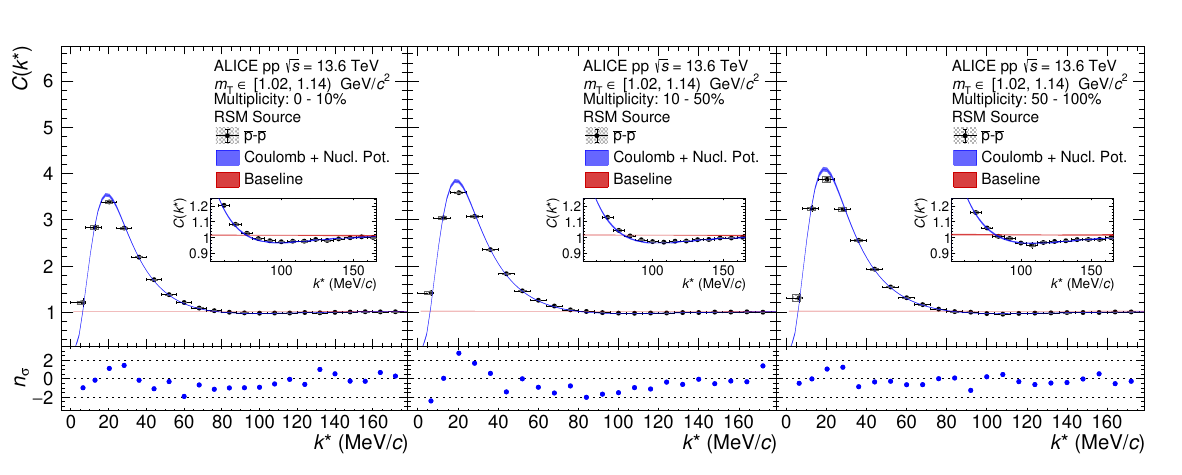}
	\hfill
	\caption{Fits of the measured \pP (upper row) and \ApAp (lower row) correlation functions in all multiplicity ranges and \mT range $[1.02, 1.14]\, \si{\gevcc}$ fitted with the RSM. The total fit with uncertainties is indicated by the blue band and the baseline with uncertainties is indicated by the red band. Statistical uncertainties are indicated by the vertical error bars and are smaller than the markers. Systematic uncertainties are indicated by the gray shaded boxes. The markers are positioned in each \kstar bin at the mean value of the underlying \kstar distribution and the horizontal lines indicate the standard deviation of the \kstar distribution within the bin. The deviation of the data points to the total fit in terms of standard deviations is shown in the lower panel.}
	\label{fig:fitsMt0}
\end{figure*}

\begin{figure*}[ht]
	\centering
	\includegraphics[width=0.98\linewidth]{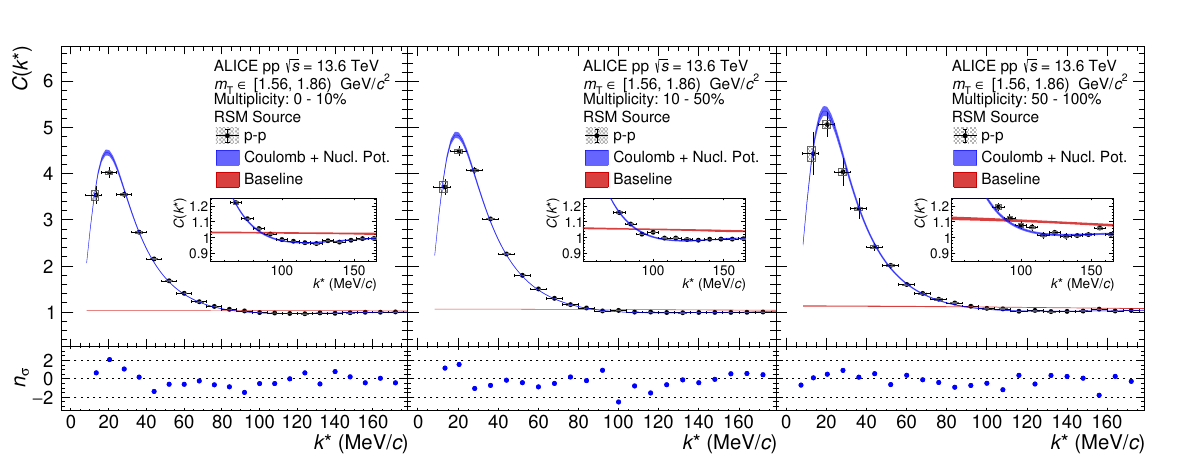}
	\includegraphics[width=0.98\linewidth]{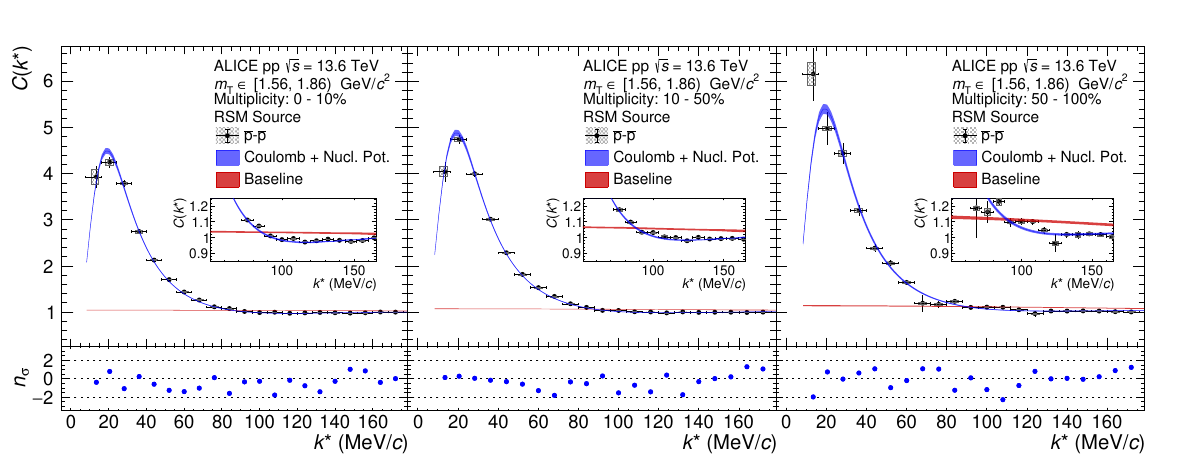}
	\hfill
	\caption{Fits of the measured \pP (upper row) and \ApAp (lower row) correlation functions in all multiplicity ranges and \mT range $[1.56, 1.86]\, \si{\gevcc}$ fitted with the RSM. For detailed description see \cref{fig:fitsMt0}.}
	\label{fig:fitsMt5}
\end{figure*}
The measured correlation functions are shown for \pP (upper row) and \ApAp (lower row) pairs in \cref{fig:fitsMt0} and \cref{fig:fitsMt5} for the first (\mT $ \in$ [\SIrange{1.02}{1.14}{\gevcc}]) and sixth (\mT $ \in$ [\SIrange{1.56}{1.86}{\gevcc}]) \mT intervals, respectively. The experimental data are shown together with the results of the fit employing the RSM. Additional figures can be found in the appendix and are organized as follows: the measurements using the RSM in the remaining five \mT intervals can be found in \cref{sec:appendix_core}. The fits using an effective Gaussian parametrization of the source are shown for all \mT and multiplicity intervals in \cref{sec:appendix_eff}. The fits for the multiplicity-integrated sample are shown in \cref{sec:appendix_MB}.

The main panels of \cref{fig:fitsMt0,fig:fitsMt5} show the measured correlation functions with the systematic uncertainties represented by the gray boxes.
The blue band represents the correlation function fitted with CATS
using a Gaussian source and different versions of the Norfolk potential for the \pP strong interaction.
Assuming a uniform distribution of the variations, the band is obtained by evaluating the maximum spread of all the variations of the fit and dividing that value by $\sqrt{12}$.
The lower panel shows the difference between the data and the fit expressed in terms of the number of standard deviations $\mathrm{n}_\sigma$.
They are evaluated by calculating the average agreement between the data and the theoretical curves, weighting the theory by the entries in the mixed-event distribution in the corresponding \kstar intervals and accounting for statistical and systematic uncertainties of both the fit and the data.

\begin{figure*}[ht]
	\centering
	\includegraphics[width=0.48\linewidth]{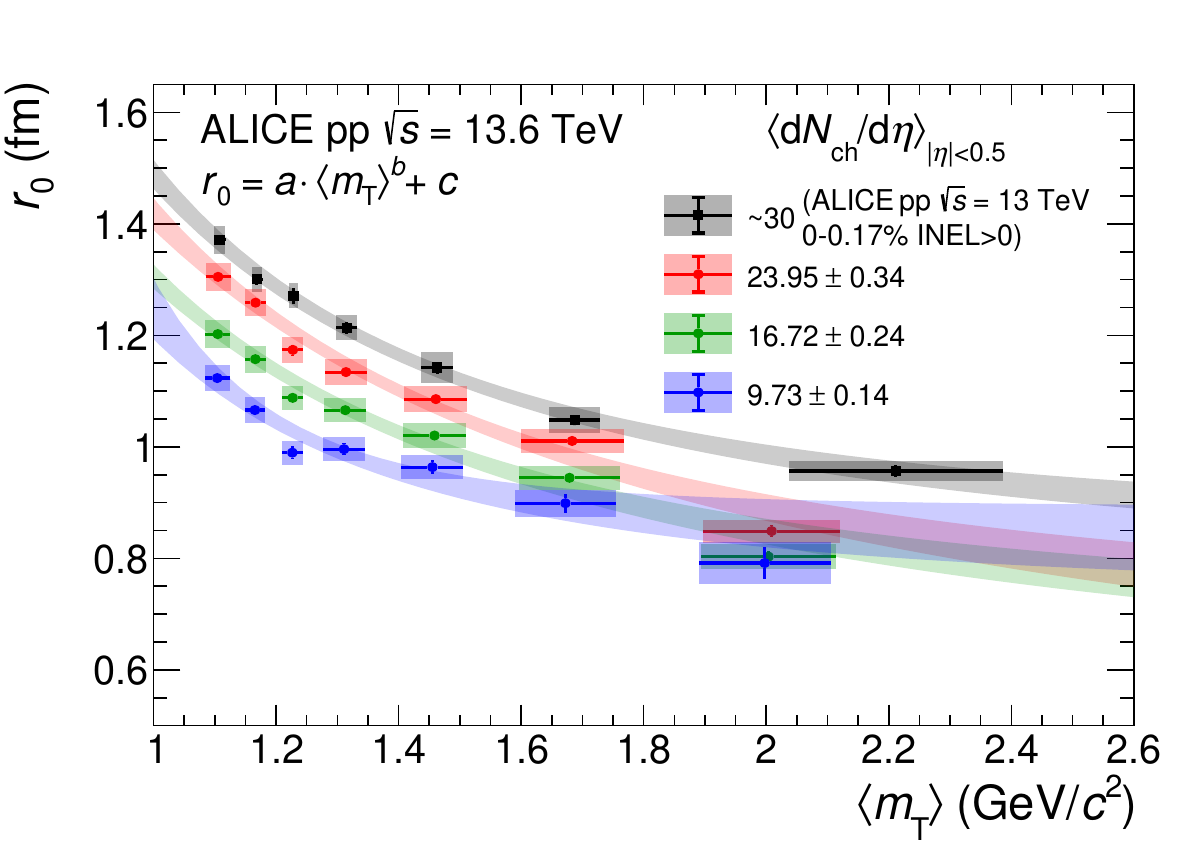}
	\includegraphics[width=0.48\linewidth]{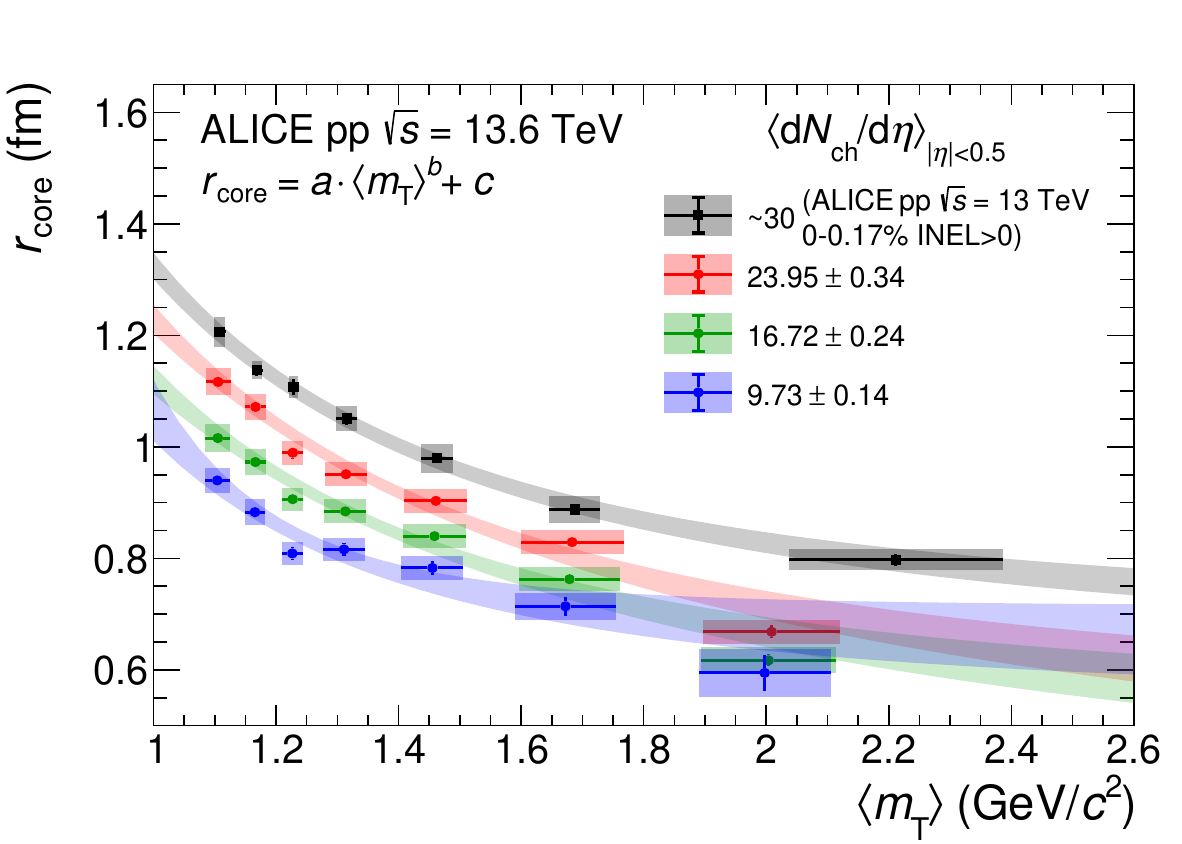}
	\hfill
	\caption{The \mT dependence of the effective radius \reff (left) and the core radius \rcore (right) for different multiplicity classes. The black data points are from a previous analysis using a high-multiplicity data set collected by ALICE at \onethree~\cite{CommonSource_pp,Erratum_CommonSourceRun2}. Here, INEL$>0$ refers to inelastic pp collisions with at least one charged particle in the event within $|\eta| < 1.0$~\cite{p-Omega_nature}, while 0–0.17\% denotes the highest-multiplicity class, corresponding to the top 0.17\% of events in the charged-particle multiplicity distribution. The bands correspond to the parametrization of the \mT dependence for the different multiplicity classes.}
	\label{fig:radiusMt1}
\end{figure*}

The same fit results are obtained when the Argonne \(v_{18}\) and  N\(^4\)LO potentials are employed.
The fits reproduce the measured correlation functions very well and the deviation from data in the lower panels does not exceed three standard deviations.
A detailed view of the intermediate \kstar range is provided in the insets, showing the depletion of the correlation function between $100$ and $150~\MeVc$ due to the inclusion of the p-wave, which is also captured by the fit. From each fit the effective source size ($r_0$) and the core source size ($r_\mathrm{core}$) can be extracted, and the values are shown in the left and right panels of \cref{fig:radiusMt1}, respectively. As before, the box represents the systematic uncertainty of the radius, while the lines show the statistical uncertainty. Both uncertainties are obtained by employing the bootstrap procedure as has been done previously in Ref.~\cite{CommonSource_2023}. The total uncertainties are estimated by resampling each \kstar bin based on the statistical and systematic uncertainties and combining them randomly with variations of the fitting procedure. The statistical uncertainties are given by resampling the datapoints only according to their statistical uncertainties and fitting them without considering systematic variations of the fit or the data.
The horizontal positions of the markers have been placed at the mean of the underlying \mT distribution, which is obtained for each multiplicity class separately.

\begin{figure*}[ht]
	\centering
	\includegraphics[width=0.48\linewidth]{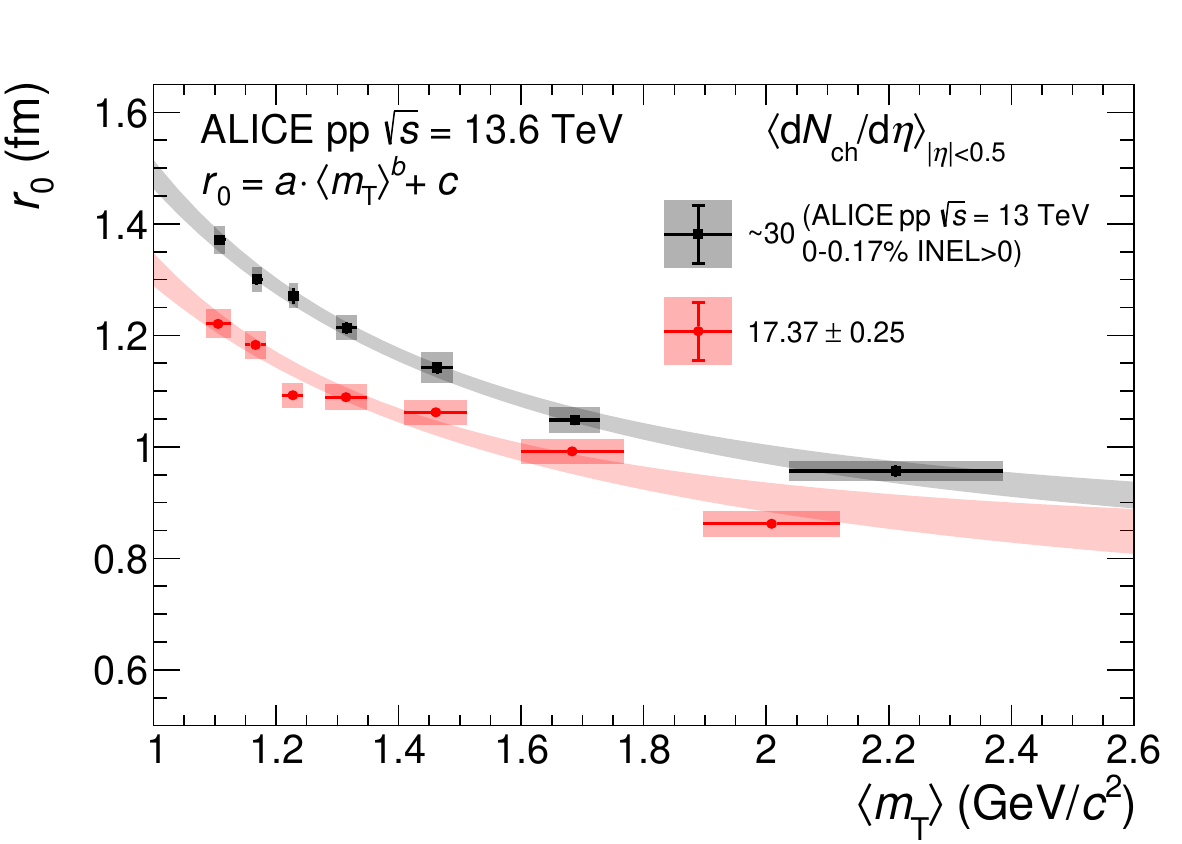}
	\includegraphics[width=0.48\linewidth]{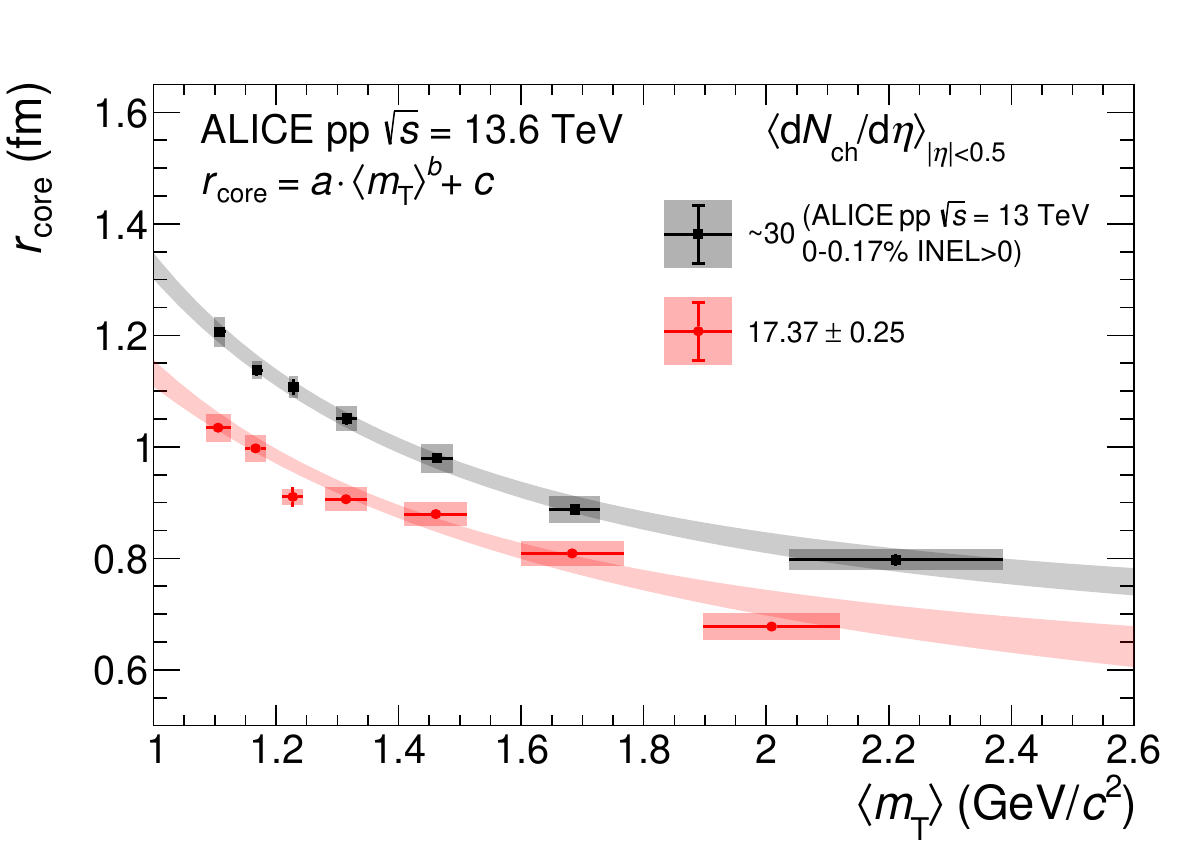}
	\hfill
	\caption{The \mT dependence of the effective radius \reff (left) and the core radius \rcore (right) for two different multiplicity classes. The red data points correspond to the extraction of the radii using this analysis's full MB data set. The black data points are from a previous analysis using a high-multiplicity data set collected by ALICE at \onethree~\cite{CommonSource_pp,Erratum_CommonSourceRun2}. The bands correspond to the parametrization of the \mT dependence for the different multiplicity classes.}
	\label{fig:radiusMt2}
\end{figure*}

Both panels in \cref{fig:radiusMt1} also include, the results at $\sqrt{s} = \SI{13}{\TeV}$ from the first study in which the RSM was applied \cite{CommonSource_pp,Erratum_CommonSourceRun2}. A direct comparison across collision energies is justified, as no explicit dependence of the source radius on the collision energy is expected, in particular for such a small difference in collision energy. This is consistent with recent findings from coalescence models applied at LHC energies~\cite{ToMCCA}.
The existing measurements, based on high-multiplicity data collected by ALICE during Run 2, provided only \mT-dependent radii and did not include a multiplicity-differential analysis \cite{CommonSource_pp,Erratum_CommonSourceRun2,CommonSource_2023,pi-p_analysis}. They have served as a reference for subsequent femtoscopic analyses by ALICE in Run 2. The current measurement expands upon this by providing results as a function of both \mT and multiplicity. In particular, we now access low-multiplicity events that were previously unexplored. This enables a more comprehensive characterization of the source dynamics.
While the \mT intervals have been chosen to coincide with those used in Refs.~\cite{CommonSource_pp,Erratum_CommonSourceRun2}, the central value of the largest \mT bin is shifted to lower values in this analysis. This shift is due to the tighter upper \pt cut of \SI{2.2}{\GeVc} for proton identification, which limits the maximum value of the pair \mT to \SI{2.40}{\GeVmass}.
The \mT dependence can be fitted by a power-law function of the form $\reff = a \cdot \langle\mT\rangle^b + c$. The corresponding bands are shown in \cref{fig:radiusMt1}. Each band is obtained by refitting the radii in each \mT interval 1000 times. For each fit, the source size value is resampled to account for both statistical and systematic uncertainties. The width of each band is given by the $1\sigma$ spread of the fitted parameterizations. The parameter values of the bands are reported in \cref{tab:fit_results_mt_core,tab:fit_results_mt_eff} for \rcore and \reff, respectively.
Similarly, \reff and \rcore are obtained from the integrated analysis of the full MB dataset. They are shown in the left and right panels of \cref{fig:radiusMt2}, respectively, along with the results from \cite{CommonSource_pp,Erratum_CommonSourceRun2} and the power-law fit using the same parametrization as before. The parameter values of the power-law fit are also reported in \cref{tab:fit_results_mt_core,tab:fit_results_mt_eff}.

In each multiplicity class we observe a power-law \mT dependence, which was also previously observed in the high-multiplicity (HM) measurement performed at \onethree. The different multiplicity classes display an expected hierarchy: the source size is larger in events with a larger average multiplicity. The shapes of the power-law parametrization exhibit small differences between the multiplicity classes. A detailed understanding of these results requires a theoretical modeling including possible collective effects. The results in this paper can inform such calculations. Another observation is that both \rcore and \reff appear to approach a lower limit at very large \mT, consistent with the expectation that particles of finite size cannot be emitted from an infinitely small region.

\begin{figure*}[ht]
	\centering
	\includegraphics[width=0.9\linewidth]{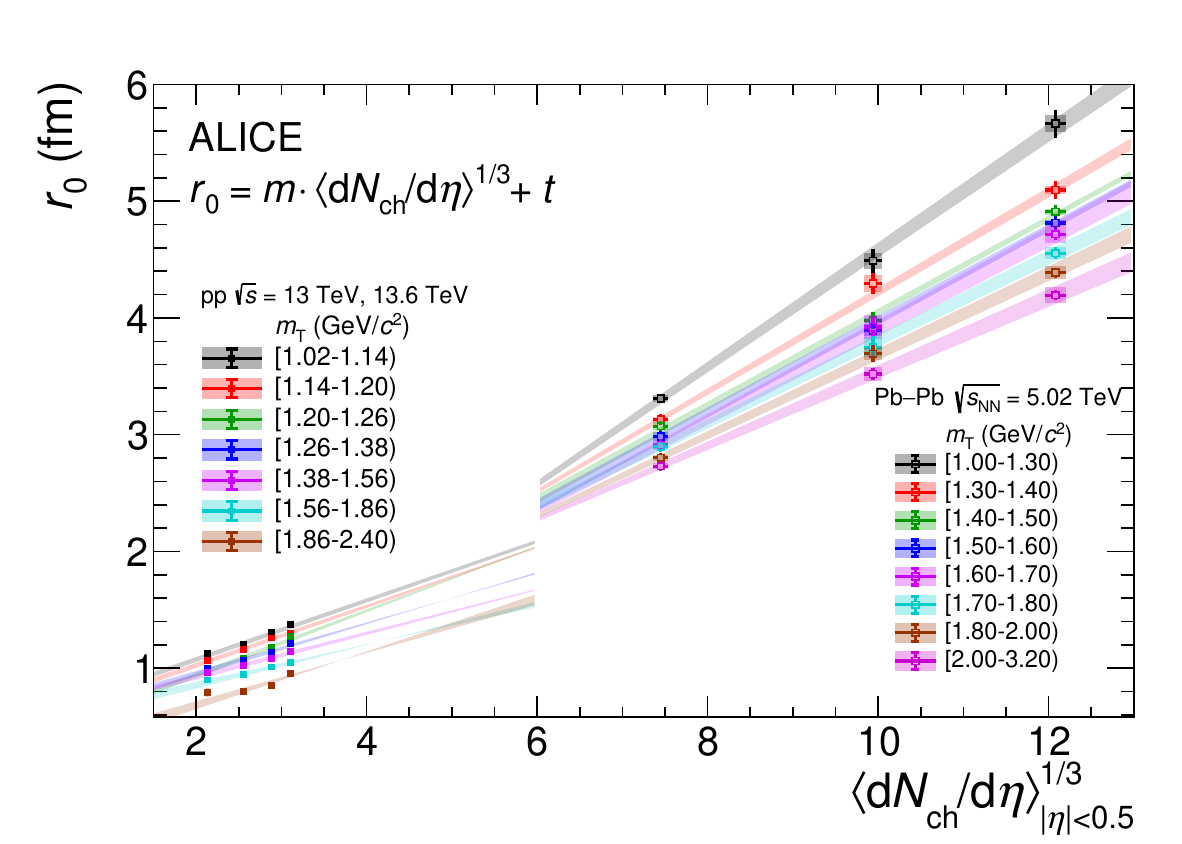}
	\hfill
	\caption{Multiplicity dependence of the effective radius \(\reff\) for various \(\mT\) intervals measured in pp and Pb--Pb collisions~\cite{ALICE:2025wuy}. The bands represent the parametrization for different multiplicity classes, shown with a \(1\sigma\) uncertainty. Note that the \(\mT\) intervals differ between the two collision systems.
	}
	\label{fig:radiusPPvsPBPB}
\end{figure*}

From relativistic heavy-ion collisions, one expects the source size to increase linearly with the cube root of the charged particle multiplicity \cite{ALICE:2025wuy}, i.e. assuming that the multiplicity and the volume of the emission region are proportional to each other. To study this assumption in \pp collisions, the \reff is shown in \cref{fig:radiusPPvsPBPB} as a function of the cube root of the charged-particle multiplicity at mid-rapidity ($|\eta|<0.5$). In each \mT class, the multiplicity dependence is fitted with a linear function, which is represented by the bands in \cref{fig:radiusPPvsPBPB}, where the width is again given by the statistical and systematic uncertainties of the measured radii. The values of the parametrization are reported in \cref{tab:fit_results_mult_eff} for \reff.
Extending this analysis, we compare the measured multiplicity dependence in \pp\ collisions with recently published source-size measurements in Pb--Pb collisions~\cite{ALICE:2025wuy}. The comparison, shown in \cref{fig:radiusPPvsPBPB}, reveals a striking difference in slope. In the Pb--Pb measurement, centrality classes extend up to 50\%, where the formation of a quark--gluon plasma (QGP) is well established. In contrast, the existence of a QGP in \pp\ collisions remains an open question.
The multiplicity dependence of the source size appears steeper in Pb--Pb than in \pp, indicating a stronger dependence with multiplicity; however, no quantitative threshold or established theoretical expectation currently links this behavior directly to QGP formation.
Further theoretical input, together with measurements in intermediate collision systems such as O--O or Ne--Ne, will be essential to clarify whether this behavior can be attributed to the presence of a quark-gluon medium or arises from other mechanisms.

\begin{table}[h]
	\centering
	\caption{Values for the parametrization of \rcore as a function of \mT in different multiplicity classes as shown in left plots of \cref{fig:radiusMt1,fig:radiusMt2}. The uncertainties correspond to the 1$\sigma$ spread obtained by repeating the fit while varying the data points within their uncertainties.}
	\label{tab:fit_results_mt_core}
	\begin{tabular}{c|c|c|c}
		\avgdndeta                     & \textit{\textbf{a}} [$\text{fm}\cdot(\text{\GeVmass})^{-b}$] & \textit{\textbf{b}}         & \textit{\textbf{c}} [fm]  \\ \hline
		$\approx 30$ (Run~2 HM)        & $0.684^{+0.032}_{-0.012}$                                    & $-1.859^{+0.252}_{-0.094}$  & $0.637^{+0.059}_{-0.032}$ \\ \hline
		\avgdndetaPairzeroTen          & $0.818^{+0.089}_{-0.145}$                                    & $-1.426^{+0.331}_{-0.332}$  & $0.412^{+0.114}_{-0.171}$ \\ \hline
		\avgdndetaPairtenFifty         & $0.706^{+0.087}_{-0.152}$                                    & $-1.476^{+0.401}_{-0.392}$  & $0.414^{+0.112}_{-0.178}$ \\ \hline
		\avgdndetaPairfiftyHundred     & $0.411^{+0.006}_{-0.058}$                                    & $-3.473^{+1.569}_{-1.141}$  & $0.644^{+0.070}_{-0.103}$ \\   \hline
		\avgdndetaPairinclusive{} (MB) & $0.672^{+0.081}_{-0.125}$                                    & $ -1.364^{+0.359}_{-0.321}$ & $0.458^{+0.106}_{-0.148}$
	\end{tabular}
\end{table}

\begin{table}[h]
	\centering
	\caption{Values for the parametrization of \reff as a function of \mT in different multiplicity classes as shown in left plots of \cref{fig:radiusMt1,fig:radiusMt2}. The uncertainties correspond to the 1$\sigma$ spread obtained by repeating the fit while varying the data points within their uncertainties.}
	\label{tab:fit_results_mt_eff}
	\begin{tabular}{c|c|c|c}
		\avgdndeta                     & \textit{\textbf{a}} [$\text{fm}\cdot(\text{\GeVmass})^{-b}$] & \textit{\textbf{b}}        & \textit{\textbf{c}} [fm]   \\ \hline
		$\approx 30$ (Run~2 HM)        & $0.686^{+0.018}_{-0.031}$                                    & $-1.897^{+0.189}_{-0.220}$ & $0.802^{+0.044}_{-0.057}$  \\ \hline
		\avgdndetaPairzeroTen          & $0.861^{+0.088}_{-0.158}$                                    & $-1.367^{+0.305}_{-0.333}$ & $ 0.557^{+0.115}_{-0.188}$ \\ \hline
		\avgdndetaPairtenFifty         & $0.694^{+0.039}_{-0.112}$                                    & $-1.557^{+0.163}_{-0.343}$ & $0.615^{+0.058}_{-0.137}$  \\ \hline
		\avgdndetaPairfiftyHundred     & $0.416^{+0.007}_{-0.047}$                                    & $-3.753^{+1.157}_{-1.387}$ & $ 0.835^{+0.058}_{-0.105}$ \\   \hline
		\avgdndetaPairinclusive{} (MB) & $0.569^{+0.052}_{-0.083}$                                    & $-1.807^{+0.492}_{-0.409}$ & $0.747^{+0.084}_{-0.111}$
	\end{tabular}
\end{table}

\begin{table}[h]
	\centering
	\caption{Values for the parametrization of \reff as a function of \avgdndeta in different \mT bins as shown in  Fig.~ \ref{fig:radiusPPvsPBPB}. The uncertainties correspond to the 1$\sigma$ spread obtained by repeating the fit while varying the data points within their uncertainties.}
	\begin{tabular}{c|c|c}
		\mT [\GeVmass] & \textit{\textbf{m}} [fm]  & \textit{\textbf{t}} [fm]   \\ \hline
		\multicolumn{3}{c}{\textbf{pp}}                                         \\ \hline
		$[1.02,1.14)$  & $0.253^{+0.002}_{-0.002}$ & $0.569^{+0.024}_{-0.024}$  \\ \hline
		$[1.14,1.20)$  & $0.253^{+0.004}_{-0.004}$ & $0.518^{+0.028}_{-0.028}$  \\ \hline
		$[1.20,1.26)$  & $0.276^{+0.004}_{-0.004}$ & $0.388^{+0.030}_{-0.020}$  \\ \hline
		$[1.26,1.38)$  & $0.218^{+0.010}_{-0.005}$ & $0.515^{+0.045}_{-0.020}$  \\ \hline
		$[1.38,1.56)$  & $0.195^{+0.012}_{-0.003}$ & $0.527^{+0.050}_{-0.009}$  \\ \hline
		$[1.56,1.86)$  & $0.173^{+0.013}_{-0.013}$ & $0.508^{+0.054}_{-0.054}$  \\ \hline
		$[1.86,2.40)$  & $0.227^{+0.020}_{-0.020}$ & $0.228^{+0.073}_{-0.073}$  \\ \hline
		\multicolumn{3}{c}{\textbf{Pb---Pb}}                                    \\ \hline
		$[1.00,1.30)$  & $0.502^{+0.009}_{-0.009}$ & $-0.434^{+0.023}_{-0.023}$ \\ \hline
		$[1.30,1.40)$  & $0.400^{+0.002}_{-0.002}$ & $0.049^{+0.053}_{-0.053}$  \\ \hline
		$[1.40,1.50)$  & $0.400^{+0.002}_{-0.002}$ & $0.049^{+0.053}_{-0.053}$  \\ \hline
		$[1.50,1.60)$  & $0.396^{+0.002}_{-0.002}$ & $0.023^{+0.062}_{-0.062}$  \\ \hline
		$[1.60,1.70)$  & $0.391^{+0.011}_{-0.011}$ & $-0.002^{+0.038}_{-0.038}$ \\ \hline
		$[1.70,1.80)$  & $0.357^{+0.006}_{-0.006}$ & $0.235^{+0.008}_{-0.008}$  \\ \hline
		$[1.80,2.00)$  & $0.346^{+0.007}_{-0.007}$ & $0.230^{+0.022}_{-0.022}$  \\ \hline
		$[2.00,2.30)$  & $0.316^{+0.010}_{-0.010}$ & $0.375^{+0.042}_{-0.042}$  \\
	\end{tabular}
	\label{tab:fit_results_mult_eff}
\end{table}

\section{Summary}
This work investigates differential \pP and \ApAp correlations in seven \mT intervals for MB events and for three different multiplicity classes of pp collisions at $\sqrt{s} =$ 13.6 TeV measured by ALICE during the Run 3 data taking. The aim is to study the common particle-emitting source and the robustness of the source size measurement with respect to different nucleon--nucleon potentials.
All correlation functions are fitted using two source models (Gaussian and RSM), and the extracted \reff and \rcore parameters show a decrease as a function of \mT for all the multiplicity classes, and larger radii are observed for events with higher multiplicities.

For the modeling of the strong interaction between the protons, the Argonne $\nu_{18}$, the  N\(^4\)LO and the Norfolk potentials have been considered. Since the wave functions are all identical for intra-particle distances above 1.2 fm and the measured source radii are sufficiently large, no differences in the predicted correlation functions and in the extracted radii are found.

The differential measurement of the radii extracted from \pp data enables the first comparison to Pb--Pb data.  It is shown that one can not smoothly interpolate between the two colliding systems. The source size scales linearly with the cube root of multiplicity for both systems, consistent with the idea that it reflects the system’s volume, but the slopes are different. Further studies are necessary to understand this effect.

The presented results provide new constraints to our understanding of particle production and will allow precise determination of the source size for various particle pairs in Run~3, supporting interaction studies and light nuclei formation models at accelerators and in the Galaxy.
%


\newenvironment{acknowledgement}{\relax}{\relax}
\begin{acknowledgement}
	\section*{Acknowledgements}

The ALICE Collaboration would like to thank all its engineers and technicians for their invaluable contributions to the construction of the experiment and the CERN accelerator teams for the outstanding performance of the LHC complex.
The ALICE Collaboration gratefully acknowledges the resources and support provided by all Grid centres and the Worldwide LHC Computing Grid (WLCG) collaboration.
The ALICE Collaboration acknowledges the following funding agencies for their support in building and running the ALICE detector:
A. I. Alikhanyan National Science Laboratory (Yerevan Physics Institute) Foundation (ANSL), State Committee of Science and World Federation of Scientists (WFS), Armenia;
Austrian Academy of Sciences, Austrian Science Fund (FWF): [M 2467-N36] and Nationalstiftung f\"{u}r Forschung, Technologie und Entwicklung, Austria;
Ministry of Communications and High Technologies, National Nuclear Research Center, Azerbaijan;
Rede Nacional de Física de Altas Energias (Renafae), Financiadora de Estudos e Projetos (Finep), Funda\c{c}\~{a}o de Amparo \`{a} Pesquisa do Estado de S\~{a}o Paulo (FAPESP) and The Sao Paulo Research Foundation  (FAPESP), Brazil;
Bulgarian Ministry of Education and Science, within the National Roadmap for Research Infrastructures 2020-2027 (object CERN), Bulgaria;
Ministry of Education of China (MOEC) , Ministry of Science \& Technology of China (MSTC) and National Natural Science Foundation of China (NSFC), China;
Ministry of Science and Education and Croatian Science Foundation, Croatia;
Centro de Aplicaciones Tecnol\'{o}gicas y Desarrollo Nuclear (CEADEN), Cubaenerg\'{\i}a, Cuba;
Ministry of Education, Youth and Sports of the Czech Republic, Czech Republic;
The Danish Council for Independent Research | Natural Sciences, the VILLUM FONDEN and Danish National Research Foundation (DNRF), Denmark;
Helsinki Institute of Physics (HIP), Finland;
Commissariat \`{a} l'Energie Atomique (CEA) and Institut National de Physique Nucl\'{e}aire et de Physique des Particules (IN2P3) and Centre National de la Recherche Scientifique (CNRS), France;
Bundesministerium f\"{u}r Forschung, Technologie und Raumfahrt (BMFTR) and GSI Helmholtzzentrum f\"{u}r Schwerionenforschung GmbH, Germany;
National Research, Development and Innovation Office, Hungary;
Department of Atomic Energy Government of India (DAE), Department of Science and Technology, Government of India (DST), University Grants Commission, Government of India (UGC) and Council of Scientific and Industrial Research (CSIR), India;
National Research and Innovation Agency - BRIN, Indonesia;
Istituto Nazionale di Fisica Nucleare (INFN), Italy;
Japanese Ministry of Education, Culture, Sports, Science and Technology (MEXT) and Japan Society for the Promotion of Science (JSPS) KAKENHI, Japan;
Consejo Nacional de Ciencia (CONACYT) y Tecnolog\'{i}a, through Fondo de Cooperaci\'{o}n Internacional en Ciencia y Tecnolog\'{i}a (FONCICYT) and Direcci\'{o}n General de Asuntos del Personal Academico (DGAPA), Mexico;
Nederlandse Organisatie voor Wetenschappelijk Onderzoek (NWO), Netherlands;
The Research Council of Norway, Norway;
Pontificia Universidad Cat\'{o}lica del Per\'{u}, Peru;
Ministry of Science and Higher Education, National Science Centre and WUT ID-UB, Poland;
Korea Institute of Science and Technology Information and National Research Foundation of Korea (NRF), Republic of Korea;
Ministry of Education and Scientific Research, Institute of Atomic Physics, Ministry of Research and Innovation and Institute of Atomic Physics and Universitatea Nationala de Stiinta si Tehnologie Politehnica Bucuresti, Romania;
Ministerstvo skolstva, vyskumu, vyvoja a mladeze SR, Slovakia;
National Research Foundation of South Africa, South Africa;
Swedish Research Council (VR) and Knut \& Alice Wallenberg Foundation (KAW), Sweden;
European Organization for Nuclear Research, Switzerland;
Suranaree University of Technology (SUT), National Science and Technology Development Agency (NSTDA) and National Science, Research and Innovation Fund (NSRF via PMU-B B05F650021), Thailand;
Turkish Energy, Nuclear and Mineral Research Agency (TENMAK), Turkey;
National Academy of  Sciences of Ukraine, Ukraine;
Science and Technology Facilities Council (STFC), United Kingdom;
National Science Foundation of the United States of America (NSF) and United States Department of Energy, Office of Nuclear Physics (DOE NP), United States of America.
In addition, individual groups or members have received support from:
Czech Science Foundation (grant no. 23-07499S), Czech Republic;
FORTE project, reg.\ no.\ CZ.02.01.01/00/22\_008/0004632, Czech Republic, co-funded by the European Union, Czech Republic;
European Research Council (grant no. 101220549), European Union;
Deutsche Forschungs Gemeinschaft (DFG, German Research Foundation) ``Neutrinos and Dark Matter in Astro- and Particle Physics'' (grant no. SFB 1258), Germany;
CONVECS project, CUP C97H23001700002 FESR 2021-2027 program, Italy.

\end{acknowledgement}

\bibliographystyle{utphys}   
\bibliography{bibliography}

\newpage
\appendix

\section{Norfolk potentials}
\label{sec:norfolk_details}
The Norfolk nucleon--nucleon potentials considered to evaluate the systematics of the fit correspond to two choices of the scale regulation \cite{Norfolk_01, Norfolk_02}.
These potentials are local chiral EFT potentials, formulated in coordinate space and developed up to N\(^3\)LO. The long-range part of the interaction stems from one-pion-exchange and
two-pion-exchange contributions, while the short-range part is given by
contact-like interactions.
The low-energy constants have been fixed by fitting to np and pp
phase shifts \cite{Norfolk_scattering_data}, the deuteron binding energy and the \(^1S_0\) nn scattering length~\cite{Norfolk_deuteron_energy}.

Two classes of Norfolk potentials are used: NV2-I where phase shifts are fitted up to \(T_{\mathrm{lab}}\) of \SI{125}{\MeV}, and NV2-II fitted up to $T_{\mathrm{lab}} = \SI{200}{\MeV}$. Each class
has A, B, and C variants, which differ in the coordinate-space regulator radii. Variants A and B are commonly used in systematic analyses.
The A variant employs a long-range regulation scale of 1.2~fm and a short-range regulation scale of 0.8~fm, while for the B variant the scales are 1.0~fm and 0.7~fm.
For context, these potentials have been successfully employed in a wide range of calculations, including studies of the energy levels of light nuclei \cite{Norfolk_03} as well as investigations of the magnetic moments and form factors of light nuclei \cite{Norfolk_04}. The Norfolk potentials, together with Argonne $\nu_{18}$, have also been used recently to examine the sensitivity of the correlation function to different interaction models \cite{Norfolk_05}. In that study, for relative momenta $k \leq 500$~MeV in the pp system and a source radius of 1.249~fm, variations in the interaction were found to induce changes in the correlation function of up to 1.4\%.

\section{Extracting the corrected data}
\label{sec:corrected_data}
Assuming that the decomposition of the data is given by \cref{eq:ck_total} with inserting \cref{eq:ck_model}, one can exchange $C_\mathrm{fit}(k^*)$ with $C_\mathrm{data}(k^*)$ and $C_\mathrm{p\text{--}p}(k^*)$ with $C_\mathrm{data}^\mathrm{corrected}(k^*)$ and solve for $C_\mathrm{data}^\mathrm{corrected}(k^*)$:

\begin{equation}
	\begin{aligned}
		C_\mathrm{data}^\mathrm{corrected}(k^*)
		= \frac{1}{\lambda_\mathrm{p\text{--}p}(k^*)} \times & \Bigl[ \frac{C_\mathrm{data}(k^*)}{B_\mathrm{non-femto}(k^*)}
		- \lambda_\mathrm{p\text{--}p_\Lambda}(k^*) C_\mathrm{p\text{--}p_\Lambda}(k^*)                                                                    \\
		                                                     & \quad - \lambda_\mathrm{p\text{--}p_{\Sigma^+}}(k^*) C_\mathrm{p\text{--}p_{\Sigma^+}}(k^*)
			- \lambda_\mathrm{feed}(k^*) - \lambda_\mathrm{misid}(k^*) \Bigr]
	\end{aligned}
	\label{eq:cf_corrected}
\end{equation}

The fits with the RSM were used for this procedure. Having saved each component of the right-hand side of \cref{eq:cf_corrected} during each iteration of the bootstrapping procedure, it is possible to compute $C_\mathrm{data}^\mathrm{corrected}(k^*)$ directly for each iteration. This procedure delivers two distributions of the value for the corrected data points, based on the total uncertainty and the statistical uncertainty.
The central value and the uncertainty of the corrected correlation function are then obtained by the mean and the standard deviation of the distribution in each \kstar bin, respectively. Since this correction procedure relies on the fitted values of the \rcore and the non-femtoscopic baseline, its validity relies on the assumption of the potential used in this analysis and the parametrization of the baseline, namely the Norfolk potential and the third-order polynomial, respectively. However, the studies presented in this paper have shown that all the state-of-the-art models for the interaction yield compatible results, and the influence of the baseline is on a few percent level. The corrected data are available on HEPdata. The corrected data from the \pP and \ApAp correlation functions are compatible.

\clearpage
\newpage

\section{Decomposition of the correlation function}
\label{sec:decompCor}
\begin{figure}[h!]
	\centering
	\includegraphics[width=0.49\linewidth]{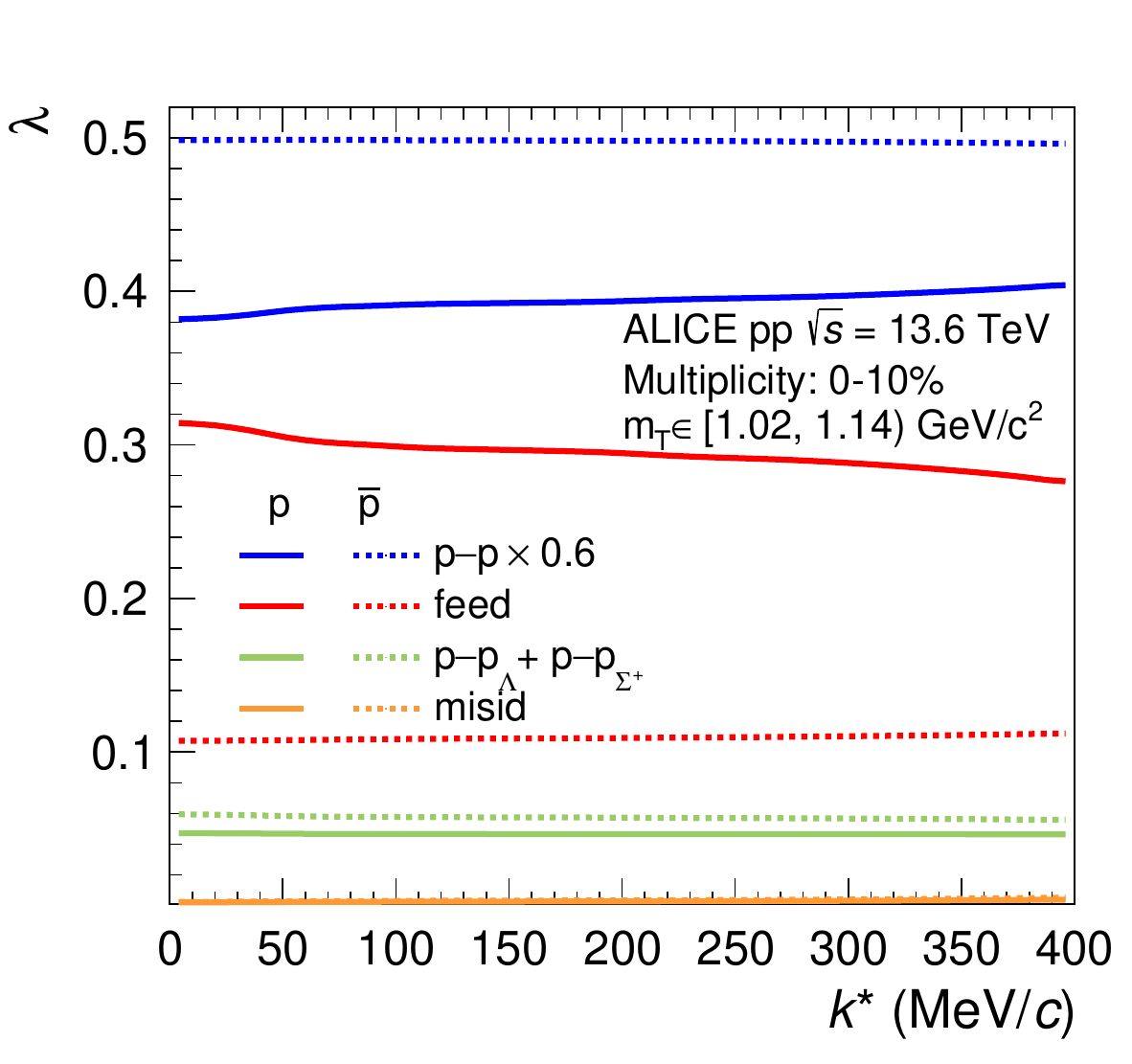}
	\includegraphics[width=0.49\linewidth]{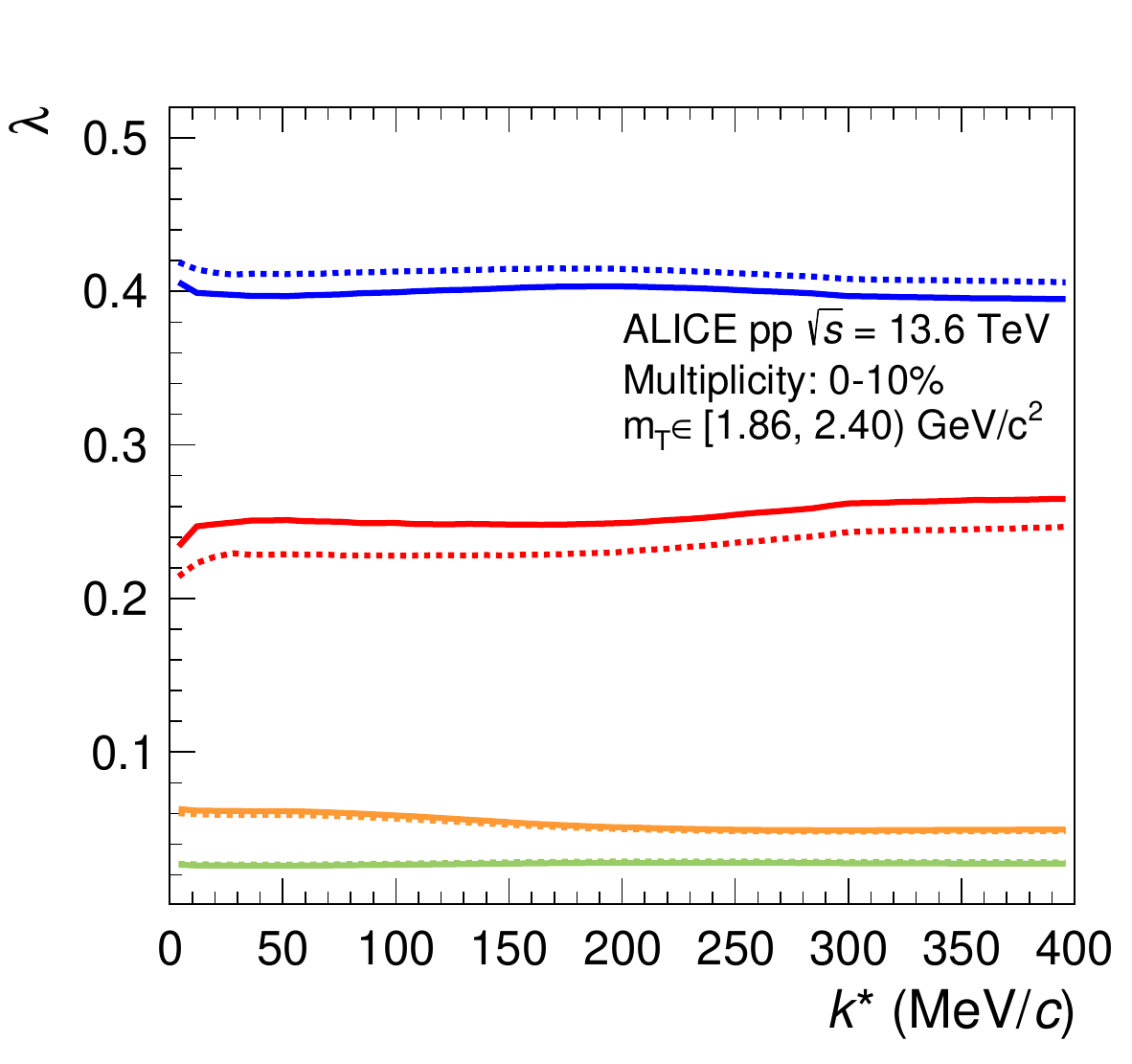}
	\caption{\kstar dependent $\lambda$ parameters for the 0--10\% multiplicity class and the lowest \mT interval (left panel) and the highest \mT interval (right panel). The $\lambda$ parameters for the other \mT intervals can be found on HEPData. The $\lambda$ parameters for protons and antiprotons are shown in a solid and dotted line, respectively. The genuine contributions were scaled down by a factor of 0.6 in order to ensure a better visibility of all lines. }
	\label{fig:lambda_parameters}
\end{figure}

\Cref{fig:lambda_parameters} shows the \kstar dependence of the $\lambda$ parameters in \cref{eq:ck_model} for \pP (solid line) and \ApAp pairs (dotted line) in the 0--10\% multiplicity class and for the lowest and highest \mT interval on the left and right panels, respectively. These are shown here as an example. The $\lambda$ parameters for all \mT and multiplicity bins are available on HEPData. For visibility, the $\lambda$ parameters for the genuine contribution are scaled down by 0.6.

The difference between protons and antiprotons is mainly due to material knock-out processes, which affect only the proton sample. These knock-out protons contribute mostly at low transverse momenta to the proton yield and therefore the differences between particle and antiparticle pairs are more pronounced in the lower \mT intervals.
The \kstar dependence of $\lambda_\mathrm{p\text{--}p_\Lambda}$ and $\lambda_\mathrm{p\text{--}p_{\Sigma^+}}$ are identical and have been combined in \cref{fig:lambda_parameters}. They can be recovered by multiplying the total contribution by 2/3 and 1/3 to obtain $\lambda_\mathrm{p\text{--}p_\Lambda}(k^*)$ and $\lambda_\mathrm{p\text{--}p_{\Sigma^+}}(k^*)$, respectively.

\begin{figure}[h!]
	\centering
	\includegraphics[width=0.49\linewidth]{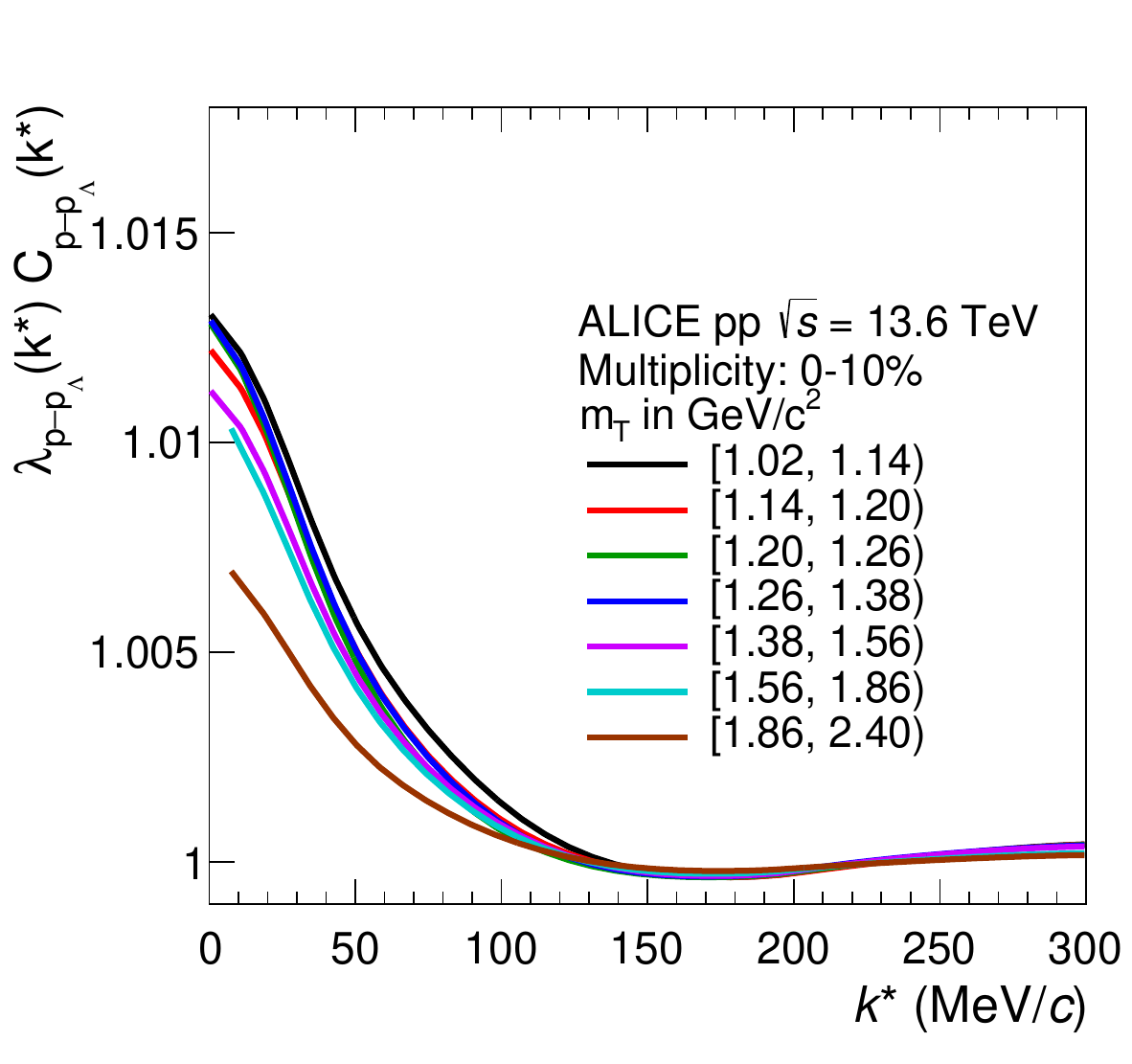}
	\includegraphics[width=0.49\linewidth]{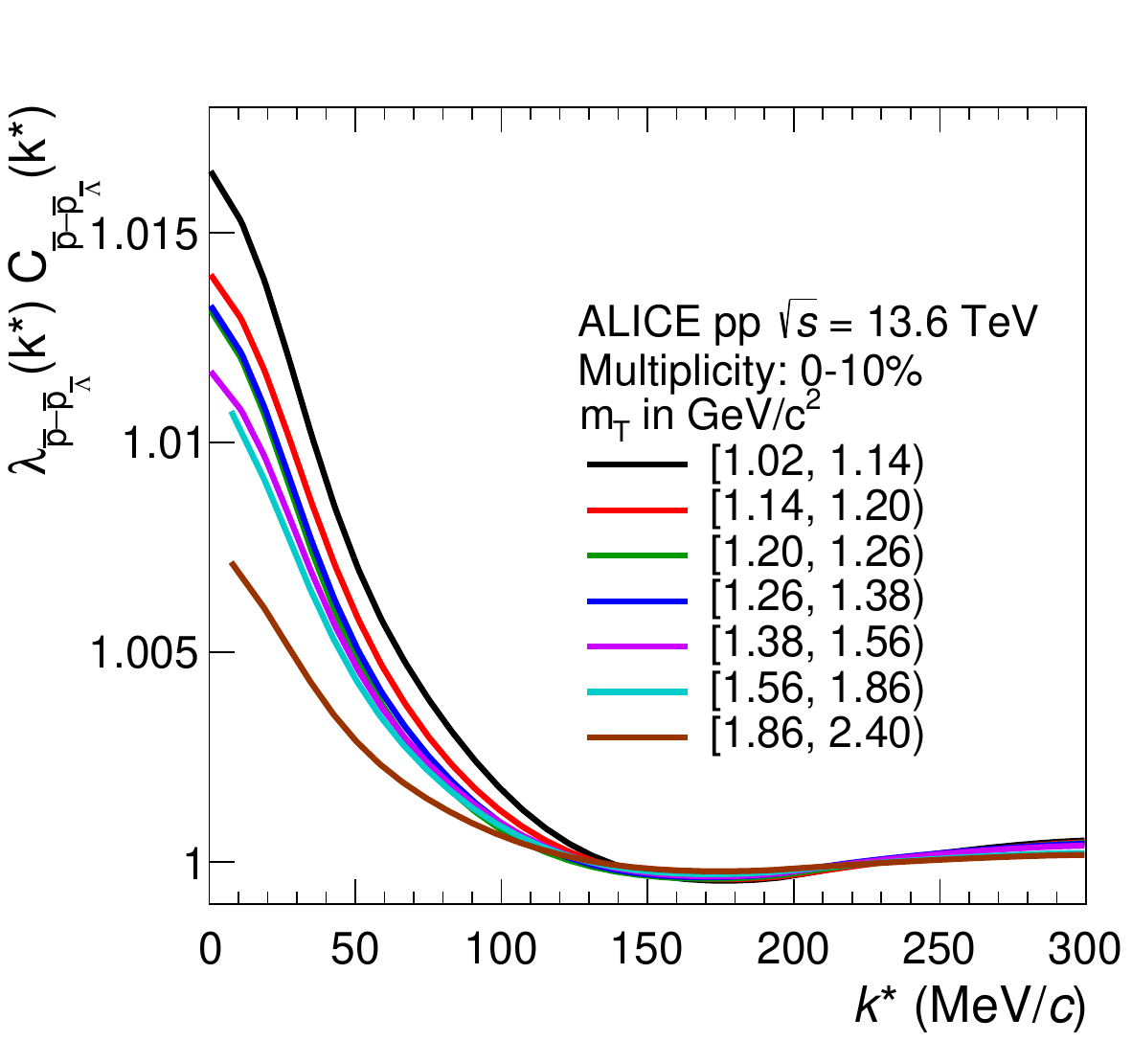}
	\caption{The shape of the $C_\mathrm{p\text{--}p_\Lambda}(k^*)$, scaled by the corresponding $\lambda$ parameters for the 0--10\% multiplicity class and all \mT intervals for particle pairs (left panel) and antiparticle pairs (right panel).}
	\label{fig:feed_down_pL}
\end{figure}

\begin{figure}[h!]
	\centering
	\includegraphics[width=0.49\linewidth]{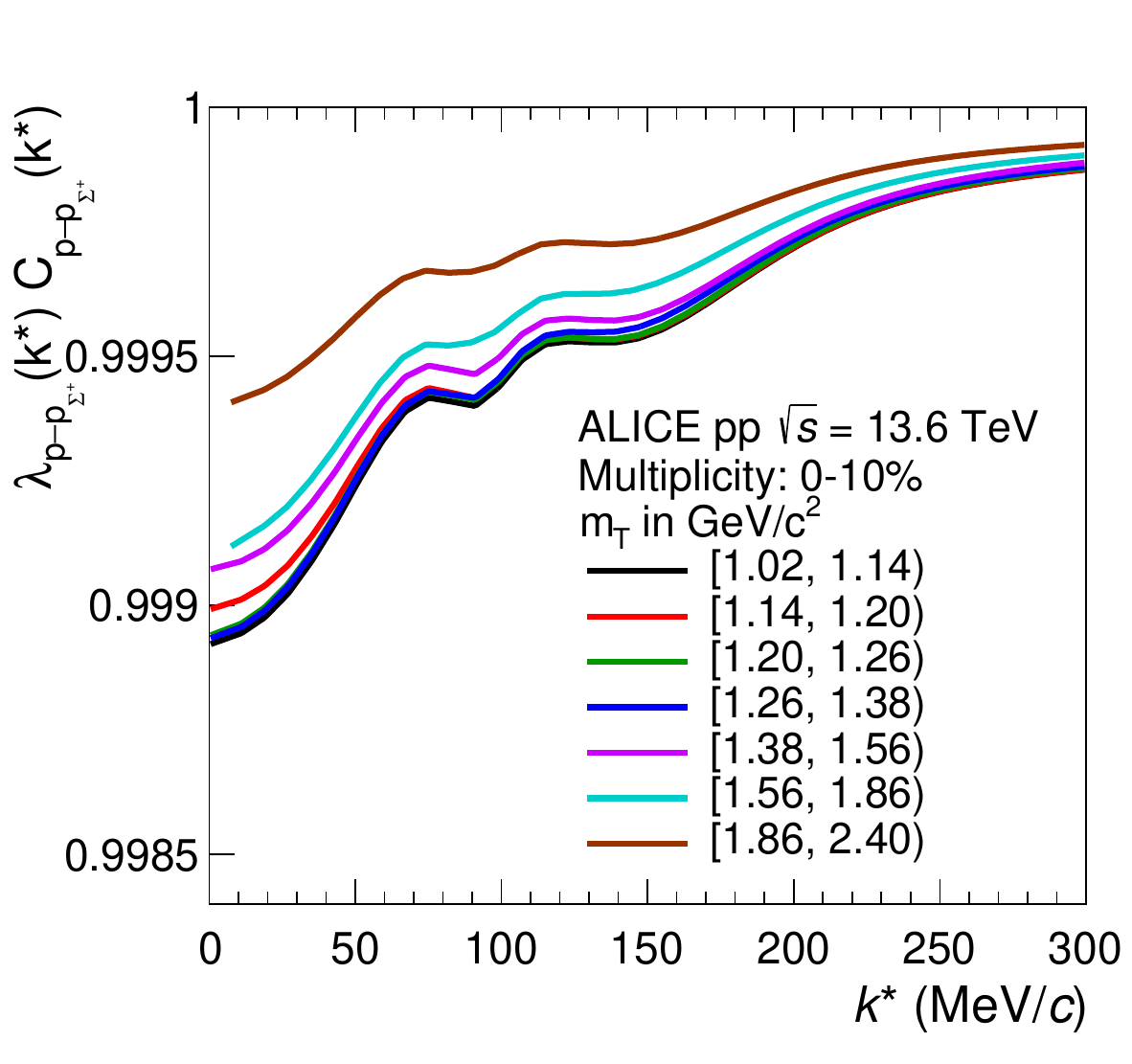}
	\includegraphics[width=0.49\linewidth]{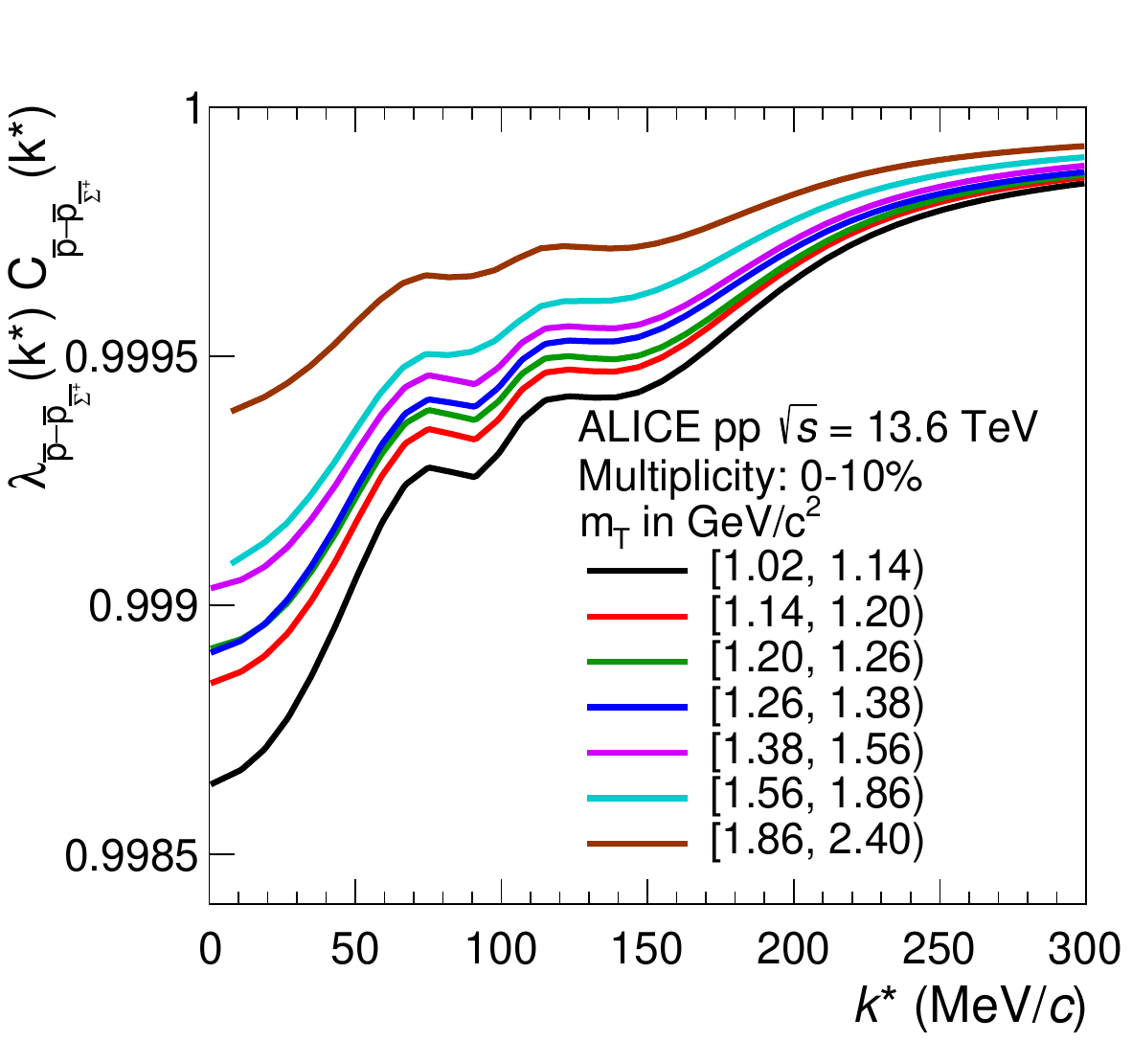}
	\caption{The shape of the $C_\mathrm{p\text{--}p_{\Sigma^{+}}}(k^*)$, scaled by the corresponding $\lambda$ parameters for the 0--10\% multiplicity class and all \mT intervals for particle pairs (left panel) and antiparticle pairs (right panel).}
	\label{fig:feed_down_pS}
\end{figure}

The scaled contributions of the feed down from \pL are shown in the left and right panels of \cref{fig:feed_down_pL} for particle and antiparticle pairs, respectively, for all \mT intervals in the 0-10\% multiplicity class. Similarly, \cref{fig:feed_down_pS} shows the feed down contributions from \pSiplus. For a better interpretation, the correlation functions have been shifted by 1 in order to converge to unity. The contribution for the other two multiplicity intervals are available on HEPData.

\clearpage
\newpage

\section{Fits of the \pP and \ApAp correlation function using the effective Gaussian source}
\label{sec:appendix_eff}
\begin{figure*}[ht]
	\begin{center}
		\includegraphics[width=0.98\linewidth]{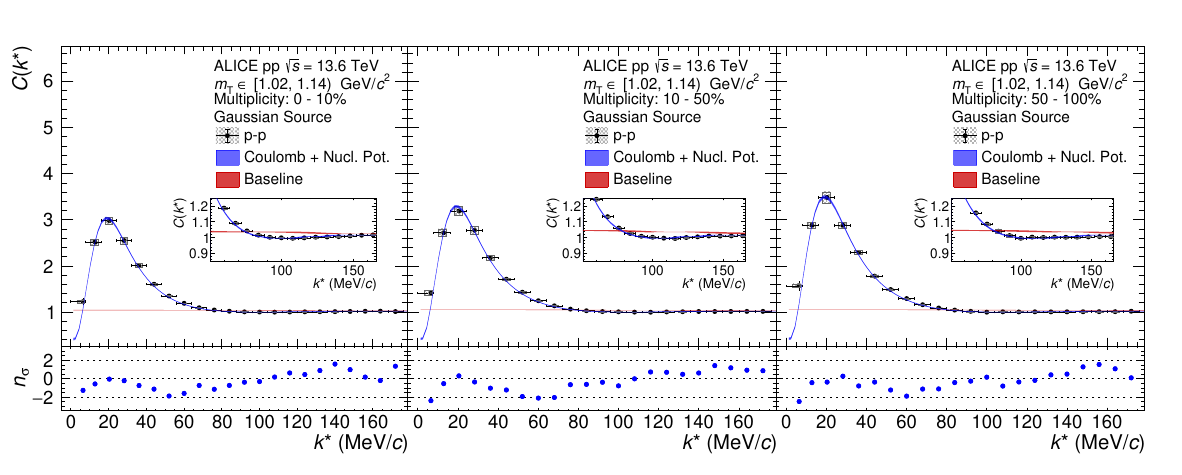}
		\includegraphics[width=0.98\linewidth]{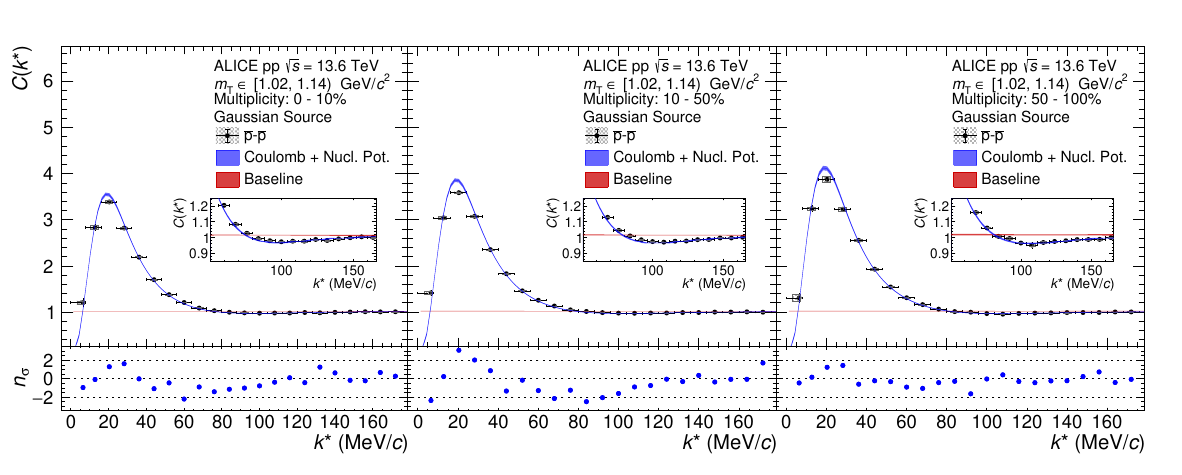}
	\end{center}
	\hfill
	\caption{Fits of the measured \pP (upper row) and \ApAp (lower row) correlation functions in all multiplicity ranges and \mT range $[1.02, 1.14]\, \si{\gevcc}$ fitted with the effective source size. For a detailed description see \cref{fig:fitsMt0}.}
	\label{fig:pp_eff_mt_0_allmult}
\end{figure*}

\begin{figure*}[ht]
	\begin{center}
		\includegraphics[width=0.98\linewidth]{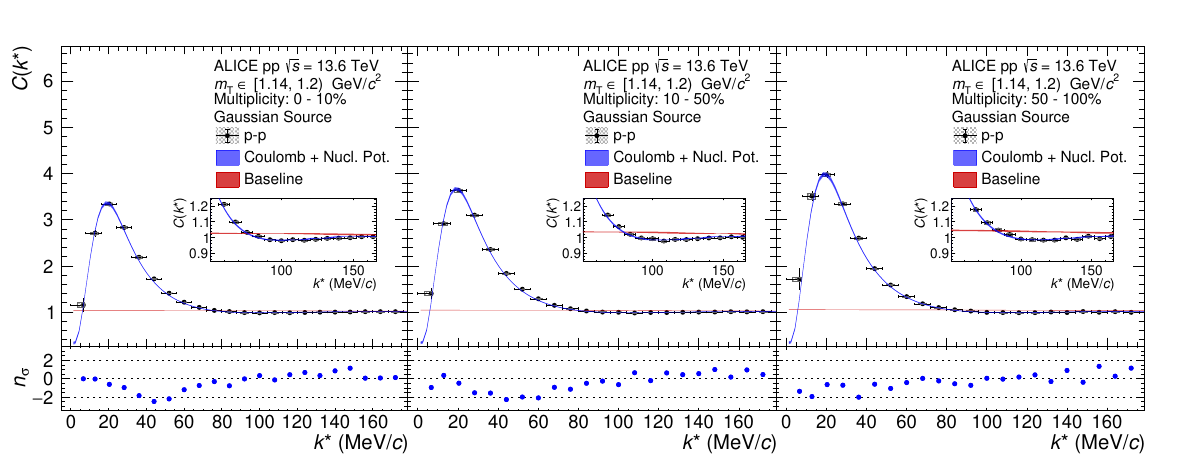}
		\includegraphics[width=0.98\linewidth]{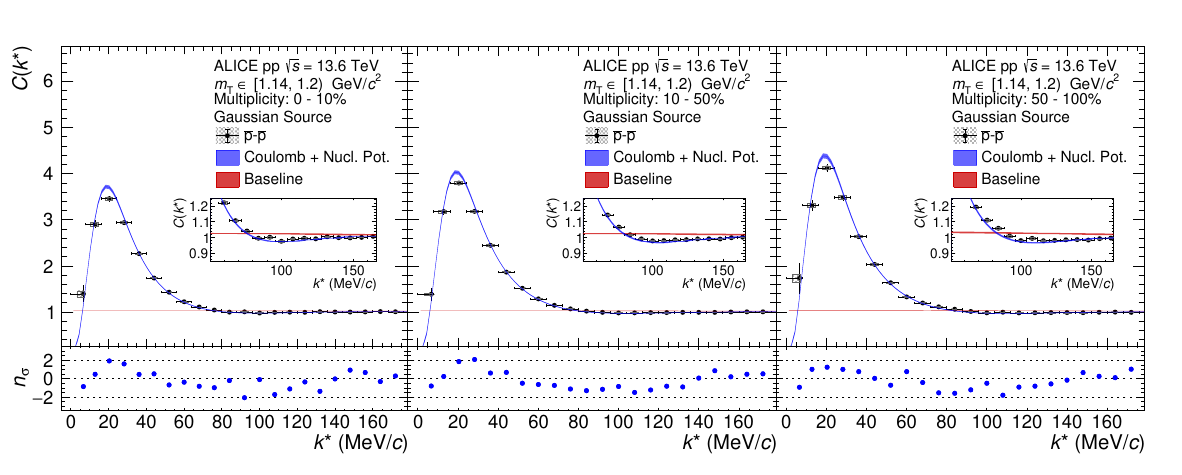}
	\end{center}
	\hfill
	\caption{Fits of the measured \pP (upper row) and \ApAp (lower row) correlation functions in all multiplicity ranges and \mT range $[1.14, 1.20]\, \si{\gevcc}$ fitted with the effective source size. For a detailed description see \cref{fig:fitsMt0}.}
	\label{fig:pp_eff_mt_1_allmult}
\end{figure*}

\begin{figure*}[ht]
	\begin{center}
		\includegraphics[width=0.98\linewidth]{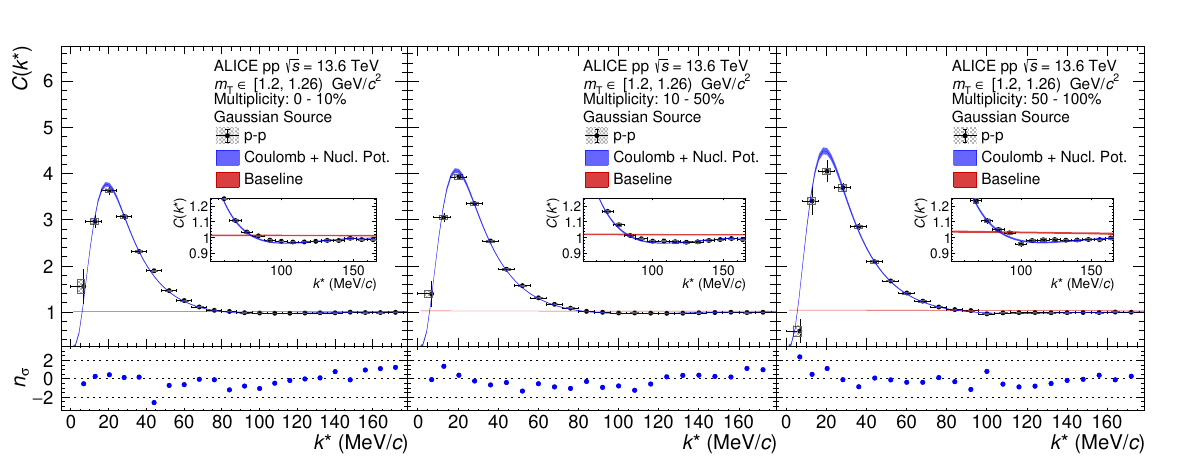}
		\includegraphics[width=0.98\linewidth]{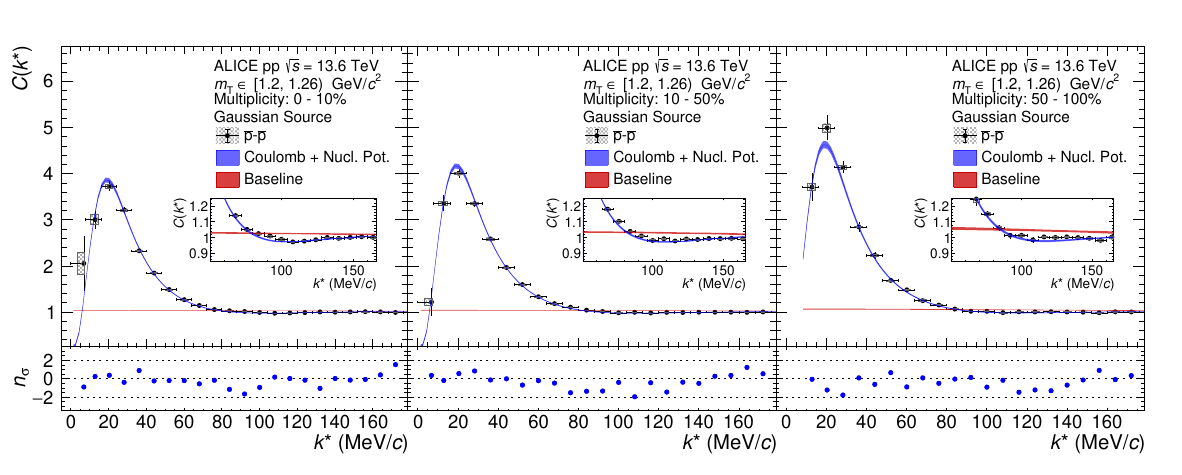}
	\end{center}
	\hfill
	\caption{Fits of the measured \pP (upper row) and \ApAp (lower row) correlation functions in all multiplicity ranges and \mT range $[1.20, 1.26]\, \si{\gevcc}$ fitted with the effective source size. For a detailed description see \cref{fig:fitsMt0}.}
	\label{fig:pp_eff_mt_2_allmult}
\end{figure*}

\begin{figure*}[ht]
	\begin{center}
		\includegraphics[width=0.98\linewidth]{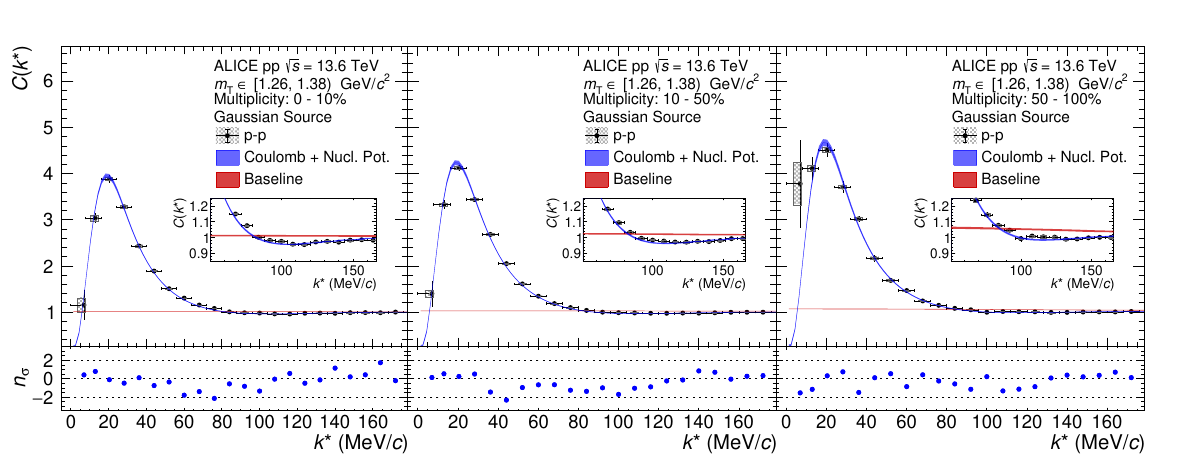}
		\includegraphics[width=0.98\linewidth]{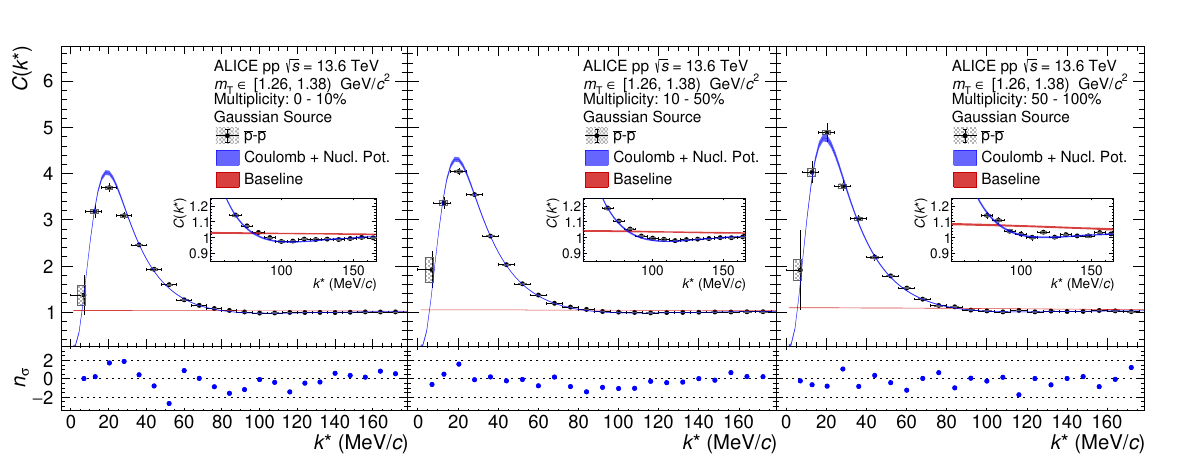}
	\end{center}
	\hfill
	\caption{Fits of the measured \pP (upper row) and \ApAp (lower row) correlation functions in all multiplicity ranges and \mT range $[1.26, 1.38]\, \si{\gevcc}$ fitted with the effective source size. For a detailed description see \cref{fig:fitsMt0}.}
	\label{fig:pp_eff_mt_3_allmult}
\end{figure*}

\begin{figure*}[ht]
	\begin{center}
		\includegraphics[width=0.98\linewidth]{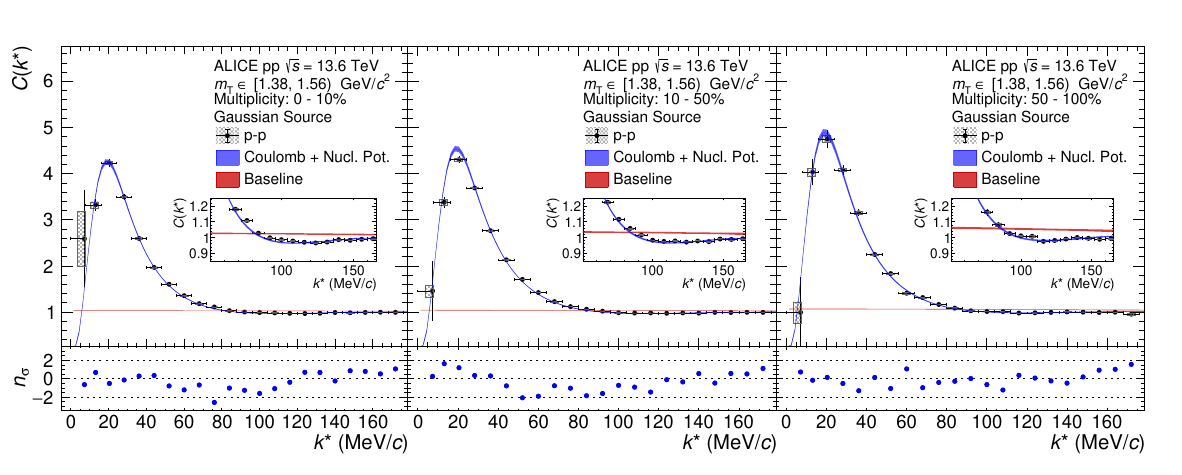}
		\includegraphics[width=0.98\linewidth]{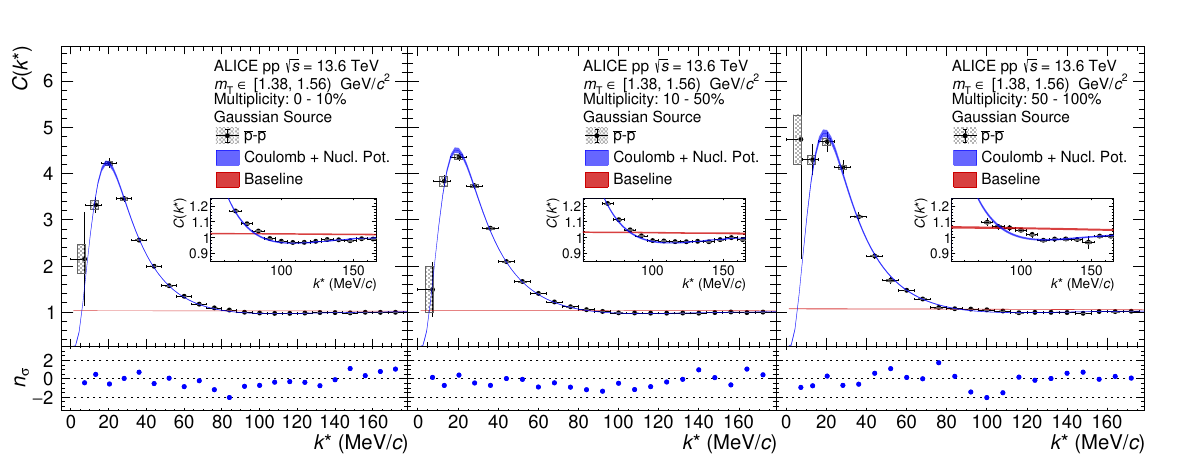}
	\end{center}
	\hfill
	\caption{Fits of the measured \pP (upper row) and \ApAp (lower row) correlation functions in all multiplicity ranges and \mT range $[1.38, 1.56]\, \si{\gevcc}$ fitted with the effective source size. For a detailed description see \cref{fig:fitsMt0}.}
	\label{fig:pp_eff_mt_4_allmult}
\end{figure*}

\begin{figure*}[ht]
	\begin{center}
		\includegraphics[width=0.98\linewidth]{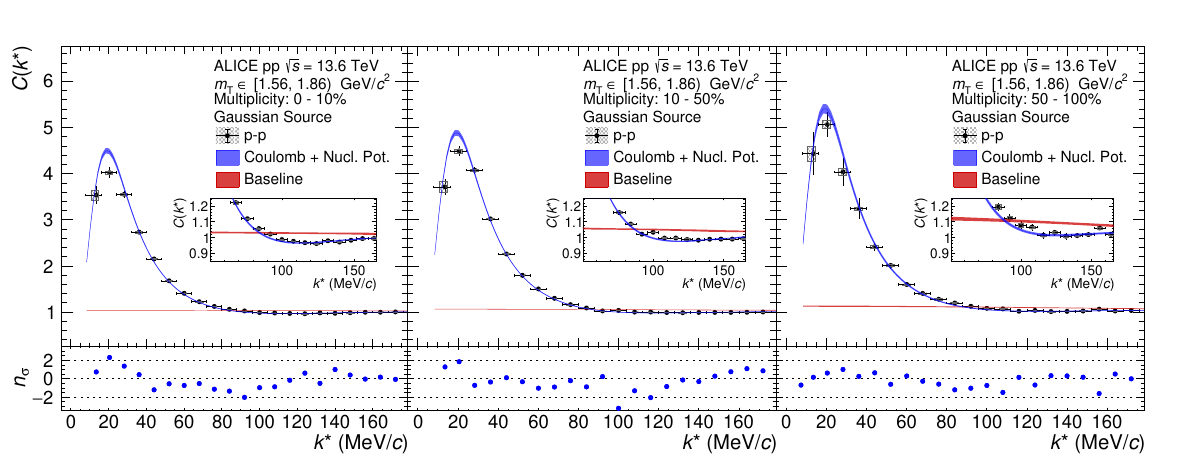}
		\includegraphics[width=0.98\linewidth]{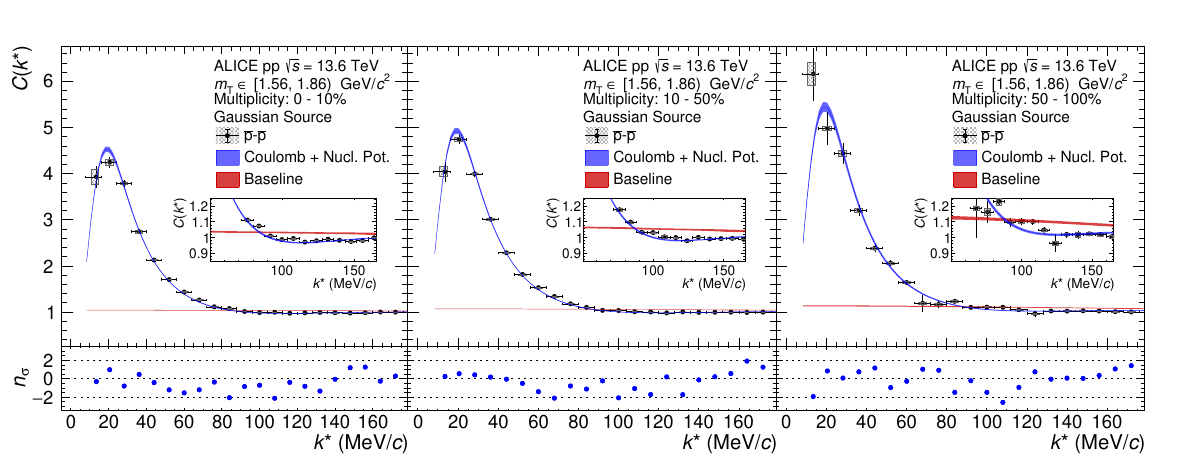}
	\end{center}
	\hfill
	\caption{Fits of the measured \pP (upper row) and \ApAp (lower row) correlation functions in all multiplicity ranges and \mT range $[1.56, 1.86]\, \si{\gevcc}$ fitted with the effective source size. For a detailed description see \cref{fig:fitsMt0}.}
	\label{fig:pp_eff_mt_5_allmult}
\end{figure*}

\begin{figure*}[ht]
	\begin{center}
		\includegraphics[width=0.98\linewidth]{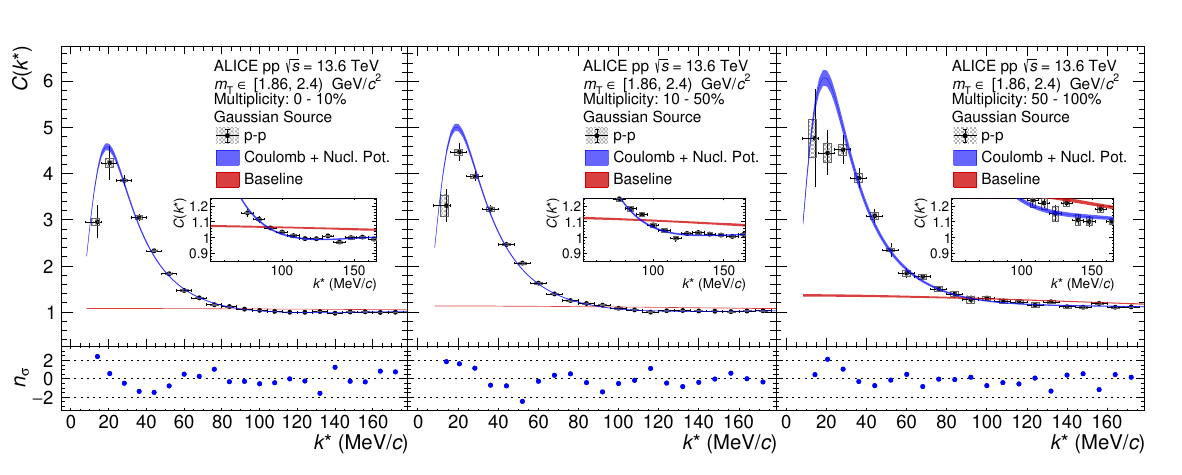}
		\includegraphics[width=0.98\linewidth]{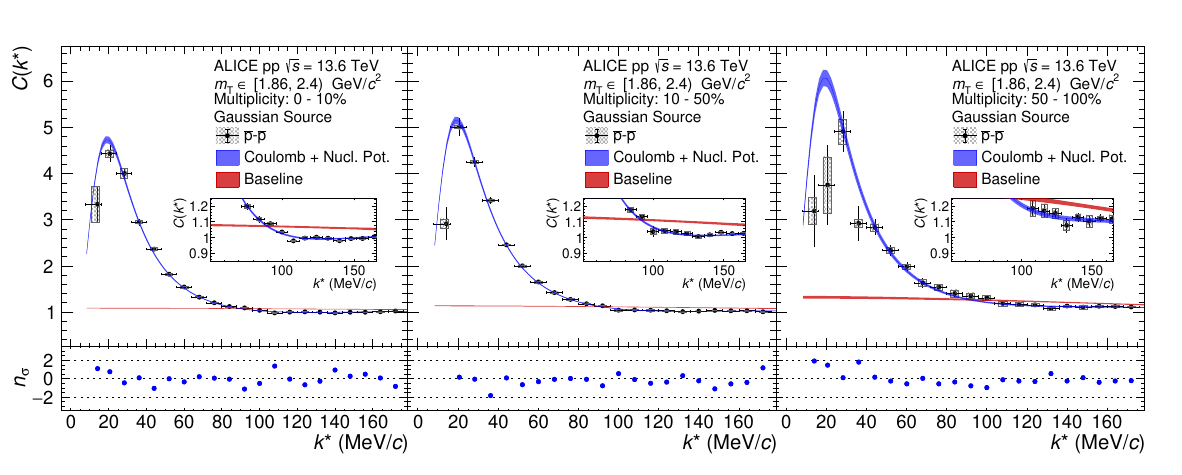}
	\end{center}
	\hfill
	\caption{Fits of the measured \pP (upper row) and \ApAp (lower row) correlation functions in all multiplicity ranges and \mT range $[1.86, 2.40]\, \si{\gevcc}$ fitted with the effective source size. For a detailed description see \cref{fig:fitsMt0}.}
	\label{fig:pp_eff_mt_6_allmult}
\end{figure*}

The fits to the \pP and \ApAp correlation functions using the effective Gaussian source in all remaining \mT bins and multiplicity intervals which were not shown in the main text, are shown in \cref{fig:pp_eff_mt_0_allmult,fig:pp_eff_mt_1_allmult,fig:pp_eff_mt_2_allmult,fig:pp_eff_mt_3_allmult,fig:pp_eff_mt_4_allmult,fig:pp_eff_mt_5_allmult,fig:pp_eff_mt_6_allmult}.

\clearpage
\newpage
\section{Fits of the \pP and \ApAp correlation function using the Resonance Source model}
\label{sec:appendix_core}
\begin{figure*}[ht]
	\begin{center}
		\includegraphics[width=0.98\linewidth]{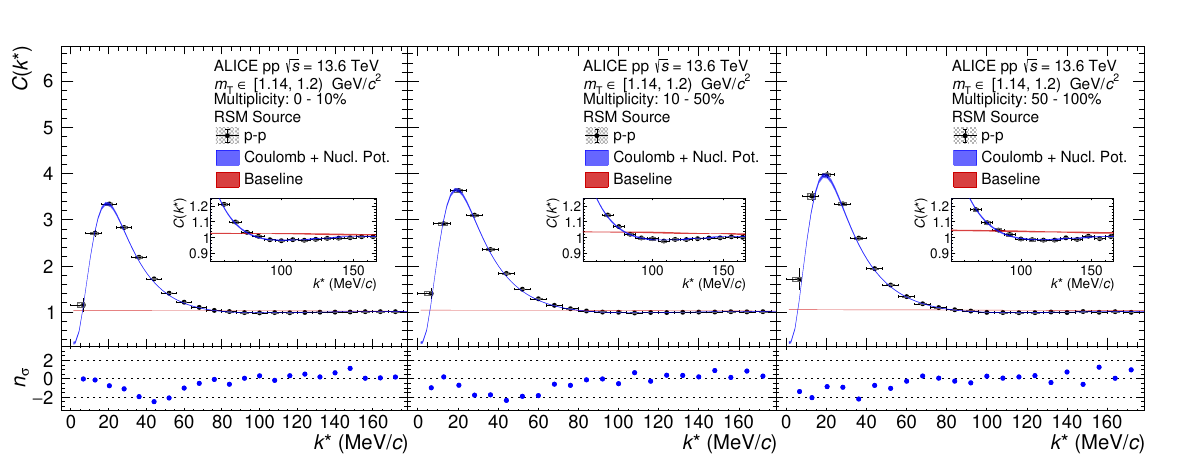}
		\includegraphics[width=0.98\linewidth]{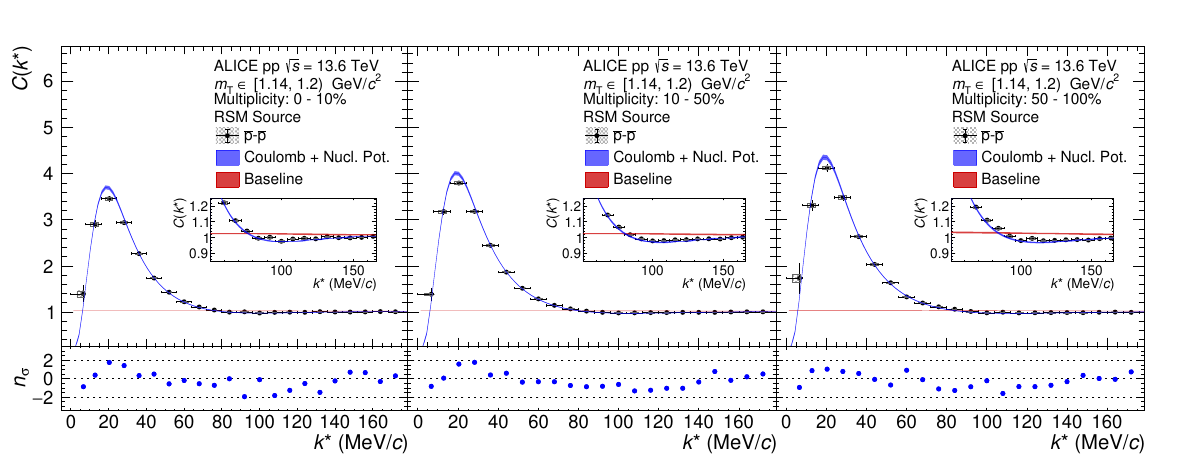}
	\end{center}
	\hfill
	\caption{Fits of the measured \pP (upper row) and \ApAp (lower row) correlation functions in all multiplicity ranges and \mT range $[1.14, 1.20]\, \si{\gevcc}$ fitted with the RSM. For a detailed description see \cref{fig:fitsMt0}.}
	\label{fig:pp_core_mt_1_allmult}
\end{figure*}

\begin{figure*}[ht]
	\begin{center}
		\includegraphics[width=0.98\linewidth]{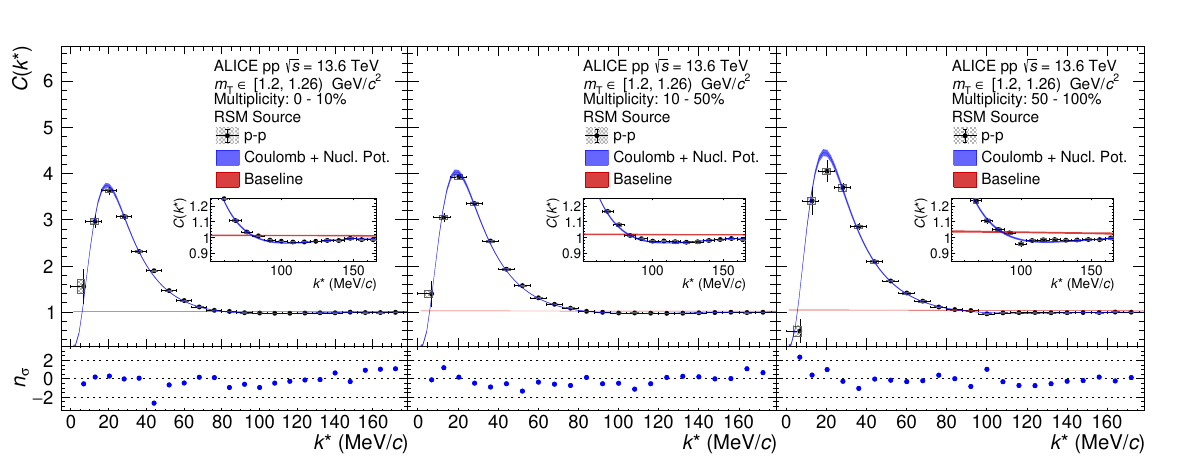}
		\includegraphics[width=0.98\linewidth]{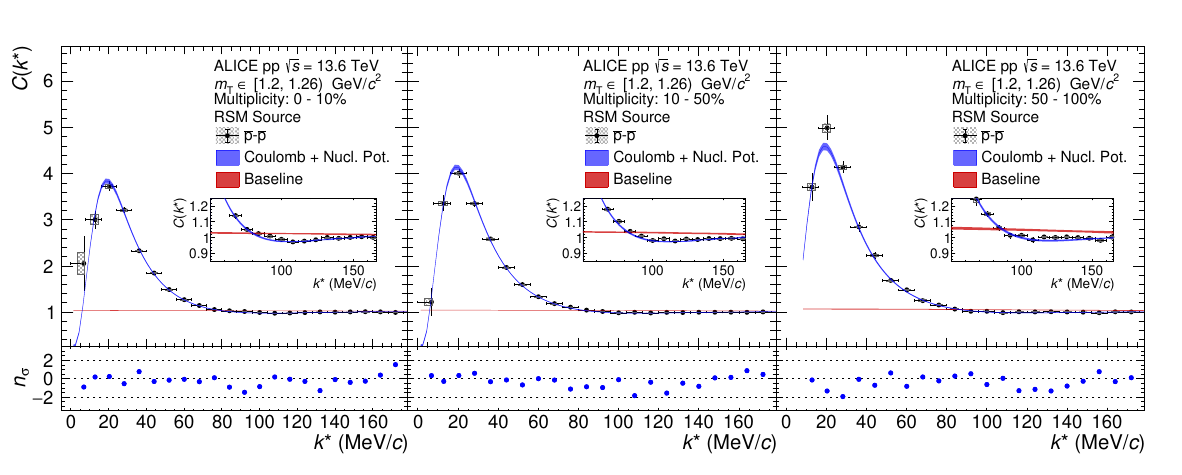}
	\end{center}
	\hfill
	\caption{Fits of the measured \pP (upper row) and \ApAp (lower row) correlation functions in all multiplicity ranges and \mT range $[1.20, 1.26]\, \si{\gevcc}$ fitted with the RSM. For a detailed description see \cref{fig:fitsMt0}.}
	\label{fig:pp_core_mt_2_allmult}
\end{figure*}

\begin{figure*}[ht]
	\begin{center}
		\includegraphics[width=0.98\linewidth]{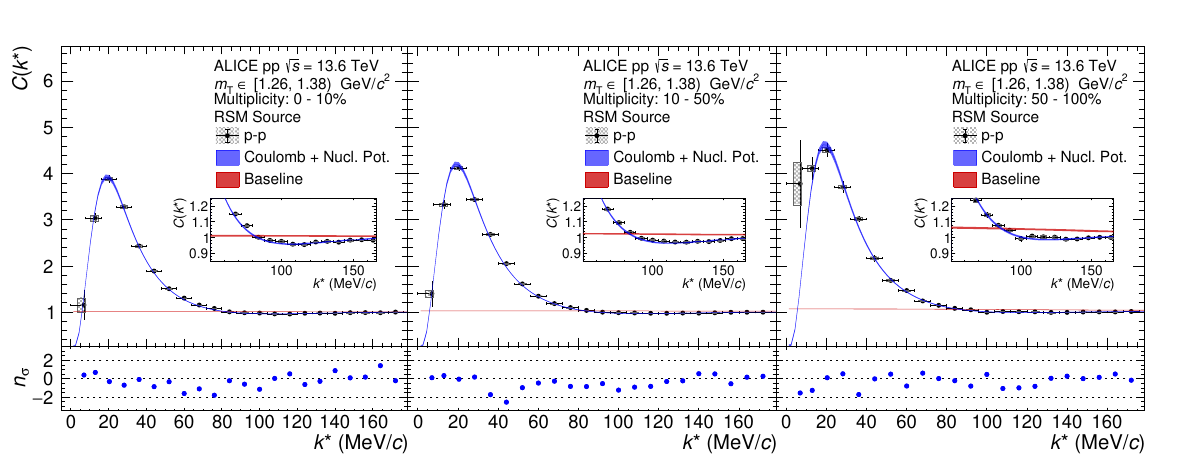}
		\includegraphics[width=0.98\linewidth]{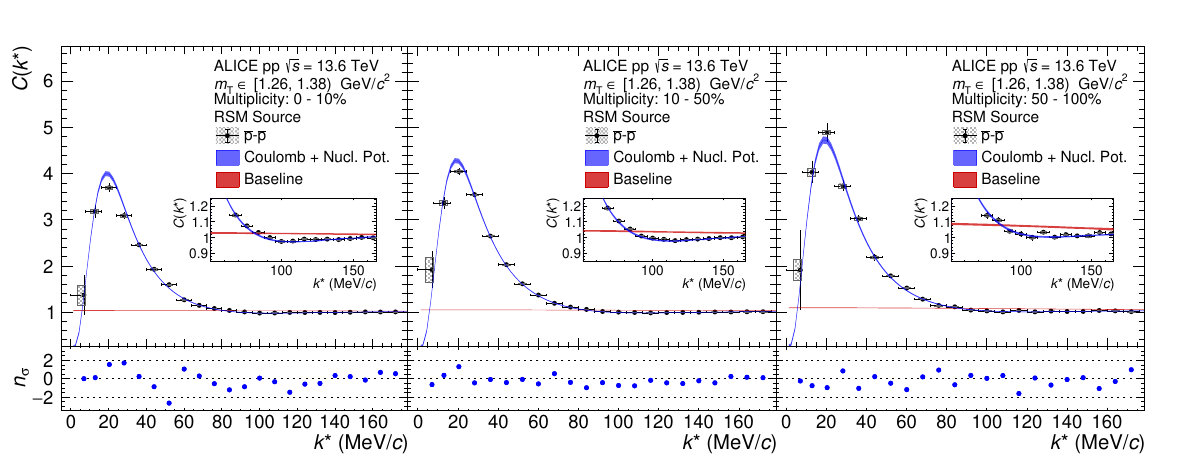}
	\end{center}
	\hfill
	\caption{Fits of the measured \pP (upper row) and \ApAp (lower row) correlation functions in all multiplicity ranges and \mT range $[1.26, 1.38]\, \si{\gevcc}$ fitted with the RSM. For a detailed description see \cref{fig:fitsMt0}.}
	\label{fig:pp_core_mt_3_allmult}
\end{figure*}

\begin{figure*}[ht]
	\begin{center}
		\includegraphics[width=0.98\linewidth]{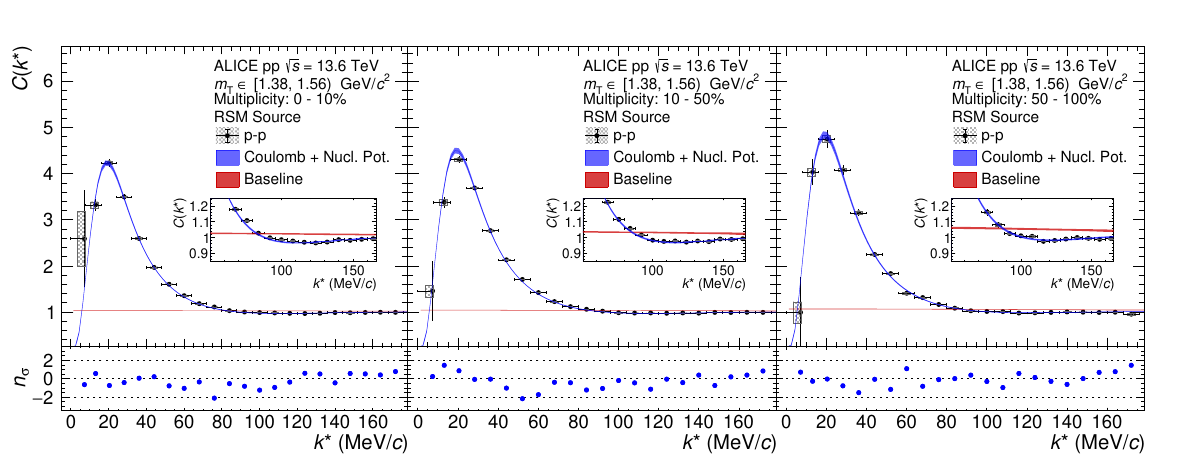}
		\includegraphics[width=0.98\linewidth]{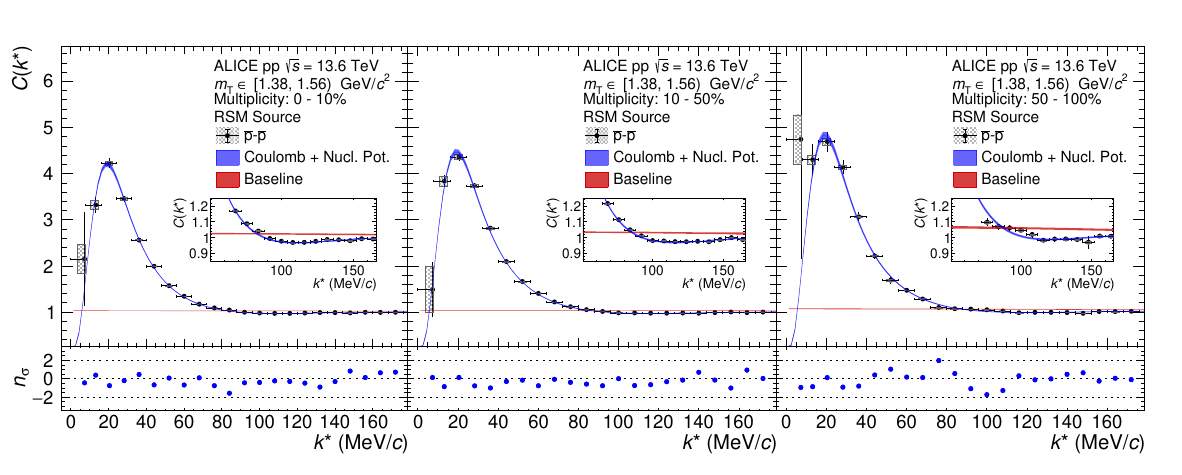}
	\end{center}
	\hfill
	\caption{Fits of the measured \pP (upper row) and \ApAp (lower row) correlation functions in all multiplicity ranges and \mT range $[1.38, 1.56]\, \si{\gevcc}$ fitted with the RSM. For a detailed description see \cref{fig:fitsMt0}.}
	\label{fig:pp_core_mt_4_allmult}
\end{figure*}

\begin{figure*}[ht]
	\begin{center}
		\includegraphics[width=0.98\linewidth]{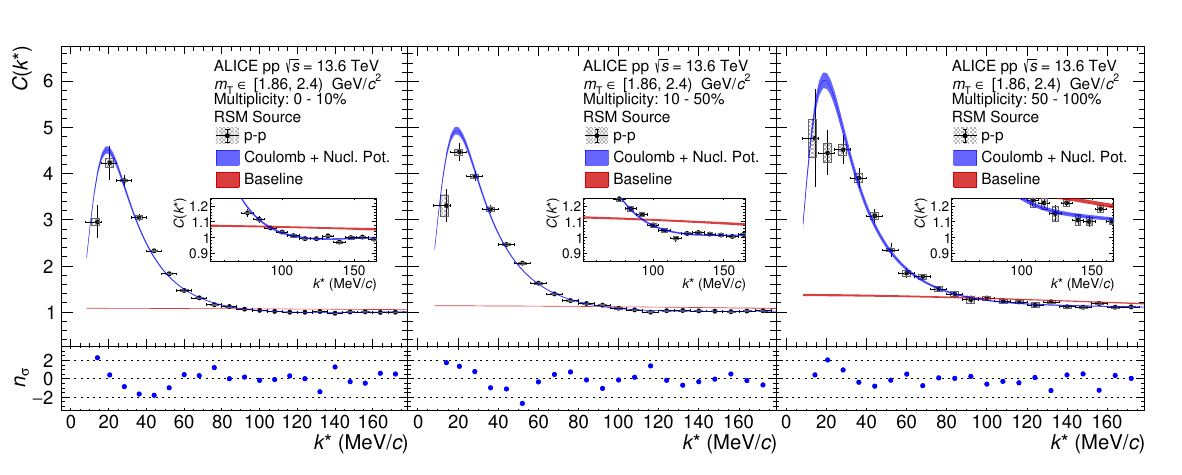}
		\includegraphics[width=0.98\linewidth]{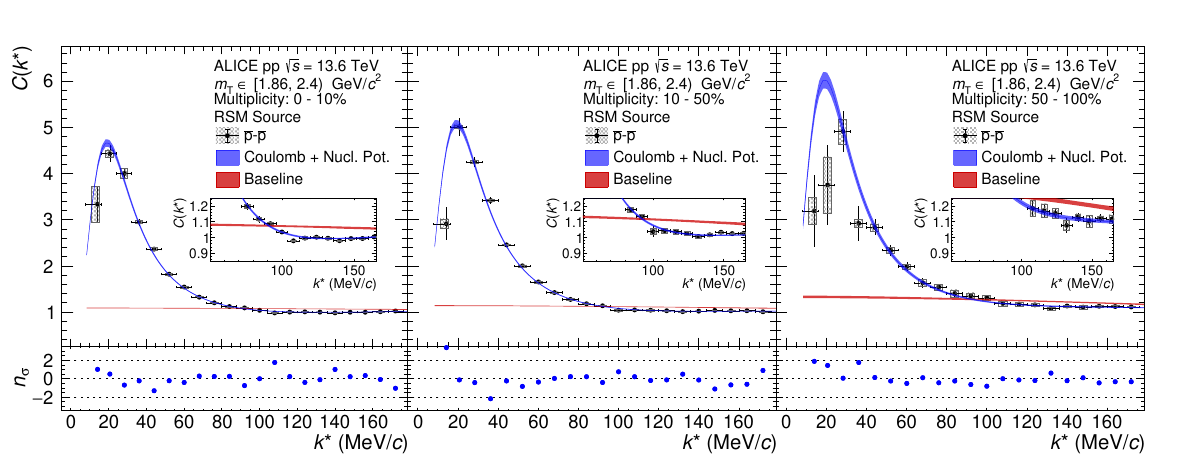}
	\end{center}
	\hfill
	\caption{Fits of the measured \pP (upper row) and \ApAp (lower row) correlation functions in all multiplicity ranges and \mT range $[1.86, 2.40]\, \si{\gevcc}$ fitted with the RSM. For a detailed description see \cref{fig:fitsMt0}.}
	\label{fig:pp_core_mt_6_allmult}
\end{figure*}

The fits to the \pP and \ApAp correlation functions using the RSM in all remaining \mT bins and multiplicity intervals which were not shown in the main text, are shown in \cref{fig:pp_core_mt_1_allmult,fig:pp_core_mt_2_allmult,fig:pp_core_mt_3_allmult,fig:pp_core_mt_4_allmult,fig:pp_core_mt_6_allmult}.

\clearpage
\newpage
\section{Fits of the \pP and \ApAp multiplicity inclusive correlation function using the effective Gaussian and the Resonance Source model}
\label{sec:appendix_MB}

\begin{figure*}[ht]
	\begin{center}
		\includegraphics[width=0.98\linewidth]{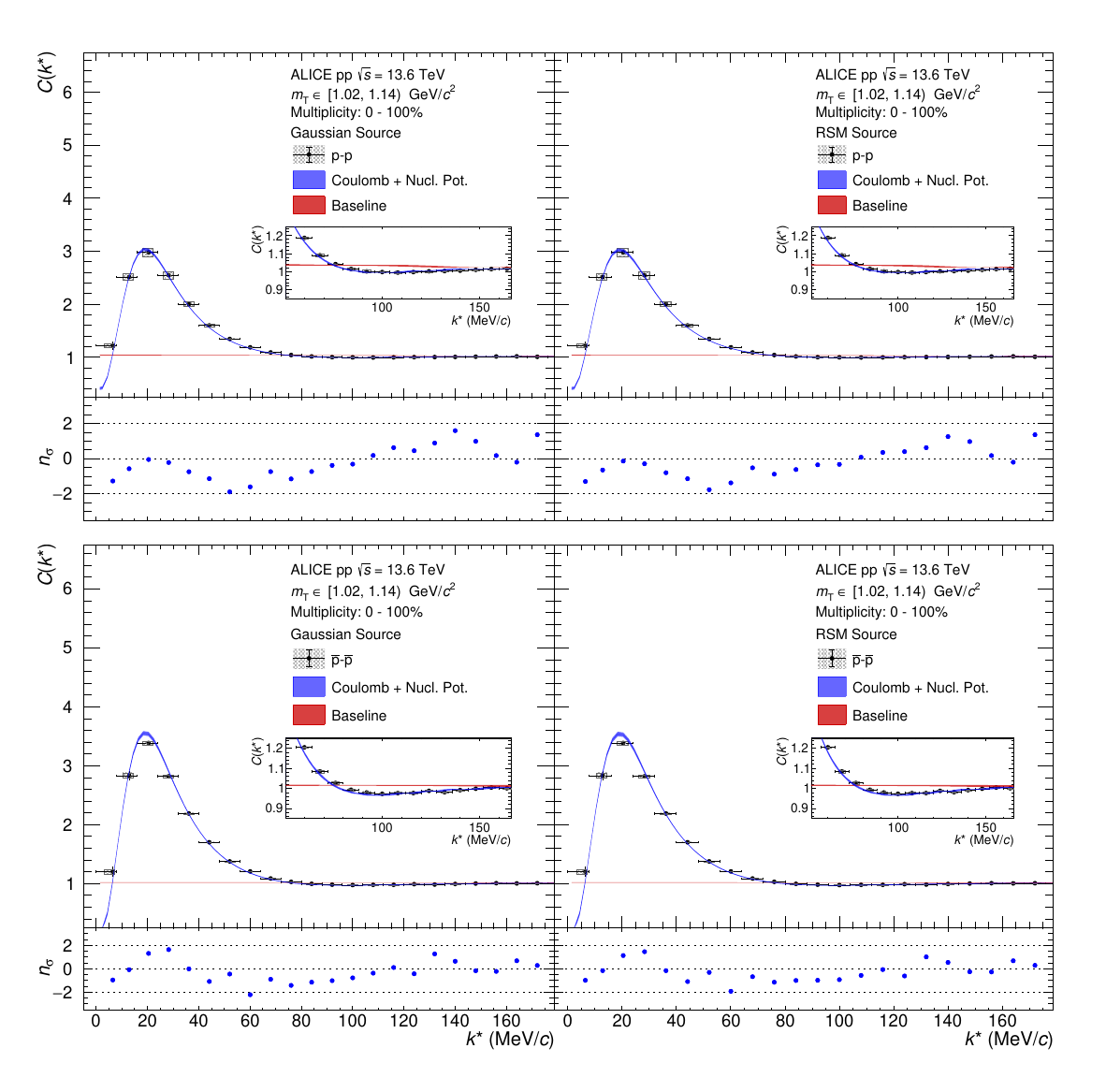}
	\end{center}
	\hfill
	\caption{Fits of the multiplicity integrated \pP (upper row) and \ApAp (lower row) correlation functions using the effective Gaussian source (left column) and the RSM (right column) in the \mT range $[1.02, 1.14]\, \si{\gevcc}$. For a detailed description of the panels see \cref{fig:fitsMt0}.}
	\label{fig:MB_mt_0}
\end{figure*}

\begin{figure*}[ht]
	\begin{center}
		\includegraphics[width=0.98\linewidth]{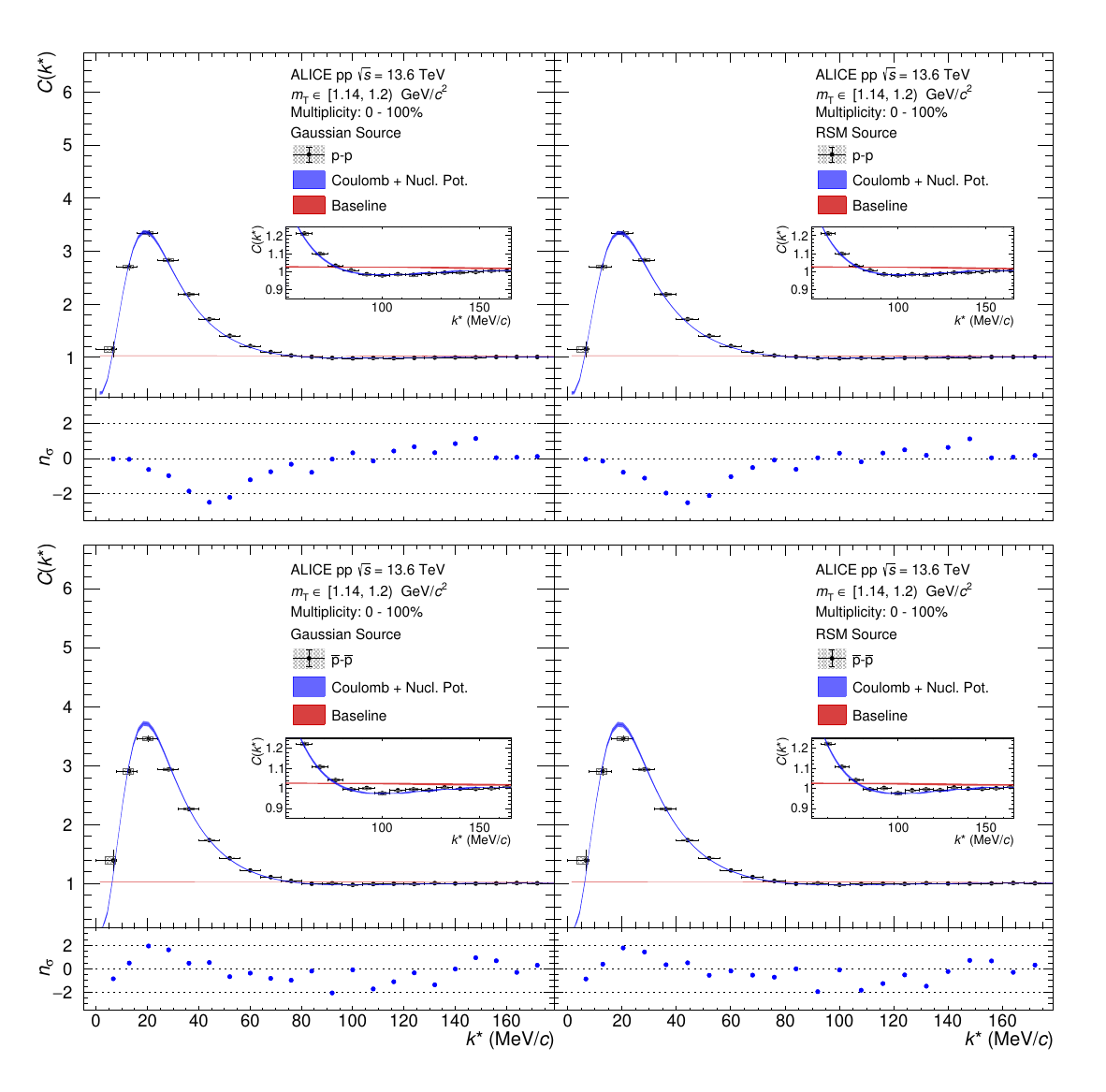}
	\end{center}
	\hfill
	\caption{Fits of the multiplicity integrated \pP (upper row) and \ApAp (lower row) correlation functions using the effective Gaussian source (left column) and the RSM (right column) in the \mT range $[1.14, 1.20]\, \si{\gevcc}$. For a detailed description of the panels see \cref{fig:fitsMt0}.}
	\label{fig:MB_mt_1}
\end{figure*}

\begin{figure*}[ht]
	\begin{center}
		\includegraphics[width=0.98\linewidth]{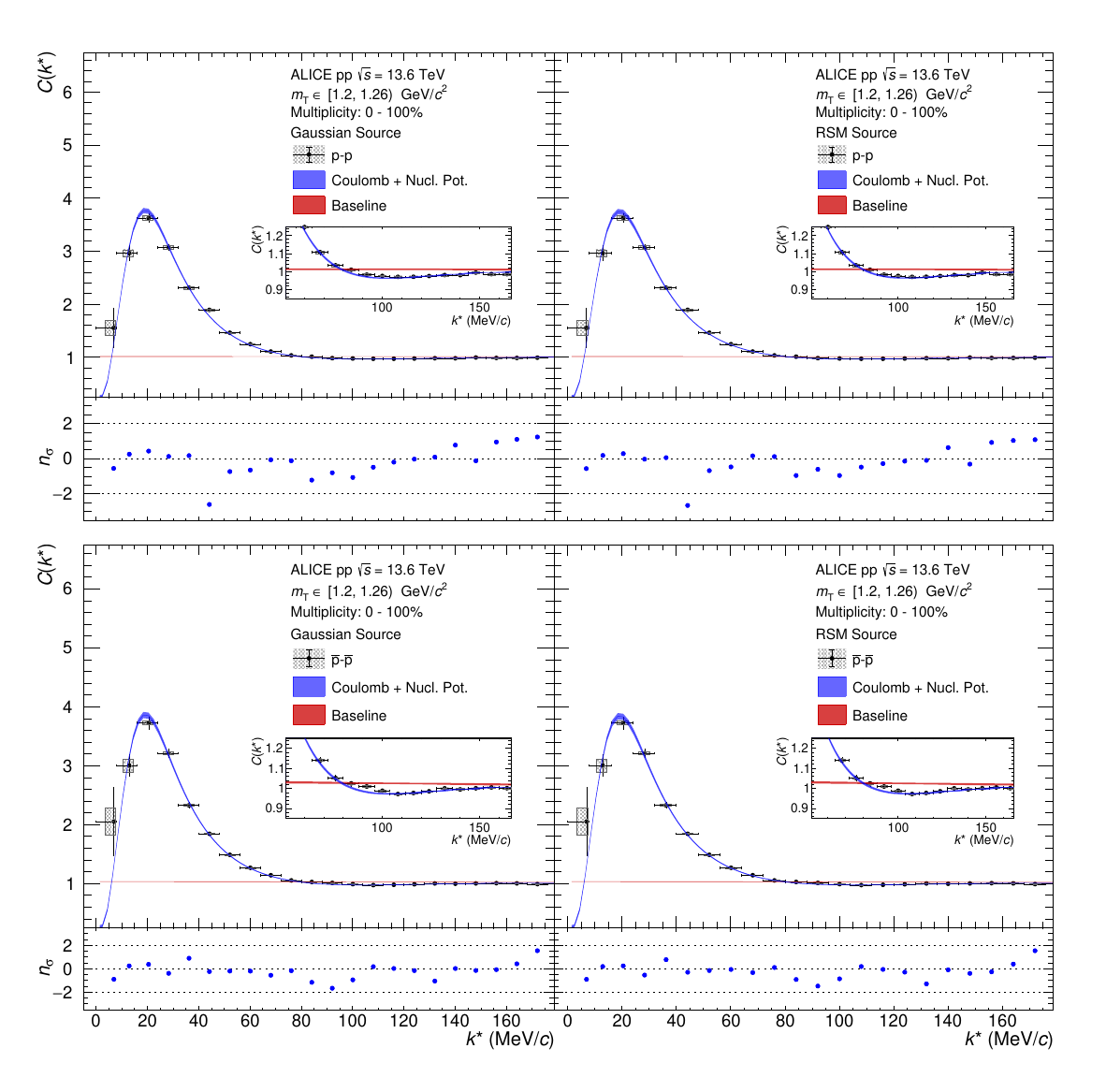}
	\end{center}
	\hfill
	\caption{Fits of the multiplicity integrated \pP (upper row) and \ApAp (lower row) correlation functions using the effective Gaussian source (left column) and the RSM (right column) in the \mT range $[1.20, 1.26]\, \si{\gevcc}$. For a detailed description of the panels see \cref{fig:fitsMt0}.}
	\label{fig:MB_mt_2}
\end{figure*}

\begin{figure*}[ht]
	\begin{center}
		\includegraphics[width=0.98\linewidth]{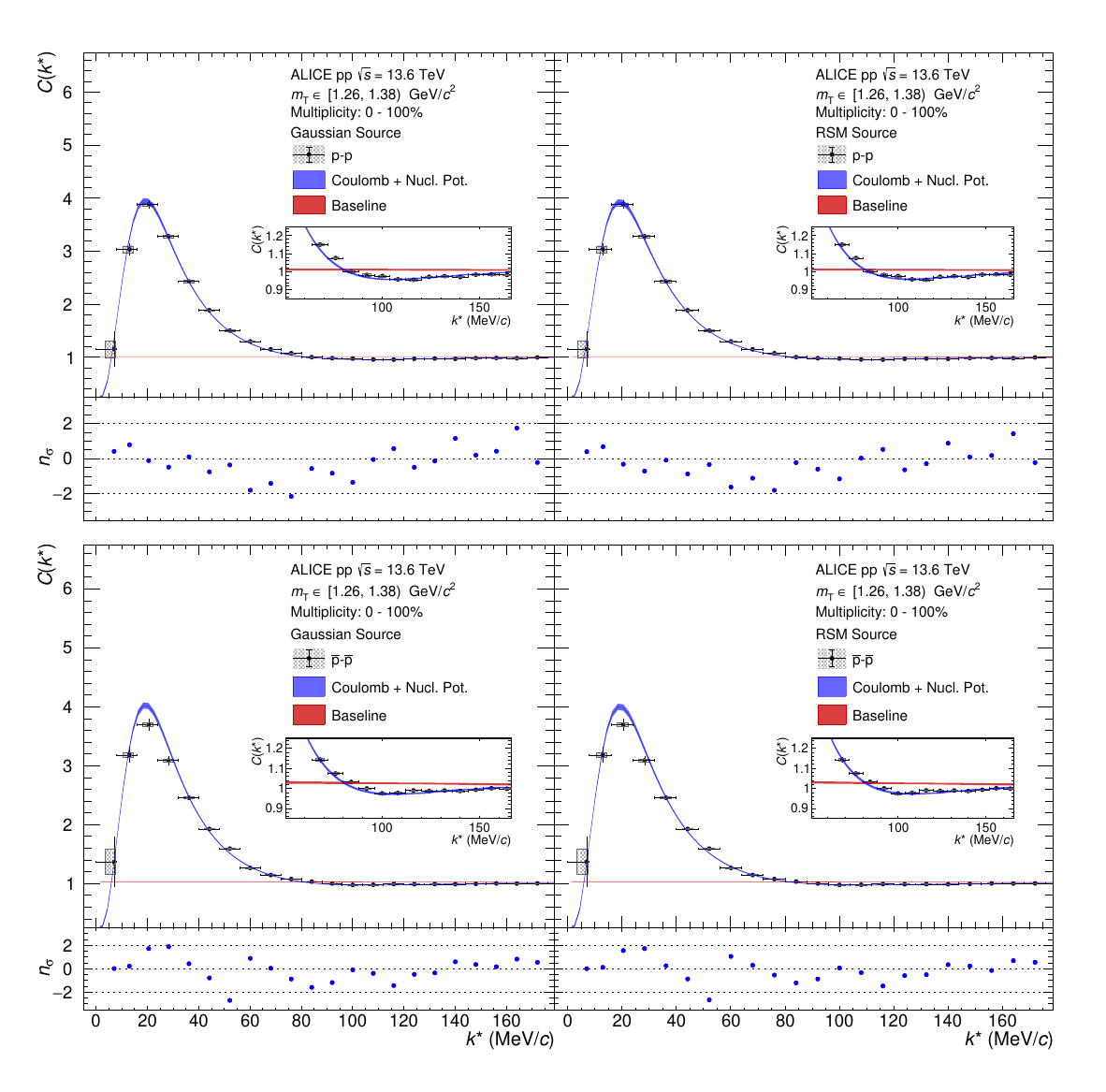}
	\end{center}
	\hfill
	\caption{Fits of the multiplicity integrated \pP (upper row) and \ApAp (lower row) correlation functions using the effective Gaussian source (left column) and the RSM (right column) in the \mT range $[1.26, 1.38]\, \si{\gevcc}$. For a detailed description of the panels see \cref{fig:fitsMt0}.}
	\label{fig:MB_mt_3}
\end{figure*}

\begin{figure*}[ht]
	\begin{center}
		\includegraphics[width=0.98\linewidth]{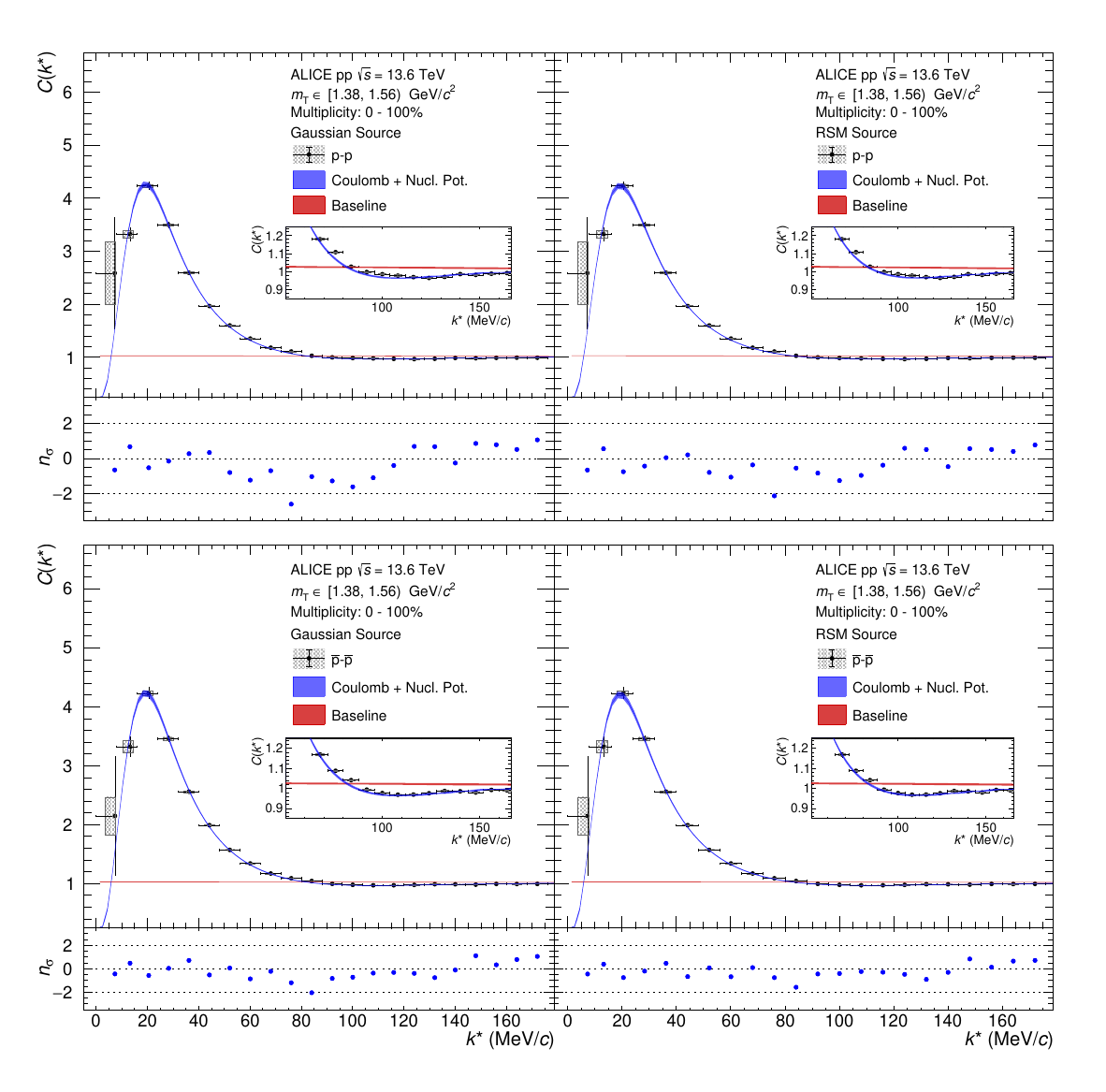}
	\end{center}
	\hfill
	\caption{Fits of the multiplicity integrated \pP (upper row) and \ApAp (lower row) correlation functions using the effective Gaussian source (left column) and the RSM (right column) in the \mT range $[1.38, 1.56]\, \si{\gevcc}$. For a detailed description of the panels see \cref{fig:fitsMt0}.}
	\label{fig:MB_mt_4}
\end{figure*}

\begin{figure*}[ht]
	\begin{center}
		\includegraphics[width=0.98\linewidth]{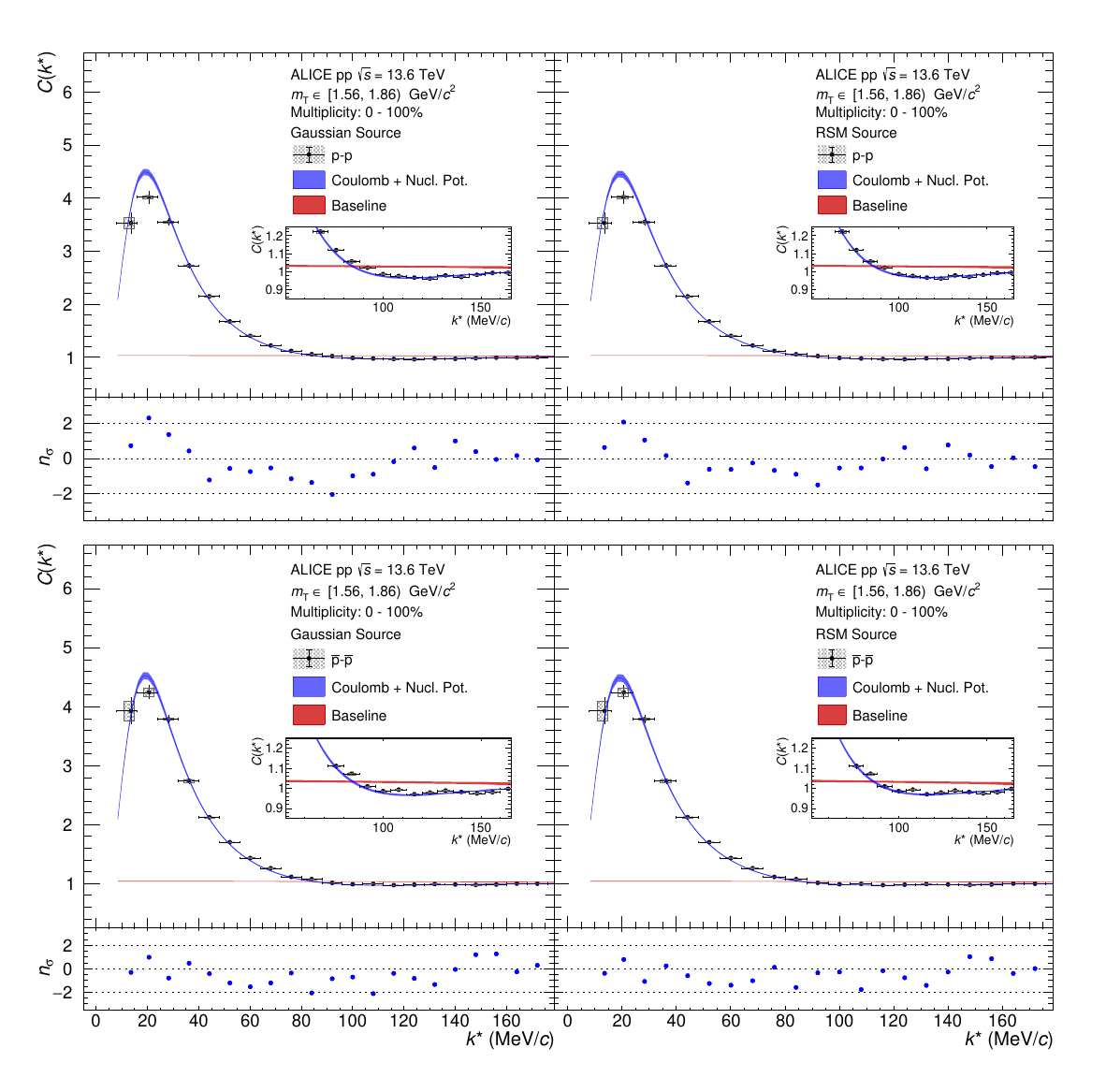}
	\end{center}
	\hfill
	\caption{Fits of the multiplicity integrated \pP (upper row) and \ApAp (lower row) correlation functions using the effective Gaussian source (left column) and the RSM (right column) in the \mT range $[1.56, 1.86]\, \si{\gevcc}$. For a detailed description of the panels see \cref{fig:fitsMt0}.}
	\label{fig:MB_mt_5}
\end{figure*}

\begin{figure*}[ht]
	\begin{center}
		\includegraphics[width=0.98\linewidth]{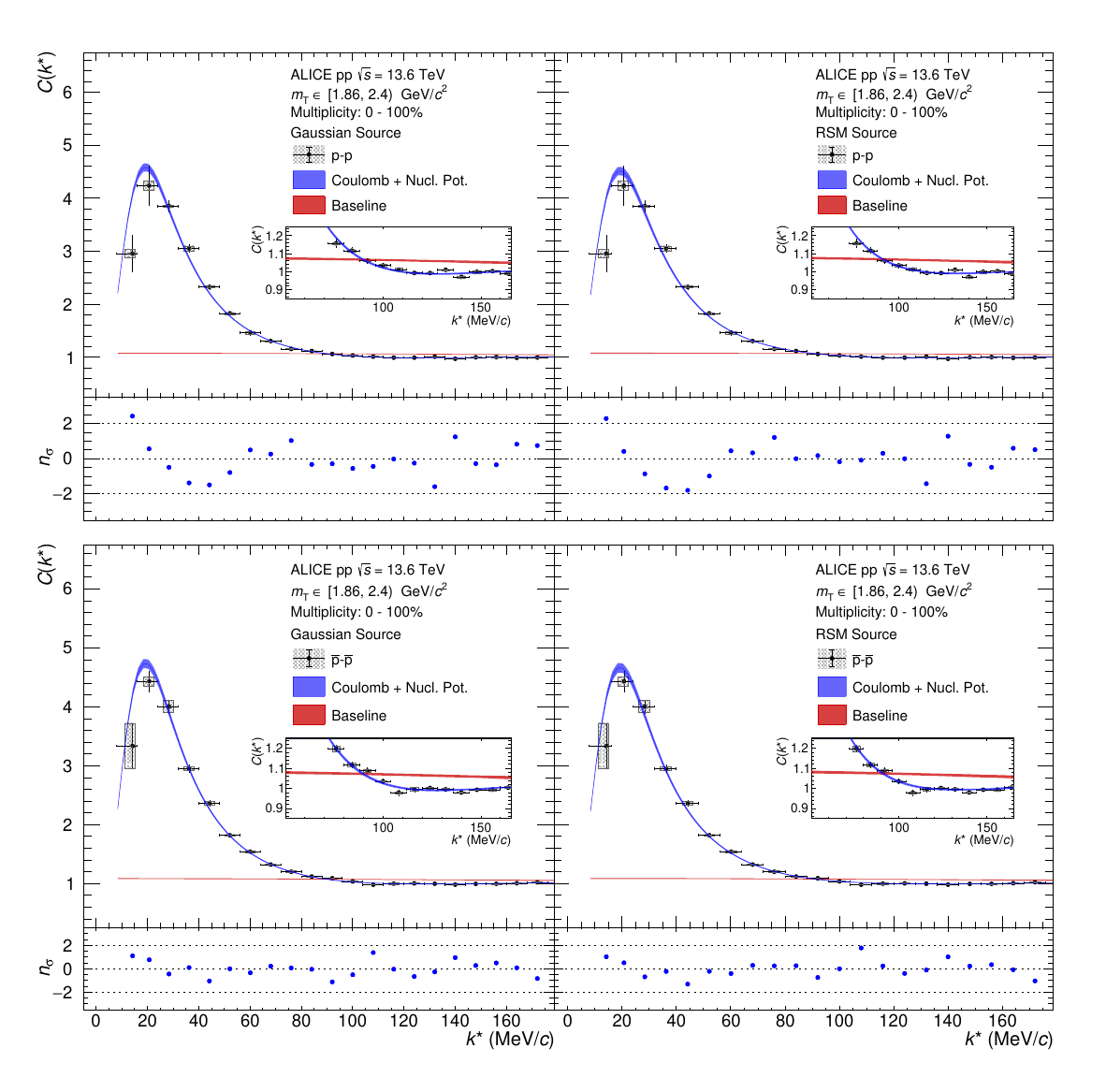}
	\end{center}
	\hfill
	\caption{Fits of the multiplicity integrated \pP (upper row) and \ApAp (lower row) correlation functions using the effective Gaussian source (left column) and the RSM (right column) in the \mT range $[1.86, 2.40]\, \si{\gevcc}$. For a detailed description of the panels see \cref{fig:fitsMt0}.}
	\label{fig:MB_mt_6}
\end{figure*}

The fits to the \pP and \ApAp multiplicity inclusive correlation functions in all \mT bins using the effective Gaussian source and the RSM are shown in \cref{fig:MB_mt_0,fig:MB_mt_1,fig:MB_mt_2,fig:MB_mt_3,fig:MB_mt_4,fig:MB_mt_5,fig:MB_mt_6}.

\clearpage
%
%

\section{The ALICE Collaboration}
\label{app:collab}
\begin{flushleft} 
\small

D.A.H.~Abdallah\,\orcidlink{0000-0003-4768-2718}\,$^{\rm 134}$, 
I.J.~Abualrob\,\orcidlink{0009-0005-3519-5631}\,$^{\rm 112}$, 
S.~Acharya\,\orcidlink{0000-0002-9213-5329}\,$^{\rm 49}$, 
K.~Agarwal\,\orcidlink{0000-0001-5781-3393}\,$^{\rm II,}$$^{\rm 23}$, 
G.~Aglieri Rinella\,\orcidlink{0000-0002-9611-3696}\,$^{\rm 32}$, 
L.~Aglietta\,\orcidlink{0009-0003-0763-6802}\,$^{\rm 24}$, 
N.~Agrawal\,\orcidlink{0000-0003-0348-9836}\,$^{\rm 25}$, 
Z.~Ahammed\,\orcidlink{0000-0001-5241-7412}\,$^{\rm 132}$, 
S.~Ahmad\,\orcidlink{0000-0003-0497-5705}\,$^{\rm 15}$, 
Z.~Akbar$^{\rm 79}$
% ...

\section*{Affiliation Notes}

$^{\rm I}$ Deceased\\
$^{\rm II}$ Also at: INFN Trieste, Trieste, Italy\\
$^{\rm III}$ Also at: Fondazione Bruno Kessler (FBK), Trento, Italy\\
$^{\rm IV}$ Also at: Czech Technical University in Prague, Prague, Czech Republic\\
$^{\rm V}$ Also at: Instituto de Fisica da Universidade de Sao Paulo\\
$^{\rm VI}$ Also at: Dipartimento DET del Politecnico di Torino, Turin, Italy\\
$^{\rm VII}$ Also at: University College of Dublin, Dublin, Ireland\\
$^{\rm VIII}$ Also at: Department of Applied Physics, Aligarh Muslim University, Aligarh, India\\
$^{\rm IX}$ Also at: Institute of Theoretical Physics, University of Wroclaw, Wroclaw, Poland\\
$^{\rm X}$ Also at: Facultad de Ciencias, Universidad Nacional Aut\'{o}noma de M\'{e}xico, Mexico City, Mexico\\

\section*{Collaboration Institutes}

$^{1}$ A.I. Alikhanyan National Science Laboratory (Yerevan Physics Institute) Foundation, Yerevan, Armenia\\
$^{2}$ AGH University of Krakow, Cracow, Poland\\
$^{3}$ Bogolyubov Institute for Theoretical Physics, National Academy of Sciences of Ukraine, Kyiv, Ukraine\\
$^{4}$ Bose Institute, Department of Physics  and Centre for Astroparticle Physics and Space Science (CAPSS), Kolkata, India\\
$^{5}$ California Polytechnic State University, San Luis Obispo, California, United States\\
$^{6}$ Central China Normal University, Wuhan, China\\
$^{7}$ Centro de Aplicaciones Tecnol\'{o}gicas y Desarrollo Nuclear (CEADEN), Havana, Cuba\\
$^{8}$ Centro de Investigaci\'{o}n y de Estudios Avanzados (CINVESTAV), Mexico City and M\'{e}rida, Mexico\\
$^{9}$ Chicago State University, Chicago, Illinois, United States\\
$^{10}$ China Nuclear Data Center, China Institute of Atomic Energy, Beijing, China\\
$^{11}$ China University of Geosciences, Wuhan, China\\
$^{12}$ Chungbuk National University, Cheongju, Republic of Korea\\
$^{13}$ Comenius University Bratislava, Faculty of Mathematics, Physics and Informatics, Bratislava, Slovak Republic\\
$^{14}$ Creighton University, Omaha, Nebraska, United States\\
$^{15}$ Department of Physics, Aligarh Muslim University, Aligarh, India\\
$^{16}$ Department of Physics, Pusan National University, Pusan, Republic of Korea\\
$^{17}$ Department of Physics, Sejong University, Seoul, Republic of Korea\\
$^{18}$ Department of Physics, University of California, Berkeley, California, United States\\
$^{19}$ Department of Physics, University of Oslo, Oslo, Norway\\
$^{20}$ Department of Physics and Technology, University of Bergen, Bergen, Norway\\
$^{21}$ Dipartimento di Fisica, Universit\`{a} di Pavia, Pavia, Italy\\
$^{22}$ Dipartimento di Fisica dell'Universit\`{a} and Sezione INFN, Cagliari, Italy\\
$^{23}$ Dipartimento di Fisica dell'Universit\`{a} and Sezione INFN, Trieste, Italy\\
$^{24}$ Dipartimento di Fisica dell'Universit\`{a} and Sezione INFN, Turin, Italy\\
$^{25}$ Dipartimento di Fisica e Astronomia dell'Universit\`{a} and Sezione INFN, Bologna, Italy\\
$^{26}$ Dipartimento di Fisica e Astronomia dell'Universit\`{a} and Sezione INFN, Catania, Italy\\
$^{27}$ Dipartimento di Fisica e Astronomia dell'Universit\`{a} and Sezione INFN, Padova, Italy\\
$^{28}$ Dipartimento di Fisica `E.R.~Caianiello' dell'Universit\`{a} and Gruppo Collegato INFN, Salerno, Italy\\
$^{29}$ Dipartimento DISAT del Politecnico and Sezione INFN, Turin, Italy\\
$^{30}$ Dipartimento di Scienze MIFT, Universit\`{a} di Messina, Messina, Italy\\
$^{31}$ Dipartimento Interateneo di Fisica `M.~Merlin' and Sezione INFN, Bari, Italy\\
$^{32}$ European Organization for Nuclear Research (CERN), Geneva, Switzerland\\
$^{33}$ Faculty of Electrical Engineering, Mechanical Engineering and Naval Architecture, University of Split, Split, Croatia\\
$^{34}$ Faculty of Nuclear Sciences and Physical Engineering, Czech Technical University in Prague, Prague, Czech Republic\\
$^{35}$ Faculty of Physics, Sofia University, Sofia, Bulgaria\\
$^{36}$ Faculty of Science, P.J.~\v{S}af\'{a}rik University, Ko\v{s}ice, Slovak Republic\\
$^{37}$ Faculty of Technology, Environmental and Social Sciences, Bergen, Norway\\
$^{38}$ Frankfurt Institute for Advanced Studies, Johann Wolfgang Goethe-Universit\"{a}t Frankfurt, Frankfurt, Germany\\
$^{39}$ Fudan University, Shanghai, China\\
$^{40}$ Gauhati University, Department of Physics, Guwahati, India\\
$^{41}$ Helmholtz-Institut f\"{u}r Strahlen- und Kernphysik, Rheinische Friedrich-Wilhelms-Universit\"{a}t Bonn, Bonn, Germany\\
$^{42}$ Helsinki Institute of Physics (HIP), Helsinki, Finland\\
$^{43}$ High Energy Physics Group,  Universidad Aut\'{o}noma de Puebla, Puebla, Mexico\\
$^{44}$ Horia Hulubei National Institute of Physics and Nuclear Engineering, Bucharest, Romania\\
$^{45}$ HUN-REN Wigner Research Centre for Physics, Budapest, Hungary\\
$^{46}$ Indian Institute of Technology Bombay (IIT), Mumbai, India\\
$^{47}$ Indian Institute of Technology Indore, Indore, India\\
$^{48}$ INFN, Laboratori Nazionali di Frascati, Frascati, Italy\\
$^{49}$ INFN, Sezione di Bari, Bari, Italy\\
$^{50}$ INFN, Sezione di Bologna, Bologna, Italy\\
$^{51}$ INFN, Sezione di Cagliari, Cagliari, Italy\\
$^{52}$ INFN, Sezione di Catania, Catania, Italy\\
$^{53}$ INFN, Sezione di Padova, Padova, Italy\\
$^{54}$ INFN, Sezione di Pavia, Pavia, Italy\\
$^{55}$ INFN, Sezione di Torino, Turin, Italy\\
$^{56}$ INFN, Sezione di Trieste, Trieste, Italy\\
$^{57}$ Inha University, Incheon, Republic of Korea\\
$^{58}$ Institute for Gravitational and Subatomic Physics (GRASP), Utrecht University/Nikhef, Utrecht, Netherlands\\
$^{59}$ Institute of Experimental Physics, Slovak Academy of Sciences, Ko\v{s}ice, Slovak Republic\\
$^{60}$ Institute of Physics, Homi Bhabha National Institute, Bhubaneswar, India\\
$^{61}$ Institute of Physics of the Czech Academy of Sciences, Prague, Czech Republic\\
$^{62}$ Institute of Space Science (ISS), Bucharest, Romania\\
$^{63}$ Institut f\"{u}r Kernphysik, Johann Wolfgang Goethe-Universit\"{a}t Frankfurt, Frankfurt, Germany\\
$^{64}$ Instituto de Ciencias Nucleares, Universidad Nacional Aut\'{o}noma de M\'{e}xico, Mexico City, Mexico\\
$^{65}$ Instituto de F\'{i}sica, Universidade Federal do Rio Grande do Sul (UFRGS), Porto Alegre, Brazil\\
$^{66}$ Instituto de F\'{\i}sica, Universidad Nacional Aut\'{o}noma de M\'{e}xico, Mexico City, Mexico\\
$^{67}$ iThemba LABS, National Research Foundation, Somerset West, South Africa\\
$^{68}$ Jeonbuk National University, Jeonju, Republic of Korea\\
$^{69}$ Korea Institute of Science and Technology Information, Daejeon, Republic of Korea\\
$^{70}$ Laboratoire de Physique Subatomique et de Cosmologie, Universit\'{e} Grenoble-Alpes, CNRS-IN2P3, Grenoble, France\\
$^{71}$ Lawrence Berkeley National Laboratory, Berkeley, California, United States\\
$^{72}$ Lund University Department of Physics, Division of Particle Physics, Lund, Sweden\\
$^{73}$ Marietta Blau Institute, Vienna, Austria\\
$^{74}$ Nagasaki Institute of Applied Science, Nagasaki, Japan\\
$^{75}$ Nara Women{'}s University (NWU), Nara, Japan\\
$^{76}$ National Centre for Nuclear Research, Warsaw, Poland\\
$^{77}$ National Institute of Science Education and Research, Homi Bhabha National Institute, Jatni, India\\
$^{78}$ National Nuclear Research Center, Baku, Azerbaijan\\
$^{79}$ National Research and Innovation Agency - BRIN, Jakarta, Indonesia\\
$^{80}$ Niels Bohr Institute, University of Copenhagen, Copenhagen, Denmark\\
$^{81}$ Nikhef, National institute for subatomic physics, Amsterdam, Netherlands\\
$^{82}$ Nuclear Physics Group, STFC Daresbury Laboratory, Daresbury, United Kingdom\\
$^{83}$ Nuclear Physics Institute of the Czech Academy of Sciences, Husinec-\v{R}e\v{z}, Czech Republic\\
$^{84}$ Oak Ridge National Laboratory, Oak Ridge, Tennessee, United States\\
$^{85}$ Ohio State University, Columbus, Ohio, United States\\
$^{86}$ Physics department, Faculty of science, University of Zagreb, Zagreb, Croatia\\
$^{87}$ Physics Department, Panjab University, Chandigarh, India\\
$^{88}$ Physics Department, University of Jammu, Jammu, India\\
$^{89}$ Physics Program and International Institute for Sustainability with Knotted Chiral Meta Matter (WPI-SKCM$^{2}$), Hiroshima University, Hiroshima, Japan\\
$^{90}$ Physikalisches Institut, Eberhard-Karls-Universit\"{a}t T\"{u}bingen, T\"{u}bingen, Germany\\
$^{91}$ Physikalisches Institut, Ruprecht-Karls-Universit\"{a}t Heidelberg, Heidelberg, Germany\\
$^{92}$ Physik Department, Technische Universit\"{a}t M\"{u}nchen, Munich, Germany\\
$^{93}$ Politecnico di Bari and Sezione INFN, Bari, Italy\\
$^{94}$ Research Division and ExtreMe Matter Institute EMMI, GSI Helmholtzzentrum f\"ur Schwerionenforschung GmbH, Darmstadt, Germany\\
$^{95}$ Saga University, Saga, Japan\\
$^{96}$ Saha Institute of Nuclear Physics, Homi Bhabha National Institute, Kolkata, India\\
$^{97}$ School of Physics and Astronomy, University of Birmingham, Birmingham, United Kingdom\\
$^{98}$ Secci\'{o}n F\'{\i}sica, Departamento de Ciencias, Pontificia Universidad Cat\'{o}lica del Per\'{u}, Lima, Peru\\
$^{99}$ SUBATECH, IMT Atlantique, Nantes Universit\'{e}, CNRS-IN2P3, Nantes, France\\
$^{100}$ Sungkyunkwan University, Suwon City, Republic of Korea\\
$^{101}$ Suranaree University of Technology, Nakhon Ratchasima, Thailand\\
$^{102}$ Technical University of Ko\v{s}ice, Ko\v{s}ice, Slovak Republic\\
$^{103}$ The Henryk Niewodniczanski Institute of Nuclear Physics, Polish Academy of Sciences, Cracow, Poland\\
$^{104}$ The University of Texas at Austin, Austin, Texas, United States\\
$^{105}$ Universidad Aut\'{o}noma de Sinaloa, Culiac\'{a}n, Mexico\\
$^{106}$ Universidade de S\~{a}o Paulo (USP), S\~{a}o Paulo, Brazil\\
$^{107}$ Universidade Estadual de Campinas (UNICAMP), Campinas, Brazil\\
$^{108}$ Universidade Federal do ABC, Santo Andre, Brazil\\
$^{109}$ Universitatea Nationala de Stiinta si Tehnologie Politehnica Bucuresti, Bucharest, Romania\\
$^{110}$ University of Cape Town, Cape Town, South Africa\\
$^{111}$ University of Derby, Derby, United Kingdom\\
$^{112}$ University of Houston, Houston, Texas, United States\\
$^{113}$ University of Jyv\"{a}skyl\"{a}, Jyv\"{a}skyl\"{a}, Finland\\
$^{114}$ University of Kansas, Lawrence, Kansas, United States\\
$^{115}$ University of Liverpool, Liverpool, United Kingdom\\
$^{116}$ University of Science and Technology of China, Hefei, China\\
$^{117}$ University of Silesia in Katowice, Katowice, Poland\\
$^{118}$ University of South-Eastern Norway, Kongsberg, Norway\\
$^{119}$ University of Tennessee, Knoxville, Tennessee, United States\\
$^{120}$ University of the Witwatersrand, Johannesburg, South Africa\\
$^{121}$ University of Tokyo, Tokyo, Japan\\
$^{122}$ University of Tsukuba, Tsukuba, Japan\\
$^{123}$ Universit\"{a}t M\"{u}nster, Institut f\"{u}r Kernphysik, M\"{u}nster, Germany\\
$^{124}$ Universit\'{e} Clermont Auvergne, CNRS/IN2P3, LPC, Clermont-Ferrand, France\\
$^{125}$ Universit\'{e} de Lyon, CNRS/IN2P3, Institut de Physique des 2 Infinis de Lyon, Lyon, France\\
$^{126}$ Universit\'{e} de Strasbourg, CNRS, IPHC UMR 7178, F-67000 Strasbourg, France, Strasbourg, France\\
$^{127}$ Universit\'{e} Paris-Saclay, Centre d'Etudes de Saclay (CEA), IRFU, D\'{e}partment de Physique Nucl\'{e}aire (DPhN), Saclay, France\\
$^{128}$ Universit\'{e}  Paris-Saclay, CNRS/IN2P3, IJCLab, Orsay, France\\
$^{129}$ Universit\`{a} degli Studi di Foggia, Foggia, Italy\\
$^{130}$ Universit\`{a} del Piemonte Orientale, Vercelli, Italy\\
$^{131}$ Universit\`{a} di Brescia, Brescia, Italy\\
$^{132}$ Variable Energy Cyclotron Centre, Homi Bhabha National Institute, Kolkata, India\\
$^{133}$ Warsaw University of Technology, Warsaw, Poland\\
$^{134}$ Wayne State University, Detroit, Michigan, United States\\
$^{135}$ Yale University, New Haven, Connecticut, United States\\
$^{136}$ Yildiz Technical University, Istanbul, Turkey\\
$^{137}$ Yonsei University, Seoul, Republic of Korea\\
$^{138}$ Affiliated with an institute formerly covered by a cooperation agreement with CERN\\
$^{139}$ Affiliated with an international laboratory covered by a cooperation agreement with CERN.\\

\end{flushleft} 
  
\end{document}